%% file: main.tex
\documentclass[runningheads]{llncs}

\usepackage[T1]{fontenc}
\usepackage{subcaption}

\usepackage{import}
\import{./lib/}{boilerplate_pdflatex.tex}

\import{./lib/}{colors.tex}

\import{./lib/}{tikzpgfplot.tex}
\import{./lib/}{tikzpgfplot_parulalinestyles.tex}

\import{./lib/}{tables.tex}

\import{./lib/}{algorithms.tex}
\import{./lib/}{__spacinghacks.tex}
\import{./lib/}{lateplate_pdflatex.tex}

\usepackage{color}

\usepackage{noto-emoji-easy}
\input{custom_commands}

\newcommand{\blfootnote}[1]{%
\begingroup%
\renewcommand\thefootnote{}\footnotetext{#1}%
\endgroup%
}%

\newboolean{fullVersion}
\setboolean{fullVersion}{true}

\usepackage{xr}
\usepackage{comment}

\ifthenelse{\boolean{fullVersion}}
{}
{\externaldocument{appendices}}

\title{Consensus Under Adversary Majority Done Right}
\titlerunning{Consensus Under Adversary Majority Done Right}
\author{Srivatsan Sridhar\inst{1}%
\and Ertem Nusret Tas\inst{1}%
\and Joachim Neu\inst{2}%
\and Dionysis Zindros\inst{1,3}%
\and David Tse\inst{1}%
}
\authorrunning{S. Sridhar, E. N. Tas, J. Neu, D. Zindros, D. Tse}
\institute{Stanford University\\\email{\{svatsan,nusret,dntse\}@stanford.edu}
\and a16z Crypto Research\\\email{jneu@a16z.com}
\and Common Prefix\\\email{dionyziz@commonprefix.com}}

\begin{document}
\maketitle
\blfootnote{%
\ifthenelse{\boolean{fullVersion}}%
{}%
{Due to space constraints, some lemmas, theorems, algorithms, proofs, etc.\ are not included in this short version, but are found in the appendices of the full version~\cite{fullversion}.}
SS and ENT contributed equally and are listed alphabetically.
A part of \cref{sec:sleepy-active} 
appeared in an earlier preprint~\cite{bettersafethansorry} by SS, DZ, and DT.}%
\begin{abstract}
\import{./sections/}{00_abstract.tex}
\end{abstract}

\import{./sections/}{01_introduction.tex}
\import{./sections/}{03_model.tex}
\section{Synchrony with \StatPart}
\label{sec:synchrony}
\import{./sections/}{04_passive.tex}
\import{./sections/}{05_sleepy_active.tex}
\import{./sections/}{06_alert_active.tex}
\import{./sections/}{07_sleepy_validators.tex}
\import{./sections/}{02_related_work.tex}
\ifthenelse{\boolean{fullVersion}}
{
\subsubsection*{\ackname}
SS, ENT, and DT are funded by a Research Hub Collaboration agreement between Stanford University and Input Output Global Inc.
ENT is supported by the Stanford Center for Blockchain Research.
We thank 
Zeta Avarikioti, 
Christian Cachin, 
Jacob Leshno,
Orfeas Stefanos Thyfronitis Litos,
Tim Roughgarden,
Giulia Scaffino, 
Elaine Shi,
and 
Roger Wattenhofer
for fruitful discussions and feedback.
}
{}
\nocite{fullversion}
\bibliographystyle{splncs04}
\bibliography{references.bib}

\ifthenelse{\boolean{fullVersion}}
{
    \input{appendices.tex}

}
{
    \appendix
    \section*{Appendices}
    Full version with appendices: \url{https://eprint.iacr.org/2024/1799}    
    \cite{fullversion}
}

\end{document}

%% file: lib/boilerplate_pdflatex.tex
\usepackage[utf8]{inputenc}

\input{boilerplate_base.tex}

%% file: lib/boilerplate_base.tex
\PassOptionsToPackage{hyphens}{url}%
\usepackage{hyperref}
\usepackage{graphicx}

\usepackage{amsmath}

\usepackage{amsthm}
\usepackage{amssymb}
\usepackage{amsfonts}
\usepackage{mathtools}

\usepackage{shortsym}

\usepackage{IEEEtrantools}
\allowdisplaybreaks

\usepackage[shortlabels,inline]{enumitem}
\setlist[itemize]{leftmargin=1.25em}
\setlist[enumerate]{leftmargin=1.75em}
\usepackage{xstring}   %
\usepackage[normalem]{ulem}

\usepackage{import}
\subimport{.}{boilerplate_base_hyphenations.tex}

\subimport{.}{boilerplate_base_mathenvironments.tex}

\subimport{.}{boilerplate_base_math.tex}

\subimport{.}{boilerplate_base_symbols.tex}

\subimport{.}{boilerplate_base_textdecorations.tex}
\subimport{.}{boilerplate_base_textabbreviations.tex}

%% file: lib/boilerplate_base_hyphenations.tex
\hyphenation{a-vail-a-bil-i-ty}

\hyphenation{ac-count-a-bil-i-ty}
\hyphenation{ac-count-a-ble}
\hyphenation{ac-count-a-bly}

\hyphenation{led-ger}

\hyphenation{Na-ka-mo-to}

\hyphenation{white-pa-per}

\hyphenation{equi-vo-ca-tion}
\hyphenation{equi-vo-ca-ting}

%% file: lib/boilerplate_base_mathenvironments.tex
\theoremstyle{plain}
\newtheorem*{theorem*}{Theorem}
\newtheorem*{corollary*}{Corollary}
\newtheorem*{lemma*}{Lemma}
\newtheorem*{proposition*}{Proposition}
\newtheorem*{conjecture*}{Conjecture}
\newtheorem*{result*}{Result}

\theoremstyle{definition}
\makeatletter
\renewcommand*\env@matrix[1][*\c@MaxMatrixCols c]{%
    \hskip -\arraycolsep
    \let\@ifnextchar\new@ifnextchar
    \array{#1}}
\makeatother

%% file: lib/boilerplate_base_math.tex
\DeclarePairedDelimiter\abs{\lvert}{\rvert}
\DeclarePairedDelimiter\len{\lvert}{\rvert}
\DeclarePairedDelimiter\norm{\lVert}{\rVert}

\makeatletter
\let\oldabs\abs
\def\abs{\@ifstar{\oldabs}{\oldabs*}}
\let\oldlen\len
\def\len{\@ifstar{\oldlen}{\oldlen*}}
\let\oldnorm\norm
\def\norm{\@ifstar{\oldnorm}{\oldnorm*}}
\makeatother

%% file: lib/boilerplate_base_symbols.tex
\usepackage{pifont}

\usepackage{dsfont}

%% file: lib/boilerplate_base_textabbreviations.tex
\usepackage{xspace}

\newcommand{\cf}[0]{cf.\xspace}
\newcommand{\ie}[0]{\emph{i.e.}\xspace}
\newcommand{\eg}[0]{\emph{e.g.}\xspace}
\newcommand{\vs}[0]{vs.\xspace}
\newcommand{\aka}[0]{a.k.a.\xspace}

%% file: lib/colors.tex
\usepackage{xcolor}

\definecolor{jnSUCardinalRed}{HTML}{8c1515}
\definecolor{jnSUCardinalRedLight}{HTML}{B83A4B}
\definecolor{jnSUCardinalRedDark}{HTML}{820000}
\definecolor{jnSUWhite}{HTML}{ffffff}
\definecolor{jnSUCoolGrey}{HTML}{53565A}
\definecolor{jnSUBlack}{HTML}{2e2d29}
\definecolor{jnSUBlack100}{HTML}{2e2d29}
\definecolor{jnSUBlack90}{HTML}{43423E}
\definecolor{jnSUBlack80}{HTML}{585754}
\definecolor{jnSUBlack70}{HTML}{6D6C69}
\definecolor{jnSUBlack60}{HTML}{767674}
\definecolor{jnSUBlack50}{HTML}{979694}
\definecolor{jnSUBlack40}{HTML}{ABABA9}
\definecolor{jnSUBlack30}{HTML}{C0C0BF}
\definecolor{jnSUBlack20}{HTML}{D5D5D4}
\definecolor{jnSUBlack10}{HTML}{EAEAEA}

\definecolor{jnSUPaloAlto}{HTML}{175E54}
\definecolor{jnSUPaloAltoLight}{HTML}{2D716F}
\definecolor{jnSUPaloAltoDark}{HTML}{014240}
\definecolor{jnSUPaloVerde}{HTML}{279989}
\definecolor{jnSUPaloVerdeLight}{HTML}{59B3A9}
\definecolor{jnSUPaloVerdeDark}{HTML}{017E7C}
\definecolor{jnSUOlive}{HTML}{8F993E}
\definecolor{jnSUOliveLight}{HTML}{A6B168}
\definecolor{jnSUOliveDark}{HTML}{7A863B}
\definecolor{jnSUBay}{HTML}{6FA287}
\definecolor{jnSUBayLight}{HTML}{8AB8A7}
\definecolor{jnSUBayDark}{HTML}{417865}
\definecolor{jnSUSky}{HTML}{4298B5}
\definecolor{jnSUSkyLight}{HTML}{67AFD2}
\definecolor{jnSUSkyDark}{HTML}{016895}
\definecolor{jnSULagunita}{HTML}{007C92}
\definecolor{jnSULagunitaLight}{HTML}{009AB4}
\definecolor{jnSULagunitaDark}{HTML}{006B81}
\definecolor{jnSUPoppy}{HTML}{E98300}
\definecolor{jnSUPoppyLight}{HTML}{F9A44A}
\definecolor{jnSUPoppyDark}{HTML}{D1660F}
\definecolor{jnSUSpirited}{HTML}{E04F39}
\definecolor{jnSUSpiritedLight}{HTML}{F4795B}
\definecolor{jnSUSpiritedDark}{HTML}{C74632}
\definecolor{jnSUIlluminating}{HTML}{FEDD5C}
\definecolor{jnSUIlluminatingLight}{HTML}{FFE781}
\definecolor{jnSUIlluminatingDark}{HTML}{FEC51D}
\definecolor{jnSUPlum}{HTML}{620059}
\definecolor{jnSUPlumLight}{HTML}{734675}
\definecolor{jnSUPlumDark}{HTML}{350D36}
\definecolor{jnSUBrick}{HTML}{651C32}
\definecolor{jnSUBrickLight}{HTML}{7F2D48}
\definecolor{jnSUBrickDark}{HTML}{42081B}
\definecolor{jnSUArchway}{HTML}{5D4B3C}
\definecolor{jnSUArchwayLight}{HTML}{766253}
\definecolor{jnSUArchwayDark}{HTML}{2F2424}
\definecolor{jnSUStone}{HTML}{7F7776}
\definecolor{jnSUStoneLight}{HTML}{D4D1D1}
\definecolor{jnSUStoneDark}{HTML}{544948}
\definecolor{jnSUFog}{HTML}{DAD7CB}
\definecolor{jnSUFogLight}{HTML}{F4F4F4}
\definecolor{jnSUFogDark}{HTML}{B6B1A9}

\definecolor{jnSUDigitalRed}{HTML}{B1040E}
\definecolor{jnSUDigitalRedLight}{HTML}{E50808}
\definecolor{jnSUDigitalRedDark}{HTML}{820000}
\definecolor{jnSUDigitalBlue}{HTML}{006CB8}
\definecolor{jnSUDigitalBlueLight}{HTML}{6FC3FF}
\definecolor{jnSUDigitalBlueDark}{HTML}{00548f}
\definecolor{jnSUDigitalGreen}{HTML}{008566}
\definecolor{jnSUDigitalGreenLight}{HTML}{1AECBA}
\definecolor{jnSUDigitalGreenDark}{HTML}{006F54}

\definecolor{myParula01Blue}{RGB}{0,114,189}
\definecolor{myParula02Orange}{RGB}{217,83,25}
\definecolor{myParula03Yellow}{RGB}{237,177,32}
\definecolor{myParula04Purple}{RGB}{126,47,142}
\definecolor{myParula05Green}{RGB}{119,172,48}
\definecolor{myParula06LightBlue}{RGB}{77,190,238}
\definecolor{myParula07Red}{RGB}{162,20,47}

%% file: lib/tikzpgfplot.tex
\usepackage{tikz}
\usetikzlibrary{calc}
\usetikzlibrary{arrows}
\usetikzlibrary{arrows.meta}
\usetikzlibrary{patterns}
\usetikzlibrary{positioning}
\usetikzlibrary{decorations.pathreplacing}
\usetikzlibrary{calligraphy}
\usetikzlibrary{shapes.misc}
\usetikzlibrary{shapes.symbols}
\usetikzlibrary{shapes.arrows}
\usetikzlibrary{spy}
\usetikzlibrary{shapes.geometric}
\usetikzlibrary{intersections}
\usetikzlibrary{fit}

\usepackage{pgfplots}
\pgfplotsset{compat=1.17}
\usepgfplotslibrary{fillbetween}

\pgfplotsset{
    discard if not/.style 2 args={
            x filter/.code={
                    \edef\tempa{\thisrow{#1}}
                    \edef\tempb{#2}
                    \ifx\tempa\tempb
                    \else
                        
                    \fi
                }
        },
}

%% file: lib/tikzpgfplot_parulalinestyles.tex
\tikzset{myparula11/.style={color=myParula01Blue,solid,mark=+,mark options={solid}}}
\tikzset{myparula12/.style={color=myParula01Blue,densely dashed,mark=x,mark options={solid}}}
\tikzset{myparula13/.style={color=myParula01Blue,densely dotted,mark=o,mark options={solid}}}
\tikzset{myparula14/.style={color=myParula01Blue,dashdotted,mark=triangle,mark options={solid}}}
\tikzset{myparula15/.style={color=myParula01Blue,dashdotdotted,mark=square,mark options={solid}}}

\tikzset{myparula21/.style={color=myParula02Orange,solid,mark=+,mark options={solid}}}
\tikzset{myparula22/.style={color=myParula02Orange,densely dashed,mark=x,mark options={solid}}}
\tikzset{myparula23/.style={color=myParula02Orange,densely dotted,mark=o,mark options={solid}}}
\tikzset{myparula24/.style={color=myParula02Orange,dashdotted,mark=triangle,mark options={solid}}}
\tikzset{myparula25/.style={color=myParula02Orange,dashdotdotted,mark=square,mark options={solid}}}

\tikzset{myparula31/.style={color=myParula03Yellow,solid,mark=+,mark options={solid}}}
\tikzset{myparula32/.style={color=myParula03Yellow,densely dashed,mark=x,mark options={solid}}}
\tikzset{myparula33/.style={color=myParula03Yellow,densely dotted,mark=o,mark options={solid}}}
\tikzset{myparula34/.style={color=myParula03Yellow,dashdotted,mark=triangle,mark options={solid}}}
\tikzset{myparula35/.style={color=myParula03Yellow,dashdotdotted,mark=square,mark options={solid}}}

\tikzset{myparula41/.style={color=myParula04Purple,solid,mark=+,mark options={solid}}}
\tikzset{myparula42/.style={color=myParula04Purple,densely dashed,mark=x,mark options={solid}}}
\tikzset{myparula43/.style={color=myParula04Purple,densely dotted,mark=o,mark options={solid}}}
\tikzset{myparula44/.style={color=myParula04Purple,dashdotted,mark=triangle,mark options={solid}}}
\tikzset{myparula45/.style={color=myParula04Purple,dashdotdotted,mark=square,mark options={solid}}}

\tikzset{myparula51/.style={color=myParula05Green,solid,mark=+,mark options={solid}}}
\tikzset{myparula52/.style={color=myParula05Green,densely dashed,mark=x,mark options={solid}}}
\tikzset{myparula53/.style={color=myParula05Green,densely dotted,mark=o,mark options={solid}}}
\tikzset{myparula54/.style={color=myParula05Green,dashdotted,mark=triangle,mark options={solid}}}
\tikzset{myparula55/.style={color=myParula05Green,dashdotdotted,mark=square,mark options={solid}}}

\tikzset{myparula61/.style={color=myParula06LightBlue,solid,mark=+,mark options={solid}}}
\tikzset{myparula62/.style={color=myParula06LightBlue,densely dashed,mark=x,mark options={solid}}}
\tikzset{myparula63/.style={color=myParula06LightBlue,densely dotted,mark=o,mark options={solid}}}
\tikzset{myparula64/.style={color=myParula06LightBlue,dashdotted,mark=triangle,mark options={solid}}}
\tikzset{myparula65/.style={color=myParula06LightBlue,dashdotdotted,mark=square,mark options={solid}}}

\tikzset{myparula71/.style={color=myParula07Red,solid,mark=+,mark options={solid}}}
\tikzset{myparula72/.style={color=myParula07Red,densely dashed,mark=x,mark options={solid}}}
\tikzset{myparula73/.style={color=myParula07Red,densely dotted,mark=o,mark options={solid}}}
\tikzset{myparula74/.style={color=myParula07Red,dashdotted,mark=triangle,mark options={solid}}}
\tikzset{myparula75/.style={color=myParula07Red,dashdotdotted,mark=square,mark options={solid}}}

%% file: lib/tables.tex
\usepackage{multirow,booktabs}
\usepackage{makecell}

%% file: lib/algorithms.tex
\usepackage{algorithm}
\usepackage{algorithmicx}
\usepackage[noend]{algpseudocode}

\algnewcommand{\algorithmicswitch}{\textbf{switch}}
\algdef{SE}[SWITCH]{Switch}{EndSwitch}[1]{\algorithmicswitch\ #1\ \algorithmicdo}{\algorithmicend\ \algorithmicswitch}%
\algtext*{EndSwitch}%

\algnewcommand{\algorithmiccase}{\textbf{case}}
\algdef{SE}[CASE]{Case}{EndCase}[1]{\algorithmiccase\ #1}{\algorithmicend\ \algorithmiccase}%
\algtext*{EndCase}%

\algnewcommand{\algorithmicon}{\textbf{on}}
\algdef{SE}[ON]{On}{EndOn}[1]{\algorithmicon\ #1\ \algorithmicdo}{\algorithmicend\ \algorithmicon}%
\algtext*{EndOn}%

\algnewcommand{\algorithmicat}{\textbf{at}}
\algdef{SE}[AT]{At}{EndAt}[1]{\algorithmicat\ #1\ \algorithmicdo}{\algorithmicend\ \algorithmicat}%
\algtext*{EndAt}%

\algnewcommand{\algorithmicrealfunction}{\textbf{function}}
\algdef{SE}[REALFUNCTION]{RealFunction}{EndRealFunction}[1]{\algorithmicrealfunction\ #1\ \algorithmicdo}{\algorithmicend\ \algorithmicrealfunction}%
\algtext*{EndRealFunction}%

\algnewcommand{\algorithmicthroughout}{\textbf{do throughout}}
\algdef{SE}[Throughout]{Throughout}{EndThroughout}[1]{\algorithmicthroughout\ #1\ \algorithmicdo}{\algorithmicend\ \algorithmicthroughout}%
\algtext*{EndThroughout}%

\algrenewcommand{\algorithmicdo}{}
\algrenewcommand{\algorithmicthen}{}

\algnewcommand{\algorithmicgoto}{\textbf{goto}}%
\algnewcommand{\Goto}[1]{\algorithmicgoto~\ref{#1}}%

\algnewcommand{\algorithmicassert}{\textbf{assert}}%
\algnewcommand{\Assert}[1]{\algorithmicassert~{#1}}%

\algnewcommand{\algorithmicbreak}{\textbf{break}}%
\algnewcommand{\Break}[0]{\algorithmicbreak}%

\algnewcommand{\algorithmicwaiton}{\textbf{wait on}}%
\algnewcommand{\WaitOn}[1]{\algorithmicwaiton~{#1}}%

\algnewcommand{\LineComment}[1]{\State \(\triangleright\) \textit{#1}}
\algnewcommand{\InlineRequire}[1]{\textbf{require} {#1}}

\algrenewcommand\alglinenumber[1]{\scriptsize\textcolor{gray}{#1}}

%% file: lib/__spacinghacks.tex
\setlength{\floatsep}{0.65\baselineskip}
\setlength{\textfloatsep}{1.15\baselineskip}
\setlength{\dblfloatsep}{0.65\baselineskip}
\setlength{\dbltextfloatsep}{1\baselineskip}

\widowpenalty=0
\clubpenalty=0
\brokenpenalty=0

\displaywidowpenalty=0
\setlength{\emergencystretch}{1em}   %

\captionsetup[figure]{font=footnotesize,skip=4pt}

%% file: lib/lateplate_pdflatex.tex
\input{lateplate_base.tex}

%% file: lib/lateplate_base.tex
\makeatletter
\AddToHook{env/algorithmic/begin}{\def\@currentcounter{ALG@line}}
\renewcommand*\theHALG@line{\thealgorithm.\arabic{ALG@line}}
\makeatother

\usepackage[sort&compress,capitalize,nameinlink]{cleveref}

\AtBeginEnvironment{appendices}{%
    \crefalias{section}{appendix}%
    \crefalias{subsection}{subappendix}%
    \crefalias{subsubsection}{subsubappendix}%
    \crefalias{subsubsubsection}{subsubsubappendix}%
}
\AddToHook{cmd/appendix/after}{%
    \crefalias{section}{appendix}%
    \crefalias{subsection}{subappendix}%
    \crefalias{subsubsection}{subsubappendix}%
    \crefalias{subsubsubsection}{subsubsubappendix}%
}

\hypersetup{hypertexnames=false}

\crefalias{ALG@line}{line}

\crefname{figure}{Fig.}{Figs.}
\Crefname{figure}{Fig.}{Figs.}

\crefname{table}{Tab.}{Tabs.}
\Crefname{table}{Tab.}{Tabs.}

\crefname{section}{Sec.}{Secs.}
\Crefname{section}{Sec.}{Secs.}
\crefname{subsection}{Sec.}{Secs.}
\Crefname{subsection}{Sec.}{Secs.}
\crefname{subsubsection}{Sec.}{Secs.}
\Crefname{subsubsection}{Sec.}{Secs.}
\crefname{subsubsubsection}{Sec.}{Secs.}
\Crefname{subsubsubsection}{Sec.}{Secs.}
\crefname{appendix}{App.}{Apps.}
\Crefname{appendix}{App.}{Apps.}
\crefname{subappendix}{App.}{Apps.}
\Crefname{subappendix}{App.}{Apps.}
\crefname{subsubappendix}{App.}{Apps.}
\Crefname{subsubappendix}{App.}{Apps.}
\crefname{subsubsubappendix}{App.}{Apps.}
\Crefname{subsubsubappendix}{App.}{Apps.}

\crefname{algorithm}{Alg.}{Algs.}
\Crefname{algorithm}{Alg.}{Algs.}
\crefname{line}{l.}{ll.}
\Crefname{line}{l.}{ll.}

\crefname{proposition}{Prop.}{Props.}
\Crefname{proposition}{Prop.}{Props.}
\crefname{lemma}{Lem.}{Lems.}
\Crefname{lemma}{Lem.}{Lems.}
\crefname{theorem}{Thm.}{Thms.}
\Crefname{theorem}{Thm.}{Thms.}
\crefname{corollary}{Cor.}{Cors.}
\Crefname{corollary}{Cor.}{Cors.}
\crefname{definition}{Def.}{Defs.}
\Crefname{definition}{Def.}{Defs.}

%% file: custom_commands.tex
\newcommand{\myparagraph}[1]{\myparagraphNoDot{#1.}}
\newcommand{\mysubparagraph}[1]{\mysubparagraphNoDot{#1.}}
\newcommand{\myparagraphNoDot}[1]{\medskip\noindent\textbf{#1}~~}
\newcommand{\mysubparagraphNoDot}[1]{\smallskip\noindent\emph{#1}~~}

\newcommand{\alert}{always-on\xspace}
\newcommand{\sleepy}{sleepy\xspace}
\newcommand{\activ}{communicating\xspace}
\newcommand{\passive}{silent\xspace}
\newcommand{\dynpart}{sleepy validators\xspace}
\newcommand{\statpart}{always-on validators\xspace}

\newcommand{\smr}{state-machine replication\xspace}
\newcommand{\smrshort}{SMR\xspace}
\newcommand{\Alert}{Always-on\xspace}
\newcommand{\Sleepy}{Sleepy\xspace}
\newcommand{\Activ}{Communicating\xspace}
\newcommand{\Passive}{Silent\xspace}
\newcommand{\Dynpart}{Sleepy validators\xspace}
\newcommand{\Statpart}{Always-on validators\xspace}

\newcommand{\ALert}{Always-On\xspace}
\newcommand{\SLeepy}{Sleepy\xspace}
\newcommand{\ACtiv}{Communicating\xspace}
\newcommand{\PAssive}{Silent\xspace}
\newcommand{\DynPart}{Sleepy Validators\xspace}
\newcommand{\StatPart}{Always-On Validators\xspace}

\newcommand{\wokeness}{sleepiness\xspace}
\newcommand{\Wokeness}{Sleepiness\xspace}
\newcommand{\interactivity}{interactivity\xspace}
\newcommand{\Interactivity}{Interactivity\xspace}

\newcommand{\Weshow}{We show\xspace}
\newcommand{\weshow}{we show\xspace}

\newcommand{\TRUE}[0]{\ensuremath{\mathtt{true}}}
\newcommand{\FALSE}[0]{\ensuremath{\mathtt{false}}}

\newcommand{\iconvalidatoralertpsync}{\ensuremath{%
\tikz{ \node [inner sep=0pt] at (0,0) {$\Delta$}; \draw [red,thick] ([xshift=-0.4em,yshift=0.4em] 0,0) -- ([xshift=0.4em,yshift=-0.4em] 0,0); }%
\text{\notoemoji[height=1em]{desktop computer}}%
^{\text{\textcolor{black}{\textnormal{\textbf{!!!}}}}}%
}}
\newcommand{\iconvalidatoralertsync}{\ensuremath{%
\tikz{ \draw [opacity=0,thick] ([xshift=-0.4em,yshift=0.4em] 0,0) -- ([xshift=0.4em,yshift=-0.4em] 0,0); \node [inner sep=0pt] at (0,0) {$\Delta$}; }%
\text{\notoemoji[height=1em]{desktop computer}}%
^{\text{\textcolor{black}{\textnormal{\textbf{!!!}}}}}%
}}
\newcommand{\iconvalidatorsleepypsync}{\ensuremath{%
\tikz{ \node [inner sep=0pt] at (0,0) {$\Delta$}; \draw [red,thick] ([xshift=-0.4em,yshift=0.4em] 0,0) -- ([xshift=0.4em,yshift=-0.4em] 0,0); }%
\text{\notoemoji[height=1em]{desktop computer}}%
^{\text{\textcolor{red}{\textnormal{\textbf{zZ}}}}}%
}}
\newcommand{\iconvalidatorsleepysync}{\ensuremath{%
\tikz{ \draw [opacity=0,thick] ([xshift=-0.4em,yshift=0.4em] 0,0) -- ([xshift=0.4em,yshift=-0.4em] 0,0); \node [inner sep=0pt] at (0,0) {$\Delta$}; }%
\text{\notoemoji[height=1em]{desktop computer}}%
^{\text{\textcolor{red}{\textnormal{\textbf{zZ}}}}}%
}}
\newcommand{\iconclientalertactiv}{\ensuremath{%
\text{\notoemoji[height=1em]{person}}%
_{\text{\tikz{ \draw [opacity=0,thick] ([xshift=-0.4em,yshift=0.4em] 0,0) -- ([xshift=0.4em,yshift=-0.4em] 0,0); \node [inner sep=0pt] at (0,0) {\notoemoji[height=1em]{speech balloon}}; }}}%
^{\text{\textcolor{black}{\textnormal{\textbf{!!!}}}}}%
}}
\newcommand{\iconclientalertpassive}{\ensuremath{%
\text{\notoemoji[height=1em]{person}}%
_{\text{\tikz{ \node [inner sep=0pt] at (0,0) {\notoemoji[height=1em]{speech balloon}}; \draw [red,thick] ([xshift=-0.4em,yshift=0.4em] 0,0) -- ([xshift=0.4em,yshift=-0.4em] 0,0); }}}%
^{\text{\textcolor{black}{\textnormal{\textbf{!!!}}}}}%
}}
\newcommand{\iconclientsleepyactiv}{\ensuremath{%
\text{\notoemoji[height=1em]{person}}%
_{\text{\tikz{ \draw [opacity=0,thick] ([xshift=-0.4em,yshift=0.4em] 0,0) -- ([xshift=0.4em,yshift=-0.4em] 0,0); \node [inner sep=0pt] at (0,0) {\notoemoji[height=1em]{speech balloon}}; }}}%
^{\text{\textcolor{red}{\textnormal{\textbf{zZ}}}}}%
}}
\newcommand{\iconclientsleepypassive}{\ensuremath{%
\text{\notoemoji[height=1em]{person}}%
_{\text{\tikz{ \node [inner sep=0pt] at (0,0) {\notoemoji[height=1em]{speech balloon}}; \draw [red,thick] ([xshift=-0.4em,yshift=0.4em] 0,0) -- ([xshift=0.4em,yshift=-0.4em] 0,0); }}}%
^{\text{\textcolor{red}{\textnormal{\textbf{zZ}}}}}%
}}
\newcommand{\iconvalidatoralert}{\ensuremath{%
\text{\notoemoji[height=1em]{desktop computer}}%
^{\text{\textcolor{black}{\textnormal{\textbf{!!!}}}}}%
}}
\newcommand{\iconvalidatorpsync}{\ensuremath{%
\tikz{ \node [inner sep=0pt] at (0,0) {$\Delta$}; \draw [red,thick] ([xshift=-0.4em,yshift=0.4em] 0,0) -- ([xshift=0.4em,yshift=-0.4em] 0,0); }%
}}
\newcommand{\iconvalidatorsleepy}{\ensuremath{%
\text{\notoemoji[height=1em]{desktop computer}}%
^{\text{\textcolor{red}{\textnormal{\textbf{zZ}}}}}%
}}
\newcommand{\iconvalidatorsync}{\ensuremath{%
\tikz{ \draw [opacity=0,thick] ([xshift=-0.4em,yshift=0.4em] 0,0) -- ([xshift=0.4em,yshift=-0.4em] 0,0); \node [inner sep=0pt] at (0,0) {$\Delta$}; }%
}}
\newcommand{\iconclientalert}{\ensuremath{%
\text{\notoemoji[height=1em]{person}}%
^{\text{\textcolor{black}{\textnormal{\textbf{!!!}}}}}%
}}
\newcommand{\iconclientsleepy}{\ensuremath{%
\text{\notoemoji[height=1em]{person}}%
^{\text{\textcolor{red}{\textnormal{\textbf{zZ}}}}}%
}}
\newcommand{\iconclientactiv}{\ensuremath{%
\text{\notoemoji[height=1em]{person}}%
_{\text{\tikz{ \draw [opacity=0,thick] ([xshift=-0.4em,yshift=0.4em] 0,0) -- ([xshift=0.4em,yshift=-0.4em] 0,0); \node [inner sep=0pt] at (0,0) {\notoemoji[height=1em]{speech balloon}}; }}}%
}}
\newcommand{\iconclientpassive}{\ensuremath{%
\text{\notoemoji[height=1em]{person}}%
_{\text{\tikz{ \node [inner sep=0pt] at (0,0) {\notoemoji[height=1em]{speech balloon}}; \draw [red,thick] ([xshift=-0.4em,yshift=0.4em] 0,0) -- ([xshift=0.4em,yshift=-0.4em] 0,0); }}}%
}}

\newcommand{\LOG}[0]{\ensuremath{L}}
\newcommand{\LOGi}[2]{\ensuremath{\pmb{L}_{#1}^{#2}}}

\newcommand{\consistent}{\ensuremath{\sim}}
\newcommand{\inconsistent}{\ensuremath{\not\sim}}

\newcommand{\tx}[0]{\ensuremath{\mathsf{tx}}}
\newcommand{\txs}[0]{\ensuremath{\mathsf{txs}}}

\newcommand{\valset}[0]{\ensuremath{\mathcal{N}}}
\newcommand{\awakeset}[1]{\ensuremath{\mathsf{awake}_{#1}}}

\newcommand{\corruptset}[1]{\ensuremath{\mathsf{corrupt}_{#1}}}

\newcommand{\genesis}[0]{\ensuremath{\LOG_{\mathrm{genesis}}}}

\newcommand{\Adv}[0]{\CA}

\newcommand{\tL}{\ensuremath{t^{\mathrm{L}}}}
\newcommand{\tS}{\ensuremath{t^{\mathrm{S}}}}
\newcommand{\tC}{\ensuremath{t^{\mathrm{S}}}}
\newcommand{\betaL}{\ensuremath{\tL}}
\newcommand{\betaS}{\ensuremath{\tS}}
\newcommand{\rmaj}{\ensuremath{r_{\mathrm{maj}}}}

\newcommand{\PI}{\ensuremath{\Pi_{\mathrm{frz}}}}
\newcommand{\Piint}{\ensuremath{\Pi_{\mathrm{int}}}}
\newcommand{\LOGint}{\ensuremath{\LOG_{\mathrm{int}}}}

\newcommand{\GST}{\ensuremath{\mathsf{GST}}}

\newcommand{\ch}{\ensuremath{\mathsf{ch}}}
\newcommand{\Pisal}{\Pi_{\mathrm{live}}^*}

\newcommand{\transcribe}{\ensuremath{\mathcal{W}}}
\newcommand{\untranscribe}{\ensuremath{\mathcal{C}}}
\newcommand{\transcript}{%
  \ifmmode%
    C%
  \else%
    certificate\xspace %
  \fi%
}
\newcommand{\transcripts}{certificates\xspace}

\newcommand{\BGbroadcast}{BG-broadcast\xspace}
\newcommand{\BGbroadcasts}{BG-broadcasts\xspace}
\newcommand{\BGoutput}{BG-output\xspace}
\newcommand{\BGoutputs}{BG-outputs\xspace}

\newcommand{\msgSet}[0]{\ensuremath{\CS}}

\newcommand{\signedmsg}[2]{\ensuremath{\langle #2 \rangle_{#1}}}
\newcommand{\Vout}{\ensuremath{V_{\mathrm{out}}}}
\newcommand{\id}[0]{\mathsf{id}}

\newcommand*\DiagramStep[1]{\tikz[baseline=(char.south)]{
                \node[shape=circle,draw=none,fill=black,text=white,inner sep=0.75pt] (char) {\tiny\textbf{#1}}; }}

\algnewcommand{\Let}[2]{\State $#1 \gets #2$}

\algnewcommand{\algorithmicforever}{\textbf{forever}}
\algdef{SE}[FOREVER]{Forever}{EndForever}[0]{\algorithmicforever\ \algorithmicdo}{\algorithmicend\ \algorithmicforever}%
\algtext*{EndForever}%

\algnewcommand{\CommentLine}[1]{\State $\triangleright$ #1}

%% file: sections/00_abstract.tex
A specter is haunting consensus protocols---the specter of adversary majority.
Dolev and Strong in 1983 showed an early possibility for up to 99\%
adversaries.
Yet, other works show
impossibility results for
adversaries above 50\% under synchrony, 
seemingly the same setting as Dolev and Strong's.
What
gives? 
It is high time that we pinpoint 
a key
culprit for 
this ostensible contradiction:
the modeling details of \emph{clients}.
Are the clients
\emph{\sleepy} or \emph{\alert}? Are they \emph{\passive} or \emph{\activ}? 
Can validators be \emph{\sleepy} too?
We systematize models for consensus 
across four
dimensions (\sleepy/\alert clients, \passive/\activ clients, \sleepy/\alert validators, and synchrony/partial-synchrony),
some of which are new,
and tightly characterize the achievable safety and liveness resiliences with matching possibilities and impossibilities
for each of the sixteen models.
To this end, we unify folklore and earlier results,
and fill gaps left in the literature with new protocols and impossibility theorems.

%% file: sections/01_introduction.tex
\section{Introduction}
\label{section:introduction}

What fraction of parties running a protocol can be controlled
by an adversary while guaranteeing security? 
When it comes to \emph{Byzantine state-machine replication (SMR) consensus}
protocols, where security means safety and liveness,
the landscape is seemingly contradictory.
The oft-cited Dolev--Strong protocol~\cite{dolev-strong}, along with recent works~\cite{bcube,flint-adv-maj,vitalik-dolev-strong}
that extend it from broadcast to SMR consensus,
tolerates up to 99\% adversary parties (we say that it has 99\% \emph{resilience}).
This stands in contrast to the, perhaps equally-oft-cited, ``51\% attack''~\cite{51-attack} that renders many blockchains based on Nakamoto’s longest chain consensus protocol~\cite{david2018ouroboros,snowwhite}
insecure as soon as more than 50\% of the parties are corrupted.
What is more,
prior work (\eg, ~\cite[Thm.\ 3]{shi-rethinking}) claims 
that no protocol can achieve 50\% or higher resilience.
This impossibility 
holds
even if we assume
a known set of parties,
strong cryptographic functionalities like a public-key infrastructure (PKI), allow for randomized protocols, and consider a network model where the messages arrive within a known delay-bound (\ie, synchrony)%
\footnote{%
We repeat the proof of \cite[Thm.\ 3]{shi-rethinking} for this precise setting in \Cref{thm:sleepy-active-synchrony-converse}.%
}; thus, seemingly the same (or stronger) assumptions under which Dolev--Strong claims its 99\% resilience.
How do we explain the difference and resolve the conundrum?

\mysubparagraph{%
The Crucial Role of Clients%
}
The impossibility~\cite[Thm.\ 3]{shi-rethinking} crucially relies on the requirement that parties who join late, \ie, did not observe the protocol since its start, should still be able to output an up-to-date log of confirmed transactions.
In contrast, the 99\% resilience of Dolev--Strong~\cite{dolev-strong} and similar protocols~\cite{bcube,flint-adv-maj,vitalik-dolev-strong} is only guaranteed for parties who are always online, \ie, constantly monitor the protocol from its start to continuously learn the correct 
confirmed transaction sequence.
This key differentiating factor pertains specifically to the parties who attempt to learn the protocol's output, \ie, the \emph{clients}.
In blockchains, these clients may be merchants who monitor the chain for payments and ship merchandise in response.
They are usually
not actively running the protocol to maintain consensus---that is done by the \emph{validators} who are selected for instance through a mechanism such as proof-of-stake.%
\footnote{It may happen, in practice, that a party acts both as a validator and as a client, but conceptually these are two different roles.}
Despite the important role of clients in determining the maximum achievable resilience,
the modeling of the clients' capabilities has received little attention in the literature.
In this work, we focus on the role of clients.
To do so, we level the playing field with respect to other modeling aspects.
Specifically, for the remainder of the paper, we adopt a 
permissioned setting with PKI and allow the use of cryptographic primitives and randomization for all considered protocols---%
a setting that 
enables fair comparisons
between both Dolev--Strong and 
permissioned
instantiations of longest-chain protocols~\cite{david2018ouroboros,snowwhite},%
\footnote{%
The protocols of \cite{david2018ouroboros,snowwhite}, at their core, run a permissioned SMR protocol with a fixed validator set, and then rotate this validator set from epoch to epoch. We focus on this permissioned core protocol. Rotating the validator set is orthogonal.
}
used for instance in the Cardano blockchain system, as well as the other protocols considered in the paper.

\import{./figures/}{resilience_plots.tex}

\mysubparagraph{%
Types of Clients%
}
Two specific characteristics of clients are relevant:
(1) \emph{\Wokeness:} Clients may only follow the chain intermittently (\eg, a merchant during business hours), or may turn to a chain only long after its inception. We then call this the \emph{\sleepy} client model, in analogy to sleepy validators in~\cite{sleepy}.
In contrast, in the \emph{\alert} client model, we expect clients to follow the chain continuously, such as in the case of block explorers or wallet providers
for blockchains.
(2) \emph{\Interactivity:} In the \emph{\passive} client model, clients may be constrained to only \emph{listen} to messages from validators. 
In contrast, in the \emph{\activ} client model, they may be able to relay messages to validators or other clients, for instance, through a system-wide gossip protocol.
Consensus is easier when clients are \alert rather than \sleepy, and \activ rather than \passive.%
In synchronous networks,
longest-chain consensus makes only the weakest client assumptions, \emph{\sleepy \passive} clients, but achieves only 49\% resilience---which is optimal for that model~\cite{schneider-survey,sleepy,shi-rethinking}.
In contrast, the Dolev--Strong 99\%-resilience  holds under the assumption of \emph{\alert \activ} clients, i.e. the strongest client assumptions~\cite{vitalik-dolev-strong}. (The original Dolev--Strong work was developed in a model with only validators and no clients.) What about the intermediate client assumptions, \emph{\sleepy \activ} clients or \emph{\alert \passive} clients? What about if the validators themselves can also be \sleepy instead of being \alert as in
longest-chain consensus~\cite{david2018ouroboros,snowwhite}?
And what about if the network is partially synchronous instead of synchronous? 

\mysubparagraph{Our Contributions}
The main contribution of this paper is a full characterization of the achievable security in all such scenarios. The results are summarized in \cref{fig:resilience-plots} in terms of tight achievable \emph{safety} and \emph{liveness resiliences} under each scenario. \emph{Safety} \emph{resilience} of a protocol is the maximum fraction of adversary validators such that safety is guaranteed, and \emph{liveness} \emph{resilience} of a protocol is the maximum fraction of adversary validators such that liveness is guaranteed~\cite{mtbft,oflex}. Traditional \emph{resilience} of a protocol, the maximum fraction of adversary validators such that it is \emph{both safe and live}, is the minimum of the protocol's safety and liveness resiliences. Separate safety and liveness resiliences provide a meaningfully more fine-grained measure of a protocol's security since the impact of safety loss and liveness loss to a client is often different. 

\cref{fig:models-hierarchy} shows the relationship of all the scenarios we considered in this paper.

\import{./figures/}{models_hierarchy.tex}

\myparagraphNoDot{Synchronous network with \alert validators:}
The first column of \cref{fig:resilience-plots} shows the results in the synchronous network model. \Cref{fig:resilience-sync-alert-active} shows that one can achieve 99\% resilence when clients are \alert and \activ, i.e., the Dolev--Strong client model. \cref{fig:resilience-sync-sleepy-passive} shows that one can achieve 49\% resilence when clients are \sleepy and \passive, i.e., the longest-chain client model. \cref{fig:resilience-sync-alert-passive} and \cref{fig:resilience-sync-sleepy-active} are the two intermediate client settings; our impossibility results show that the achievable resilences in both settings do not improve over the longest-chain client model, i.e., 49\% resilience. However, the similarity ends when one looks at safety and liveness resiliences separately. 
In particular, \weshow a new protocol for \sleepy \activ clients that can achieve 99\% safety resilience and 49\% liveness resilience \emph{simultaneously} (\cref{fig:resilience-sync-sleepy-active}, \cref{thm:freezing-resilience}),
a resilience pair that is impossible for \sleepy \passive clients (\cref{fig:resilience-sync-sleepy-passive}),
and strictly dominates classical protocols like the longest-chain that achieve only 49\% safety resilience and 49\% liveness resilience.
\Weshow another protocol for \sleepy \activ clients that achieves 49\% safety and 99\% liveness resilience 
(\cref{fig:resilience-sync-sleepy-active}, \cref{thm:sleepy-active-synchrony-achievability-2}). On the other hand, \weshow that \passive clients do not benefit from being \alert rather than \sleepy in synchronous networks even when one considers safety and liveness resiliences separately (\cref{fig:resilience-sync-alert-passive,fig:resilience-sync-sleepy-passive}, \cref{thm:alert-passive-synchrony-converse,thm:sleepy-passive-synchrony-achievability-2}).

\myparagraphNoDot{Synchonous network with \sleepy validators:}
The concept of ``sleepy'' parties was previously introduced in the sleepy model~\cite{sleepy} where it pertains to validators rather than clients.
That model features---in addition to adversary Byzantine faults---relatively-benign mobile crash-faults. Each validator is either honest, crashed, or adversary, and liveness must hold even when many validators are crashed~\cite{ebbandflow,aa}.
\Weshow that under synchrony with \passive or with \alert \activ clients, validator \wokeness does not affect achievable safety--liveness pairs
(\cref{%
fig:resilience-sync-sleepy-passive,%
fig:resilience-dynamic-sleepy-passive,%
fig:resilience-sync-alert-passive,%
fig:resilience-dynamic-alert-passive,%
fig:resilience-sync-alert-active,%
fig:resilience-dynamic-alert-active%
}).
Validator \wokeness \emph{does} affect, though, what's achievable for \sleepy \activ clients (\cref{fig:resilience-sync-sleepy-active,fig:resilience-dynamic-sleepy-active}).
Specifically, while 49\% safety and 99\% liveness resilience remain simultaneously achievable (\cref{fig:resilience-dynamic-sleepy-active}, \cref{thm:sleepy-active-synchrony-dynamic-achievability-2}),
that is not the case for 99\% safety and 49\% liveness resilience (\cref{fig:resilience-dynamic-sleepy-active}, \cref{thm:sleepy-active-da-converse,thm:sleeepy-passive-da-achievability-1}).

\myparagraphNoDot{Partially synchronous network:}
Finally, the astute reader will note that---whether validators are \sleepy or \alert---our discussion above regarding
Dolev--Strong and longest-chain consensus assume \emph{synchrony}, where the protocol is parameterized by an upper bound $\Delta$ on the delay of message propagation among honest parties.
In contrast, PBFT-style protocols~\cite{pbft,casperffg,hotstuff,bullshark,narwhaltusk} are designed for \emph{partial synchrony},
where such a delay bound is guaranteed to hold only \emph{eventually},
after an initial period of unknown duration with arbitrary network delay.
Network delay constitutes the fourth and last aspect of our models (\cref{fig:models-hierarchy}).
Interestingly, the safety--liveness resilience pairs achievable under partial synchrony
do not depend on client \wokeness or client \interactivity (\cref{fig:resilience-psync-alert-active,fig:resilience-psync-sleepy-active,fig:resilience-psync-alert-passive,fig:resilience-psync-sleepy-passive}---with \sleepy validators, safety under partial synchrony is impossible
~\cite{ebbandflow,gilbert-lynch-cap-theorem,pye-roughgarden-cap-theorem}; \cf \cref{fig:models-hierarchy}).
This ``robustness'' of the partially synchronous model
to different client assumptions
is perhaps why 
clients have so far often been an afterthought in the distributed-systems literature.

%% file: figures/resilience_plots.tex
\tikzset{resilience_plots/.style={
            x=2.5cm,
            y=2.5cm,
            boundary/.style={
                black,
                thick,
            },
            achieveprev/.style={
                fill=jnSUDigitalBlue!50,
                draw=white,
            },
            achievenew/.style={
                fill=jnSUDigitalBlue!50,
                draw=white,
            },
            impossibleprev/.style={
                fill=jnSUDigitalRed!50,
                draw=white,
            },
            impossiblenew/.style={
                fill=jnSUDigitalRed!50,
                draw=white,
            },
}}

\newcommand{\axesbase}[2]{
    \scriptsize
    \pgfdeclarelayer{background}
    \pgfdeclarelayer{foreground}
    \pgfsetlayers{background,main,foreground}
    \draw [-Latex] (0,0) -- (1.15,0) node [below] {#1};
    \draw [-Latex] (0,0) -- (0,1.1) node [left] {#2};
    \draw (0.5,0.03) -- (0.5,-0.03);
    \draw (1,0.03) -- (1,-0.03);
    \draw (0.03,0.5) -- (-0.03,0.5);
    \draw (0.03,1) -- (-0.03,1);
}

\newcommand{\axesstatic}{
    \axesbase{$\tL$}{$\tS$}
    \node at (0.5,-0.13) {$1/2$};
    \node at (1,-0.13) {$1$};
    \node at (-0.1,0.5) {$\frac{1}{2}$};
    \node at (-0.1,1) {$1$};
}

\newcommand{\axesdynamic}{
    \axesbase{$\betaL$}{$\betaS$}
    \node at (0.5,-0.13) {$1/2$};
    \node at (1,-0.13) {$1$};
    \node at (-0.1,0.5) {$\frac{1}{2}$};
    \node at (-0.1,1) {$1$};
}

\newcommand{\figcolumnheading}[1]{\parbox[b]{0.3\textwidth}{\centering\footnotesize\textbf{#1}}}
\newcommand{\figrowheading}[1]{\makebox[0.04\textwidth][c]{\rotatebox{90}{\parbox{4.5cm}{\centering\footnotesize\textbf{#1}}}}}

\begin{figure}[ptbh]
    \centering
    \footnotesize
    \makebox[0.04\textwidth]{}
    \hfill
    \figcolumnheading{Synchronous,\\\statpart}
    \hfill
    \figcolumnheading{Synchronous,\\\dynpart}
    \hfill
    \figcolumnheading{Partially synchronous,\\\statpart}
    \figrowheading{\Sleepy \passive clients}
    \hfill
    \begin{subfigure}[b]{0.3\textwidth}
        \centering
        \begin{tikzpicture}[resilience_plots]%
            \axesstatic
            \draw [boundary] (0,1) -- (1,0);
            \begin{pgfonlayer}{background}
                \draw [achieveprev] (0,0) -- (0,0.5) -- (0.5,0.5) -- (0.5,0) -- cycle;
            \end{pgfonlayer}
            \node at (0.25,0.25) {\cite{synchotstuff}, \cref{thm:sleepy-passive-synchrony-achievability-1}};
            \begin{pgfonlayer}{background}
                \draw [achieveprev] (0,0.5) -- (0,1) -- (0.5,0.5) -- cycle;
            \end{pgfonlayer}
            \node [anchor=north,rotate=-45] at (0.75,0.25) {\Cref{thm:sleepy-passive-synchrony-achievability-2}};
            \begin{pgfonlayer}{background}
                \draw [achievenew] (0.5,0) -- (1,0) -- (0.5,0.5) -- cycle;
            \end{pgfonlayer}
            \node [anchor=north,rotate=-45] at (0.25,0.75) {\cite{synchotstuff}, \cref{thm:sleepy-passive-synchrony-achievability-1}};
            \begin{pgfonlayer}{background}
                \draw [impossiblenew] (0,1) -- (1,1) -- (1,0) -- cycle;
            \end{pgfonlayer}
            \node at (0.75,0.75) {\cite{mtbft}, \cref{thm:sleepy-passive-synchrony-converse}};
        \end{tikzpicture}%
        \caption{\iconvalidatoralertsync{}\iconclientsleepypassive{} (\cref{sec:passive})}
        \label{fig:resilience-sync-sleepy-passive}
    \end{subfigure}
    \hfill
    \begin{subfigure}[b]{0.3\textwidth}
        \centering
        \begin{tikzpicture}[resilience_plots]%
            \axesdynamic
            \draw [boundary] (0,1) -- (1,0);
            \begin{pgfonlayer}{background}
                \draw [achieveprev] (0,0) -- (0,0.5) -- (0.5,0.5) -- (0.5,0) -- cycle;
            \end{pgfonlayer}
            \node at (0.25,0.25) {\cite{sleepy}, \cref{thm:sleeepy-passive-da-achievability-0}};
            \begin{pgfonlayer}{background}
                \draw [achievenew] (0.5,0) -- (1,0) -- (0.5,0.5) -- cycle;
            \end{pgfonlayer}
            \node [anchor=north,rotate=-45] at (0.75,0.25) {\Cref{thm:sleepy-passive-da-achievability-2}};
            \begin{pgfonlayer}{background}
                \draw [achievenew] (0,0.5) -- (0,1) -- (0.5,0.5) -- cycle;
            \end{pgfonlayer}
            \node [anchor=north,rotate=-45] at (0.25,0.75) {\Cref{thm:sleeepy-passive-da-achievability-1}};
            \begin{pgfonlayer}{background}
                \draw [impossiblenew] (0,1) -- (1,1) -- (1,0) -- cycle;
            \end{pgfonlayer}
            \node at (0.75,0.75) {\cite{mtbft}, \cref{thm:sleepy-passive-synchrony-converse}};
        \end{tikzpicture}%
        \caption{\iconvalidatorsleepysync{}\iconclientsleepypassive{} (\cref{sec:passive-dynamic})}
        \label{fig:resilience-dynamic-sleepy-passive}
    \end{subfigure}
    \hfill
    \begin{subfigure}[b]{0.3\textwidth}
        \centering
        \begin{tikzpicture}[resilience_plots]%
            \axesstatic
            \draw [boundary] (0,1) -- (0.5,0);
            \begin{pgfonlayer}{background}
                \draw [achieveprev] (0,0) -- (0,1) -- (0.5,0) -- cycle;
            \end{pgfonlayer}
            \node at (0.175,0.2) [align=center] {\cite{hotstuff},\\\cref{thm:psync-achievability}};
            \begin{pgfonlayer}{background}
                \draw [impossibleprev] (0,1) -- (1,1) -- (1,0) -- (0.5,0) -- cycle;
            \end{pgfonlayer}
            \node at (0.6,0.6) {\cite{dls88,mtbft}, \cref{thm:psync-converse}};
        \end{tikzpicture}%
        \caption{\iconvalidatoralertpsync{}\iconclientsleepypassive{} (\cref{sec:partial-synchrony})}
        \label{fig:resilience-psync-sleepy-passive}
    \end{subfigure}
    \figrowheading{\Alert \passive clients}
    \hfill
    \begin{subfigure}[b]{0.3\textwidth}
        \centering
        \begin{tikzpicture}[resilience_plots]%
            \axesstatic
            \draw [boundary] (0,1) -- (1,0);
            \begin{pgfonlayer}{background}
                \draw [achieveprev] (0,0) -- (0,0.5) -- (0.5,0.5) -- (0.5,0) -- cycle;
            \end{pgfonlayer}
            \node at (0.25,0.25) {\cite{synchotstuff}, \cref{thm:sleepy-passive-synchrony-achievability-1}};
            \begin{pgfonlayer}{background}
                \draw [achievenew] (0,0.5) -- (0,1) -- (0.5,0.5) -- cycle;
            \end{pgfonlayer}
            \node [anchor=north,rotate=-45] at (0.75,0.25) {\Cref{thm:sleepy-passive-synchrony-achievability-2}};
            \begin{pgfonlayer}{background}
                \draw [achievenew] (0.5,0) -- (1,0) -- (0.5,0.5) -- cycle;
            \end{pgfonlayer}
            \node [anchor=north,rotate=-45] at (0.25,0.75) {\cite{synchotstuff}, \cref{thm:sleepy-passive-synchrony-achievability-1}};
            \begin{pgfonlayer}{background}
                \draw [impossiblenew] (0,1) -- (1,1) -- (1,0) -- cycle;
            \end{pgfonlayer}
            \node at (0.75,0.75) {\Cref{thm:alert-passive-synchrony-converse}};
        \end{tikzpicture}%
        \caption{\iconvalidatoralertsync{}\iconclientalertpassive{} (\cref{sec:passive})}
        \label{fig:resilience-sync-alert-passive}
    \end{subfigure}
    \hfill
    \begin{subfigure}[b]{0.3\textwidth}
        \centering
        \begin{tikzpicture}[resilience_plots]%
            \axesdynamic
            \draw [boundary] (0,1) -- (1,0);
            \begin{pgfonlayer}{background}
                \draw [achieveprev] (0,0) -- (0,0.5) -- (0.5,0.5) -- (0.5,0) -- cycle;
            \end{pgfonlayer}
            \node at (0.25,0.25) {\cite{sleepy}, \cref{thm:sleeepy-passive-da-achievability-0}};
            \begin{pgfonlayer}{background}
                \draw [achievenew] (0.5,0) -- (1,0) -- (0.5,0.5) -- cycle;
            \end{pgfonlayer}
            \node [anchor=north,rotate=-45] at (0.75,0.25) {\Cref{thm:sleepy-passive-da-achievability-2}};
            \begin{pgfonlayer}{background}
                \draw [achievenew] (0,0.5) -- (0,1) -- (0.5,0.5) -- cycle;
            \end{pgfonlayer}
            \node [anchor=north,rotate=-45] at (0.25,0.75) {\Cref{thm:sleeepy-passive-da-achievability-1}};
            \begin{pgfonlayer}{background}
                \draw [impossiblenew] (0,1) -- (1,1) -- (1,0) -- cycle;
            \end{pgfonlayer}
            \node at (0.75,0.75) {\Cref{thm:alert-passive-synchrony-converse}};
        \end{tikzpicture}%
        \caption{\iconvalidatorsleepysync{}\iconclientalertpassive{} (\cref{sec:passive-dynamic})}
        \label{fig:resilience-dynamic-alert-passive}
    \end{subfigure}
    \hfill
    \begin{subfigure}[b]{0.3\textwidth}
        \centering
        \begin{tikzpicture}[resilience_plots]%
            \axesstatic
            \draw [boundary] (0,1) -- (0.5,0);
            \begin{pgfonlayer}{background}
                \draw [achieveprev] (0,0) -- (0,1) -- (0.5,0) -- cycle;
            \end{pgfonlayer}
            \node at (0.175,0.2) [align=center] {\cite{hotstuff},\\\cref{thm:psync-achievability}};
            \begin{pgfonlayer}{background}
                \draw [impossibleprev] (0,1) -- (1,1) -- (1,0) -- (0.5,0) -- cycle;
            \end{pgfonlayer}
            \node at (0.6,0.6) {\cite{dls88,mtbft}, \cref{thm:psync-converse}};
        \end{tikzpicture}%
        \caption{\iconvalidatoralertpsync{}\iconclientalertpassive{} (\cref{sec:partial-synchrony})}
        \label{fig:resilience-psync-alert-passive}
    \end{subfigure}
    \figrowheading{\Sleepy \activ clients}
    \hfill
    \begin{subfigure}[b]{0.3\textwidth}
        \centering
        \begin{tikzpicture}[resilience_plots]%
            \axesstatic
            \draw [boundary] (0,1) -- (0.5,1) -- (0.5,0.5) -- (1,0.5) -- (1,0);
            \begin{pgfonlayer}{background}
                \draw [achieveprev] (0,0) -- (0,0.5) -- (0.5,0.5) -- (0.5,0) -- cycle;
            \end{pgfonlayer}
            \node at (0.25,0.25) {\cite{synchotstuff}, \cref{thm:sleepy-passive-synchrony-achievability-1}};
            \begin{pgfonlayer}{background}
                \draw [achievenew] (0,0.5) -- (0,1) -- (0.5,1) -- (0.5,0.5) -- cycle;
            \end{pgfonlayer}
            \node at (0.25,0.75) {\Cref{thm:freezing-resilience}};
            \begin{pgfonlayer}{background}
                \draw [achievenew] (0.5,0) -- (1,0) -- (1,0.5) -- (0.5,0.5) -- cycle;
            \end{pgfonlayer}
            \node at (0.75,0.25) {\Cref{thm:sleepy-active-synchrony-achievability-2}};
            \begin{pgfonlayer}{background}
                \draw [impossiblenew] (0.5,0.5) -- (1,0.5) -- (1,1) -- (0.5,1) -- cycle;
            \end{pgfonlayer}
            \node at (0.75,0.75) {\Cref{thm:sleepy-active-synchrony-converse}};
        \end{tikzpicture}%
        \caption{\iconvalidatoralertsync{}\iconclientsleepyactiv{} (\cref{sec:sleepy-active})}
        \label{fig:resilience-sync-sleepy-active}
    \end{subfigure}
    \hfill
    \begin{subfigure}[b]{0.3\textwidth}
        \centering
        \begin{tikzpicture}[resilience_plots]%
            \axesdynamic
            \draw [boundary] (0,1) -- (0.5,0.5) -- (1,0.5) -- (1,0);
            \begin{pgfonlayer}{background}
                \draw [achieveprev] (0,0) -- (0,0.5) -- (0.5,0.5) -- (0.5,0) -- cycle;
            \end{pgfonlayer}
            \node at (0.25,0.25) {\cite{sleepy}, \cref{thm:sleeepy-passive-da-achievability-0}};
            \begin{pgfonlayer}{background}
                \draw [achievenew] (0.5,0) -- (1,0) -- (1,0.5) -- (0.5,0.5) -- cycle;
            \end{pgfonlayer}
            \node at (0.75,0.25) {\Cref{thm:sleepy-active-synchrony-dynamic-achievability-2}};
            \begin{pgfonlayer}{background}
                \draw [achievenew] (0,0.5) -- (0,1) -- (0.5,0.5) -- cycle;
            \end{pgfonlayer}
            \node [anchor=north,rotate=-45] at (0.25,0.75) {\Cref{thm:sleeepy-passive-da-achievability-1}};
            \begin{pgfonlayer}{background}
                \draw [impossiblenew] (0,1) -- (0.5,1) -- (0.5,0.5) -- cycle;
            \end{pgfonlayer}
            \node [anchor=south,rotate=-45] at (0.25,0.75) {\Cref{thm:sleepy-active-da-converse}};
            \begin{pgfonlayer}{background}
                \draw [impossiblenew] (0.5,0.5) -- (0.5,1) -- (1,1) -- (1,0.5) -- cycle;
            \end{pgfonlayer}
            \node at (0.75,0.75) {\Cref{thm:sleepy-active-synchrony-converse}};
        \end{tikzpicture}%
        \caption{\iconvalidatorsleepysync{}\iconclientsleepyactiv{} (\cref{sec:sleepy-active-dynamic})}
        \label{fig:resilience-dynamic-sleepy-active}
    \end{subfigure}
    \hfill
    \begin{subfigure}[b]{0.3\textwidth}
        \centering
        \begin{tikzpicture}[resilience_plots]%
            \axesstatic
            \draw [boundary] (0,1) -- (0.5,0);
            \begin{pgfonlayer}{background}
                \draw [achieveprev] (0,0) -- (0,1) -- (0.5,0) -- cycle;
            \end{pgfonlayer}
            \node at (0.175,0.2) [align=center] {\cite{hotstuff},\\\cref{thm:psync-achievability}};
            \begin{pgfonlayer}{background}
                \draw [impossibleprev] (0,1) -- (1,1) -- (1,0) -- (0.5,0) -- cycle;
            \end{pgfonlayer}
            \node at (0.6,0.6) {\cite{dls88,mtbft}, \cref{thm:psync-converse}};
        \end{tikzpicture}%
        \caption{\iconvalidatoralertpsync{}\iconclientsleepyactiv{} (\cref{sec:partial-synchrony})}
        \label{fig:resilience-psync-sleepy-active}
    \end{subfigure}
    \figrowheading{\Alert \activ clients}
    \hfill
    \begin{subfigure}[b]{0.3\textwidth}
        \centering
        \begin{tikzpicture}[resilience_plots]%
            \axesstatic
            \draw [boundary] (0,1) -- (1,1) -- (1,0);
            \begin{pgfonlayer}{background}
                \draw [achieveprev] (0,0) -- (0,1) -- (1,1) -- (1,0) -- cycle;
            \end{pgfonlayer}
            \node at (0.5,0.5){\cite{vitalik-dolev-strong}, \cref{thm:alert-active-static-ach}};
        \end{tikzpicture}%
        \caption{\iconvalidatoralertsync{}\iconclientalertactiv{} (\cref{sec:alert-active})}
        \label{fig:resilience-sync-alert-active}
    \end{subfigure}
    \hfill
    \begin{subfigure}[b]{0.3\textwidth}
        \centering
        \begin{tikzpicture}[resilience_plots]%
            \axesdynamic
            \draw [boundary] (0,1) -- (1,1) -- (1,0);
            \begin{pgfonlayer}{background}
                \draw [achievenew] (0,0) -- (0,1) -- (1,1) -- (1,0) -- cycle;
            \end{pgfonlayer}
            \node at (0.5,0.5){\Cref{thm:alert-active-dynamic-ach}};
        \end{tikzpicture}%
        \caption{\iconvalidatorsleepysync{}\iconclientalertactiv{} (\cref{sec:alert-active-dynamic})}
        \label{fig:resilience-dynamic-alert-active}
    \end{subfigure}
    \hfill
    \begin{subfigure}[b]{0.3\textwidth}
        \centering
        \begin{tikzpicture}[resilience_plots]%
            \axesstatic
            \draw [boundary] (0,1) -- (0.5,0);
            \begin{pgfonlayer}{background}
                \draw [achieveprev] (0,0) -- (0,1) -- (0.5,0) -- cycle;
            \end{pgfonlayer}
            \node at (0.175,0.2) [align=center] {\cite{hotstuff},\\\cref{thm:psync-achievability}};
            \begin{pgfonlayer}{background}
                \draw [impossibleprev] (0,1) -- (1,1) -- (1,0) -- (0.5,0) -- cycle;
            \end{pgfonlayer}
            \node at (0.6,0.5) {\cite{dls88,mtbft}, \cref{thm:psync-converse}};
        \end{tikzpicture}%
        \caption{\iconvalidatoralertpsync{}\iconclientalertactiv{} (\cref{sec:partial-synchrony})}
        \label{fig:resilience-psync-alert-active}
    \end{subfigure}
    \caption[]{%
    Tight achievable
    (\tikz[resilience_plots]{ \draw [achievenew] ([xshift=-0.35em,yshift=-0.35em] 0,0) rectangle ([xshift=0.35em,yshift=0.35em] 0,0); })
    and impossible
    (\tikz[resilience_plots]{ \draw [impossiblenew] ([xshift=-0.35em,yshift=-0.35em] 0,0) rectangle ([xshift=0.35em,yshift=0.35em] 0,0); })
    safety resilience $\tS$ and liveness resilience $\tL$ bounds for different models (\cf~\Cref{fig:models-hierarchy}),
    each with
    four aspects:
    \emph{Network delay:} synchrony \iconvalidatorsync{} \vs partial synchrony \iconvalidatorpsync{};
    \emph{Validator \wokeness:} \statpart \iconvalidatoralert{} \vs \dynpart \iconvalidatorsleepy{};
    \emph{Client \wokeness:} \alert clients \iconclientalert{} \vs \sleepy clients \iconclientsleepy{}; 
    \emph{Client \interactivity:} \activ \iconclientactiv{} \vs \passive \iconclientpassive{}.
    Citations with 
    corollaries, or theorems, indicate previously known, or new results, respectively.
    }
    \label{fig:resilience-plots}
\end{figure}

%% file: figures/models_hierarchy.tex
\begin{figure}[tbp]
    \centering
    \begin{tikzpicture}[
            x=1cm,
            y=1cm,
            easierthanrelation/.style={
                -latex
            },
            easierthanrelationlocal/.style={},
            easierthanrelationglobal/.style={},
            model/.style={
                draw=none,
                fill=white,
                inner sep=2pt,
                rounded corners=2pt
            },
            groupofmodels/.style={
                draw=none,
                fill=black!10
            }
        ]

        \pgfdeclarelayer{backerground}
        \pgfdeclarelayer{background}
        \pgfsetlayers{backerground,background,main}
        \begin{scope}[yshift=-2cm]
            \node [model] (vAScAA) at (0,-1) {\iconvalidatoralertsync{}\iconclientalertactiv{}};
            \node [model] (vAScAP) at (-1,0) {\iconvalidatoralertsync{}\iconclientalertpassive{}};
            \node [model] (vAScSA) at (1,0) {\iconvalidatoralertsync{}\iconclientsleepyactiv{}};
            \node [model] (vAScSP) at (0,1) {\iconvalidatoralertsync{}\iconclientsleepypassive{}};
            \draw [easierthanrelation,easierthanrelationlocal] (vAScAA) -- (vAScAP);
            \draw [easierthanrelation,easierthanrelationlocal] (vAScAA) -- (vAScSA);
            \draw [easierthanrelation,easierthanrelationlocal] (vAScAP) -- (vAScSP);
            \draw [easierthanrelation,easierthanrelationlocal] (vAScSA) -- (vAScSP);
            \begin{pgfonlayer}{backerground}
                \draw [groupofmodels] (0,0) ellipse (2 and 1.5) node [below,yshift=-1.6cm,align=center,font=\tiny] {Synchronous,\\\statpart};
            \end{pgfonlayer}
        \end{scope}
        \begin{scope}[xshift=-4cm]
            \node [model] (vAPcAA) at (0,-1) {\iconvalidatoralertpsync{}\iconclientalertactiv{}};
            \node [model] (vAPcAP) at (-1,0) {\iconvalidatoralertpsync{}\iconclientalertpassive{}};
            \node [model] (vAPcSA) at (1,0) {\iconvalidatoralertpsync{}\iconclientsleepyactiv{}};
            \node [model] (vAPcSP) at (0,1) {\iconvalidatoralertpsync{}\iconclientsleepypassive{}};
            \draw [easierthanrelation,easierthanrelationlocal] (vAPcAA) -- (vAPcAP);
            \draw [easierthanrelation,easierthanrelationlocal] (vAPcAA) -- (vAPcSA);
            \draw [easierthanrelation,easierthanrelationlocal] (vAPcAP) -- (vAPcSP);
            \draw [easierthanrelation,easierthanrelationlocal] (vAPcSA) -- (vAPcSP);
            \begin{pgfonlayer}{backerground}
                \draw [groupofmodels] (0,0) ellipse (2 and 1.5) node [below,yshift=-1.6cm,align=center,font=\tiny] {Partially synchronous,\\\statpart};
            \end{pgfonlayer}
        \end{scope}
        \begin{scope}[xshift=4cm]
            \node [model] (vSScAA) at (0,-1) {\iconvalidatorsleepysync{}\iconclientalertactiv{}};
            \node [model] (vSScAP) at (-1,0) {\iconvalidatorsleepysync{}\iconclientalertpassive{}};
            \node [model] (vSScSA) at (1,0) {\iconvalidatorsleepysync{}\iconclientsleepyactiv{}};
            \node [model] (vSScSP) at (0,1) {\iconvalidatorsleepysync{}\iconclientsleepypassive{}};
            \draw [easierthanrelation,easierthanrelationlocal] (vSScAA) -- (vSScAP);
            \draw [easierthanrelation,easierthanrelationlocal] (vSScAA) -- (vSScSA);
            \draw [easierthanrelation,easierthanrelationlocal] (vSScAP) -- (vSScSP);
            \draw [easierthanrelation,easierthanrelationlocal] (vSScSA) -- (vSScSP);
            \begin{pgfonlayer}{backerground}
                \draw [groupofmodels] (0,0) ellipse (2 and 1.5) node [below,yshift=-1.6cm,align=center,font=\tiny] {Synchronous,\\\dynpart};
            \end{pgfonlayer}
        \end{scope}
        \begin{scope}[yshift=2cm,opacity=0.3]
            \node [model] (vSPcAA) at (0,-1) {\iconvalidatorsleepypsync{}\iconclientalertactiv{}};
            \node [model] (vSPcAP) at (-1,0) {\iconvalidatorsleepypsync{}\iconclientalertpassive{}};
            \node [model] (vSPcSA) at (1,0) {\iconvalidatorsleepypsync{}\iconclientsleepyactiv{}};
            \node [model] (vSPcSP) at (0,1) {\iconvalidatorsleepypsync{}\iconclientsleepypassive{}};
            \draw [easierthanrelation,easierthanrelationlocal] (vSPcAA) -- (vSPcAP);
            \draw [easierthanrelation,easierthanrelationlocal] (vSPcAA) -- (vSPcSA);
            \draw [easierthanrelation,easierthanrelationlocal] (vSPcAP) -- (vSPcSP);
            \draw [easierthanrelation,easierthanrelationlocal] (vSPcSA) -- (vSPcSP);
            \begin{pgfonlayer}{backerground}
                \draw [groupofmodels,opacity=0.3] (0,0) ellipse (2 and 1.5) node [below,yshift=-1.6cm,align=center,font=\tiny] {Partially synchronous,\\\dynpart};
            \end{pgfonlayer}
        \end{scope}

        \begin{pgfonlayer}{background}
            \draw [easierthanrelation,easierthanrelationglobal] (vAScAA) -- (vAPcAA);
            \draw [easierthanrelation,easierthanrelationglobal] (vAScAA) -- (vSScAA);
            \draw [easierthanrelation,easierthanrelationglobal] (vAScAP) -- (vAPcAP);
            \draw [easierthanrelation,easierthanrelationglobal] (vAScAP) -- (vSScAP);
            \draw [easierthanrelation,easierthanrelationglobal] (vAScSA) -- (vAPcSA);
            \draw [easierthanrelation,easierthanrelationglobal] (vAScSA) -- (vSScSA);
            \draw [easierthanrelation,easierthanrelationglobal] (vAScSP) -- (vAPcSP);
            \draw [easierthanrelation,easierthanrelationglobal] (vAScSP) -- (vSScSP);

            \draw [easierthanrelation,easierthanrelationglobal,opacity=0.3] (vAPcAA) -- (vSPcAA);
            \draw [easierthanrelation,easierthanrelationglobal,opacity=0.3] (vAPcAP) -- (vSPcAP);
            \draw [easierthanrelation,easierthanrelationglobal,opacity=0.3] (vAPcSA) -- (vSPcSA);
            \draw [easierthanrelation,easierthanrelationglobal,opacity=0.3] (vAPcSP) -- (vSPcSP);
            \draw [easierthanrelation,easierthanrelationglobal,opacity=0.3] (vSScAA) -- (vSPcAA);
            \draw [easierthanrelation,easierthanrelationglobal,opacity=0.3] (vSScAP) -- (vSPcAP);
            \draw [easierthanrelation,easierthanrelationglobal,opacity=0.3] (vSScSA) -- (vSPcSA);
            \draw [easierthanrelation,easierthanrelationglobal,opacity=0.3] (vSScSP) -- (vSPcSP);
        \end{pgfonlayer}

        \begin{scope}[xshift=4cm,yshift=-3.5cm,x=0.5cm]
            \node [font=\tiny] (x) at (-1,0) {$X$};
            \node [font=\tiny] (y) at (1,0) {$Y$};
            \draw [easierthanrelation] (x) -- (y) node (label) [midway,below,font=\tiny] {= ``$X$ is easier than $Y$''};
            \draw (x.north -| label.west) rectangle (label.south east);
        \end{scope}

    \end{tikzpicture}
    \caption[]{Hasse diagram illustrating the relative difficulty of consensus in all the different models we study (proof: \cref{lem:hierarchy}). 
    Each small white box indicates a different model (see \cref{fig:resilience-plots} for icon legend).
    Models are grouped in a shaded circle when they share validator model and network model but the client model differs.
    Group $(\iconvalidatorpsync{}, \iconvalidatorsleepy{})$ is faded out because consensus is impossible%
    ~\cite{ebbandflow,gilbert-lynch-cap-theorem,pye-roughgarden-cap-theorem}.%
    }
    \label{fig:models-hierarchy}
\end{figure}

%% file: sections/03_model.tex
\section{Model}
\label{sec:model}

We focus on the most relevant model aspects. Other standard aspects: \cref{sec:app-model-full}.

\mysubparagraph{Parties}
A fixed set $\valset$ of parties called \emph{validators} is known to all parties.
Define $n = |\valset|$.
Each validator has a secret key with which they may sign their messages, and all parties know validators' public keys.
We assume a permissioned setting~\cite{pye-roughgarden-permissionless,budish-pye-roughgarden-economic,sleepy} with a fixed and known set of validators, and defer the proof-of-stake setting to \cref{sec:discussion-pos}.
Unlike validators, the number and identities of the \emph{clients} is not known to all parties, and clients do not have public keys.

\mysubparagraph{Adversary}
At the beginning of the execution,
before any randomness is drawn,
the PPT adversary $\Adv$ controls $f$ adversary validators (\aka Byzantine faults).
$\Adv$ may also corrupt any number of clients.
All our theorems hold for any number of adversary clients.
We discuss adaptive corruption in \cref{sec:discussion-adaptive}.

\mysubparagraph{Validator \wokeness (\cf~\cite{sleepy})}
At every round $r$, a subset $\awakeset{r} \subseteq \valset$ of validators are awake while the rest are \emph{asleep}.
When asleep, validators behave like temporarily crashed nodes: they do not run computation or send messages.
Whenever a validator is awake, it knows the current round (\ie, it wakes up with a synchronized clock).
    In the \emph{\statpart} model,
    all validators are always awake, ($\forall r \colon \awakeset{r} = \valset$).
    In the \emph{\dynpart} model,
    $\Adv$ selects $\awakeset{r}$, and adversary validators are always awake.

\mysubparagraph{Client Models}
We classify clients along two orthogonal criteria.
    \emph{\Interactivity}:
    \emph{\Activ} clients may send messages
     to other parties, 
    \emph{\passive} clients do not.
    \emph{\Wokeness}:
    \emph{\Alert} clients are always awake while \emph{\sleepy} clients may be put to sleep by $\Adv$.
    When asleep, clients do not perform any computation, send messages, or output new logs. 
This results in the four client models shown in \cref{fig:models-hierarchy}.
For example, the \emph{\sleepy \activ} model means that all clients are \sleepy and \activ.

\Activ clients are new in this work
and more accurately
depict
blockchain implementations in which communication is facilitated by a non-eclipsed gossip network comprised of both clients and validators.
Since clients' messages cannot be authenticated due to their lack of PKI, really the best they can do is relay messages received from validators to other clients.
Yet, \activ clients
circumvent impossibility results for \passive clients (\cref{fig:resilience-plots}).
Relaying transactions, blocks, and votes is already in the client specifications of major blockchains like Ethereum~\cite{eth-p2p-gossip} and Tendermint~\cite{tendermint-p2p-gossip}.
Furthermore, since clients only relay messages sent by validators, the communication/computation overhead for all parties easily remains
bounded
by forwarding/processing only valid messages and only once.

\mysubparagraph{Network delay models}
We consider two standard network models.
    In the \emph{synchronous} model,
    there is a known constant $\Delta$ such that if an honest party sends a message at round $r$, then every honest party receives the message by round $r + \Delta$.%
    \footnote{Gossip networks have been shown to maintain connectivity, and thus synchrony, even under adversary majority~\cite{scuttlebutt,bftflooding,bftflooding2}.}
The partially synchronous model is described in \cref{sec:app-model-full}.

In both models, messages are delivered to asleep parties, but they can only process them after awakening.
In practice, equivalent behavior can be achieved by having the awakening party query online parties who reply with the `important' past messages (\eg, `initial block download'~\cite{btcdevp2pnetworkheadersfirst}).
Thus, although \sleepy parties receive all the same messages as \alert parties, they are less powerful since they cannot record the time of message receipt.

\mysubparagraph{\smrshort}
At the start of each round, each awake party may receive some transactions as input.
At the end of every round $r$, each awake honest client $k$ outputs a log (sequence of transactions) $\LOGi{k}{r}$.
For a client $k$ asleep at round $r$, let $\LOGi{k}{r} = \LOGi{k}{r-1}$.
For all clients $k$, $\LOGi{k}{0} = \genesis$.
We use $A \preceq B$ to denote that log $A$ is a (not necessarily strict) prefix of the log $B$. We use $A \consistent B$ (`$A$ is consistent with $B$') as a shorthand for $A \preceq B \,\lor B\, \preceq A$.

\begin{definition}[Safety]
    \label{def:safety}
    An \smrshort protocol $\Pi$ is \emph{safe} 
    iff
    for all rounds $r,s$
    and all honest clients $k,k'$,
    $\LOGi{k}{r} \consistent \LOGi{k'}{s}$.
\end{definition}

\begin{definition}[Liveness]
    \label{def:liveness}
    An \smrshort protocol $\Pi$ is \emph{live} with latency $u$
    iff
    for all rounds $r$, 
    if a transaction $\tx$ was received by an awake honest validator or \activ client before round $r - u$, 
    then for all honest clients $p$ 
    awake during rounds $[r-u,r]$,
    $\tx \in \LOGi{p}{r}$.%
    \footnote{Clients may not output new logs for a few rounds after awakening. We use a single parameter $u$ for the maximum of such delay and the protocol's latency.}
\end{definition}

\begin{definition}[Resilience]
    \label{def:resilience}
    For \statpart, a family of \smrshort protocols $\Pi(n)$ achieves \emph{safety resilience} $\tS \in [0,1]$ and \emph{liveness resilience} $\tL \in [0,1]$ if
    for all $n$,%
    \footnote{The number of parties is constrained to be polynomial in the security parameter.}
    $\Pi(n)$ is safe with overwhelming probability over executions with $f \leq \tS n$ and live with overwhelming probability over executions with $f \leq \tL n$.
    For \dynpart, denote the adversary fraction $\frac{f}{\min_{r} \awakeset{r}}$ by $\beta$. 
    Then, a protocol $\Pi$ achieves safety resilience $\betaS \in [0,1]$ and liveness resilience $\betaL \in [0,1]$ if
    $\Pi$ is safe with overwhelming probability over executions with $\beta \leq \betaS$ 
    and live with overwhelming probability over executions with $\beta \leq \betaL$.
\end{definition}

%% file: sections/04_passive.tex
\subsection{\SLeepy \PAssive, \ALert \PAssive Clients (\cref{fig:resilience-sync-sleepy-passive}, \cref{fig:resilience-sync-alert-passive})}
\label{sec:passive}

We group the protocols and impossibility results for \sleepy \passive and \alert \passive clients in this section because the results are the same for both (\cref{fig:resilience-sync-sleepy-passive,fig:resilience-sync-alert-passive}).
Due to \cref{lem:hierarchy}, we prove impossibility results for the easier \alert \passive client model and show protocols for the harder \sleepy \passive client model.

\import{./sections/}{04a_passive_imp.tex}
\import{./sections/}{04b_passive_ach.tex}

%% file: sections/04a_passive_imp.tex
\myparagraphNoDot{Impossibility for \ALert \PAssive Clients}
\label{sec:passive-imp}
\begin{theorem}
\label{thm:alert-passive-synchrony-converse}
In a synchronous network with \statpart and \alert \passive clients, no protocol can achieve resiliences $(\tL,\tS)$ such that $\tL+\tS \geq 1$.
\end{theorem}

\Cref{thm:alert-passive-synchrony-converse} is due to a split-brain attack.
Suppose a protocol has resilience $(\tL,\tS)$ such that $\tL + \tS = 1$.
Then, the protocol must remain live given $f=\tL n$ adversary validators and safe given $f = (1-\tL)n$ adversary validators.
Then, consider a set of $(1-\tL)n$ adversary validators that emulate in their heads two apparently honest executions with two different transactions.%
\footnote{Since each execution requires a polynomial amount of computation, a polynomial-time adversary can emulate both these executions.}
These validators can ensure that two clients, each hearing only one of the emulated executions, output different logs.
Thus, the protocol cannot ensure safety under $(1-\tL)n = \tS n$ adversary validators, which is a contradiction.
Note that the success of the split-brain attack crucially requires the clients to remain isolated, \ie, to be \passive.
The full proof is in \cref{sec:proofs-sync-alert-passive-imp}.
For \sleepy \passive clients, \Cref{thm:sleepy-passive-synchrony-converse} follows from \cref{thm:alert-passive-synchrony-converse} and \cref{lem:hierarchy}, and a similar proof is also in~\cite{mtbft,aav1}.

\label{sec:sleepy=passive-imp}

\begin{corollary}
\label{thm:sleepy-passive-synchrony-converse}
In a synchronous network with \statpart and \sleepy \passive clients, no protocol can achieve $(\tL,\tS)$ such that $\tL+\tS \geq 1$.
\end{corollary}

%% file: sections/04b_passive_ach.tex
\myparagraphNoDot{Achievability for \SLeepy \PAssive Clients (Safety-Favoring)}
\label{sec:passive-ach}

\label{sec:passive-ach-safe-static}

\begin{corollary}
\label{thm:sleepy-passive-synchrony-achievability-1}
In a synchronous network with \statpart and \sleepy \passive clients,
for all $(\tL,\tS)$ with $\tL + \tS < 1$ and $\tL < 1/2$, Sync HotStuff~\cite{synchotstuff} with a quorum size of $q \in (\tS n,(1-\tL)n]$ achieves $(\tL,\tS)$.
\end{corollary}

\Cref{thm:sleepy-passive-synchrony-achievability-1} follows from~\cite[Theorems 3 and 4]{synchotstuff}, by replacing the quorum sizes by $q \in (n/2,n]$.
A similar construction and its security proof can be found in \cite{mtbft}.
Other protocols such as Sync Streamlet~\cite{streamlet} can also be adapted with quorums $q \in (n/2,n]$ to achieve the same result.
Due to \cref{lem:hierarchy}, the protocol achieves the same resiliences in a synchronous network with \alert \passive clients.

\myparagraphNoDot{Achievability for \SLeepy \PAssive Clients (Liveness-Favoring)}
\label{sec:passive-ach-live-static}
We next describe a family $\Pi_{\mathrm{live}}^q$ of protocols (\cref{alg:sleepy-passive-static-live}, \cref{fig:sleepy-passive-more-live-protocol}) parameterized by the integers $q \in [0,n/2]$,
one for each resilience pair satisfying $\tL + \tS < 1$ and $\tL \geq \tS$.
The protocol $\Pi_{\mathrm{live}}^q$ consists of an \emph{internal protocol} $\Piint$ and a \emph{liveness queue}. 
The internal protocol can be any SMR protocol that achieves all $\tS < 1/2, \tL < 1/2$ under synchrony (\eg, Sync HotStuff~\cite{synchotstuff}).

\import{./algorithms/}{alg_liveness_sleepy_passive.tex}

\import{./figures/}{sleepy_passive_more_live_protocol.tex}

Each honest validator $v$ participates in the internal protocol.
Upon receiving a transaction $\tx$ 
for the first time, $v$ signs $\tx$ and sends $\tx$ and its signature to all parties (\cref{alg:sleepy-passive-static-live}~\cref{loc:sleepy-passive-live-gossip}).
Each client locally maintains a liveness queue.
If a transaction $\tx$ gathers $q$ or more signatures, it is added to the queue (\cref{alg:sleepy-passive-static-live}~\cref{loc:sleepy-passive-live-queue}).
Each client also maintains an \emph{internal log} $\LOGint$ output from the internal protocol (see \cref{fig:sleepy-passive-more-live-protocol}).
To output its log at a round $r$, a client $k$ appends transactions added to the liveness queue at rounds $r' \leq r-u_\mathrm{int}$ (where $u_\mathrm{int}$ is the internal protocol's latency)
to its internal log at round $r$,
discarding duplicates (\cref{alg:sleepy-passive-static-live}~\cref{loc:sleepy-passive-live-append}).
The augmented internal log is then output as the log at round $r$.

\begin{theorem}
\label{thm:sleepy-passive-synchrony-achievability-2}
In a synchronous network with \statpart and \sleepy \passive clients, for all $(\tL,\tS)$ with $\tL + \tS < 1$ and $\tL \geq 1/2$,  the protocol $\Pi_{\mathrm{live}}^q$ with $q \in (\tS n, (1-\tL)n]$ achieves $(\tL,\tS)$.
\end{theorem}

When $f \leq \tL n$ validators are adversary, all transactions input to an honest validator gather $q$ signatures and enter the liveness queues
and eventually enter the output log, ensuring liveness with resilience $\tL$.
When $f \leq \tS n$ (which implies $f \leq \tL n$, since $\tL + \tS < 1$ and $\tL \geq 1/2$), the internal protocol is safe \emph{and} live, and adversary validators cannot produce $q$ signatures
without an honest validator.
Therefore, any transaction $\tx$ added to the liveness queue must be known to an honest validator and processed by the internal protocol.
By the internal protocol's liveness, $\tx$ enters the internal log within $u_\mathrm{int}$ rounds.
Thus, no transaction is ever appended to the internal log.
Safety then follows from the internal protocol's safety.
The full proof is in \cref{sec:proofs-sync-sleepy-passive-ach-live}

%% file: algorithms/alg_liveness_sleepy_passive.tex
\begin{algorithm}[btp]
    \caption{Liveness-favoring \smrshort protocol $\Pi_{\mathrm{live}}^q$ for \sleepy \passive clients}
    \label{alg:sleepy-passive-static-live}
    \begin{algorithmic}[1]
    \scriptsize
    \LineComment{Code for validator $v$}
    \On{\Call{init}{$\valset,\genesis$}}
        \State $P_{\mathrm{int}} \gets \text{new } \Piint(\valset, \genesis)$ \Comment{instantiate a new $\Piint$ validator} \label{loc:sleepy-passive-live-base-protocol}
    \EndOn
    \On{receiving transaction $\tx$ or $\langle \tx \rangle_{v'}$ for some $v' \in \valset$}
        \State $\operatorname{gossip}(\langle \tx\rangle_v)$ \Comment{send $\tx$ and signature on $\tx$ to all parties} \label{loc:sleepy-passive-live-gossip}
        \State $P_{\mathrm{int}}.\operatorname{input}(\tx)$ \Comment{input $\tx$ to the internal protocol} \label{loc:sleepy-passive-live-input}
    \EndOn
    \vspace{1em}
    \LineComment{Client code}
    \On{\Call{init}{$\valset,\genesis$}}
    \label{loc:sleepy-passive-live-init}
        \State $P_{\mathrm{int}} \gets \text{new } \Piint(\valset, \genesis)$ \Comment{instantiate a new $\Piint$ client} \label{loc:sleepy-passive-live-base-client-protocol}
        \State $Q \gets \emptyset$ \Comment{liveness queue: txs seen so far}
        \State $\LOGi{}{} \gets \genesis$ \Comment{output log of the combined protocol $\Pi_{\mathrm{live}}^q$}
    \EndOn
    \On{$\{\langle \tx\rangle_v\}_{v \in V}$ such that $V \subseteq \valset$, $|V| \geq q$ at round $r$}
        \State $Q.\operatorname{enqueue}((\tx,r))$ \Comment{add $\tx$ to the liveness queue on receiving at least $q$ signatures} \label{loc:sleepy-passive-live-queue}
    \EndOn
    \On{every round $r$} \label{loc:sleepy-passive-live-on-witness}
        \State $\LOGint \gets$ output by $P_{\mathrm{int}}$ at round $r$
        \For{$(\tx,r') \in Q$ such that $r' \leq r - u_{\mathrm{int}}$ and $\tx \notin \LOGint$}
            \State $\LOGint \gets \LOGint || \tx$ \label{loc:sleepy-passive-live-append}
        \EndFor
        \State $\LOGi{}{} \gets \LOGint$ \Comment{output log}
    \EndOn
    \end{algorithmic}
\end{algorithm}

%% file: figures/sleepy_passive_more_live_protocol.tex
\begin{figure}[tb]%
    \centering%
    \begin{tikzpicture}[
            x=1.5cm,
            y=1cm,
            subprotocol/.style={
                draw,
                align=center,
                inner sep=0.5em,
            },
            messages/.style={
                -latex,
            },
            party/.style={
                draw,
                dashed,
            },
        ]
        \scriptsize

        \coordinate (txsPos) at (0,0);
        \coordinate (signPos) at ($(txsPos)+(1,0)$);
        \coordinate (PiintPos) at ($(signPos)+(1.5,0)$);
        \coordinate (waitPos) at ($(PiintPos)+(3,1)$);
        \coordinate (concatPos) at ($(PiintPos)+(4,0)$);
        \coordinate (outputPos) at ($(concatPos)+(1.2,0)$);

        \node (txs) at (txsPos) {$\txs$};
        \node [subprotocol] (sign) at (signPos) {sign};
        \node [subprotocol,minimum height=3em,minimum width=3em] (Piint) at (PiintPos) {$\Pi_{\mathrm{int}}$};
        \node [subprotocol] (wait) at (waitPos) {delay $u_{\mathrm{int}}$};
        \node [subprotocol] (concat) at (concatPos) {$\LOGint||\txs$};
        \node (output) at (outputPos) {$$\LOGi{}{}$$};

        \draw [messages] (txs) -- (sign);
        \draw [messages] (sign) -- (Piint);
        \draw [messages] (sign) -- ++(0.5,0) |- (wait) node [pos=1,above,anchor=south east,align=right] (ifqsignedlabel) {if $\geq q$\\signed};
        \draw [messages] (Piint) -- (concat) node [pos=0.2,above] {$\LOGint$};
        \draw [messages] (wait) -| (concat) node [midway,right] (txsoutputlabel) {$\txs$};
        \draw [messages] (concat) -- (output);

        \draw [party] ([xshift=-6pt,yshift=-4pt] sign.south west |- Piint.south) rectangle ([xshift=6pt,yshift=2pt] Piint.east |- ifqsignedlabel.north);
        \draw [draw=none] ([xshift=-6pt,yshift=-4pt] sign.south west |- Piint.south) |- ([xshift=6pt,yshift=2pt] Piint.east |- ifqsignedlabel.north) node [pos=0.5,anchor=south east,rotate=90] {Validators};

        \draw [party] ([xshift=-6pt,yshift=-2pt] ifqsignedlabel.south west |- Piint.south) rectangle ([xshift=6pt,yshift=2pt] concat.east |- ifqsignedlabel.north);
        \draw [draw=none] ([xshift=-6pt,yshift=-2pt] ifqsignedlabel.south west |- Piint.south) |- ([xshift=6pt,yshift=2pt] concat.east |- ifqsignedlabel.north) node [pos=1,anchor=south west,rotate=-90] {Clients};

    \end{tikzpicture}
    \caption[]{%
        A family of protocols that achieves any resilience $\tL + \tS < 1$ and $\tS < \tL$ for \sleepy \passive or \alert \passive clients (lower right triangle of \cref{fig:resilience-sync-sleepy-passive,fig:resilience-sync-alert-passive}).
        The internal protocol $\Piint$ is any \smrshort protocol achieving all resilience pairs $\tS < 1/2, \tL < 1/2$ for \sleepy \passive clients (\eg Sync HotStuff~\cite{synchotstuff}).
        On receiving transaction $\tx$, validators sign it and broadcast the signature before processing it as an input to $\Piint$.
        A client, on receiving transaction $\tx$ signed by $q > \tS n$ validators,
        and after waiting $u_{\mathrm{int}}$ rounds (where $u_{\mathrm{int}}$ is the maximum latency of $\Piint$), if $\tx$ is not included in the log $\LOGint$ output by the client from $\Pi$, concatenates $\tx$ to $\LOGint$ to output the final confirmed log $\LOGi{}{}$.
    }%
    \label{fig:sleepy-passive-more-live-protocol}%
\end{figure}%

%% file: sections/05_sleepy_active.tex
\subsection{\SLeepy \ACtiv Clients (\cref{fig:resilience-sync-sleepy-active})}
\label{sec:sleepy-active}

\import{./sections/}{05a_sleepy_active_imp.tex}

\import{./sections/}{05b_sleepy_active_ach.tex}

%% file: sections/05a_sleepy_active_imp.tex
\myparagraphNoDot{Impossibility for \SLeepy \ACtiv Clients}
\label{sec:sleepy-active-imp}

\begin{theorem}
\label{thm:sleepy-active-synchrony-converse}
Under synchrony, with \statpart and \sleepy \activ clients, no protocol can achieve resiliences $(\tL, \tS) \in [1/2,1] \times [1/2,1]$.
\end{theorem}

Suppose a protocol can achieve $\tL=\tS=1/2$.
Let $P$ and $Q$ be two disjoint sets of $n/2$ validators and $k_1,k_2$ be two clients.
Consider two worlds where the $(P,k_1)$ and $(Q,k_2)$ are adversary respectively.
In both worlds, the adversary parties initially do not communicate with honest parties.
By liveness, in world $i \in \{1,2\}$, client $k_i$ awake since the start outputs transaction $\tx_i$ by round $u$ after hearing from the honest validators.
In both worlds, a client $k_3$ awakes after round $u$ and hears from \emph{all parties} including the adversary ones.
By liveness, $k_3$ also outputs $\tx_i$ in world $i$.
However, the two worlds are indistinguishable for $k_3$ because $P-Q$ and $k_1-k_2$ exchange their roles in the two worlds, implying its log is the same and must contain both $\tx_1$ and $\tx_2$ in both worlds, leading to a safety violation in at least one world.
The full proof is in \cref{sec:proofs-sync-sleepy-active-imp}.

%% file: sections/05b_sleepy_active_ach.tex
\myparagraphNoDot{Achievability for \SLeepy \ACtiv Clients (Safety-Favoring)}
\label{sec:sleepy-active-ach}
\label{sec:sleepy-active-ach-safe}
This protocol
achieves $(\tL,\tS)$ for all $\tL < 1/2$ and $\tS = 1$. 
In particular, it is \emph{always} safe.
It uses a 
\emph{freezing gadget}
applied to an internal \smrshort protocol $\Piint$
that is \emph{certifiable}~\cite{aa,roughgarden} (\cf public verifiability~\cite{mtbft}).
Quorum-based protocols such as 
HotStuff~\cite{hotstuff,hotstuff2}, Streamlet~\cite{streamlet}, Tendermint~\cite{tendermint}, Casper~\cite{casperffg}, and their synchronous variants such as Sync HotStuff~\cite{synchotstuff} and Sync-Streamlet~\cite[Sec.~4]{streamlet}\footnote{The synchronous variants can be made certifiable by having validators broadcast a signature on their `committed'/`finalized' logs~\cite[Sec.~4.2]{mtbft}.}
are certifiable.
In these protocols, clients output a log on receiving 
enough
quorum certificates which form a \emph{\transcript}
that other clients can verify \emph{non-interactively}.
Certifiable safety means that
adversaries controlling $\leq \tC$ validators cannot forge \transcripts certifying two conflicting logs. 

\begin{definition}[{Certifiable protocol}]
\label{def:certifiable}
    An \smrshort protocol $\Pi$
    is certifiable if there exists
    a computable functionality $\transcribe$ (the \emph{\transcript producer})
    and
    a computable deterministic non-interactive function
    $\untranscribe$ (the \emph{\transcript consumer})
    such that
    when a client $p$ invokes $\transcribe()$ at round $r$, it produces a \transcript $\transcript$ such that $\untranscribe(\transcript) = \LOGi{p}{r}$.
\label{def:certifiable-safety}
    A certifiable protocol $\Pi$ 
    is certifiably safe if
    $\Pi$ is safe,
    and moreover,
    if at any round $r$,
    the adversary outputs a \transcript $\transcript$ such that $\untranscribe(\transcript) = \LOG$,
    then for all clients $q$, for all rounds $s$,
    $\LOG \consistent \LOGi{q}{s}$.
    A certifiable protocol $\Pi$ achieves certifiable safety resilience $\tC$ if $\Pi$ is certifiably safe with overwhelming probability over executions with $f \leq \tC n$.
\end{definition}

\import{./algorithms/}{alg_freezing.tex}

\import{./figures/}{freezing_protocol.tex}
The protocol is described in \cref{alg:freezing} and is illustrated as a block diagram in \cref{fig:freezing-protocol}.
Each client runs a client for the internal protocol $\Piint$
(see $\Piint$ in \cref{fig:freezing-protocol}, \cref{alg:freezing}~\cref{loc:freezing-base-protocol}),
periodically 
outputs a \transcript $\transcript$ for the internal log $\LOGint$, and gossips it to the network.
It similarly processes \transcripts received from other clients.
After waiting $\Delta$ rounds (\cref{alg:freezing}~\cref{loc:freezing-wait}), the client outputs the log $\LOGint$ iff it has seen no conflicting logs (\cref{alg:freezing}~\cref{loc:freezing-check-conflict,loc:freezing-output}).
Applying this gadget to an internal  protocol $\Piint$ with
certifiable safety and liveness resilience $\tL_\mathrm{int} < 1/2, \tC_\mathrm{int} < 1/2$ (\eg, Sync HotStuff~\cite{synchotstuff}) results in a protocol $\PI$ (\cref{fig:freezing-protocol}) with resilience $\tL < 1/2, \tS = 1$.
Safety of $\PI$ is ensured by the freezing gadget.
See \cref{fig:freezing-safety-proof} for a visual safety proof.
Liveness of $\PI$ comes from the internal protocol's liveness and certifiable safety,
which guarantee that new transactions are included in the internal log
and no conflicting \transcripts are seen.
The full proof is in \cref{sec:proofs-sync-sleepy-active-ach-safe}.

Similar to our freezing protocol, the Tendermint client implementation `panics'~\cite{tendermint-panic} upon detecting~\cite{tendermint-detect-fork} two blocks confirmed at the same height.
But the client may have already confirmed one of the blocks before panicking, potentially causing a safety violation.
If Tendermint clients instead waited for $\Delta$ time before confirming blocks (and remained online after panicking to relay evidence of misbehavior), they would have implemented our freezing protocol. This further shows that the client protocols we study in the paper are practical and reflect real implementations.

\import{./figures/}{freezing_safety_proof.tex}

\begin{theorem}
\label{thm:freezing-resilience}
    In a synchronous network with \statpart and \sleepy \activ clients,
    for all $(\tL,\tS)$ with $\tL < 1/2$ and $\tS \leq 1$,
    $\PI$
    with Sync HotStuff as its internal protocol
    achieves $(\tL,\tS)$.
\end{theorem}

\myparagraphNoDot{Achievability for \SLeepy \ACtiv Clients (Liveness-Favoring)}
\label{sec:sleepy-active-ach-live}
The protocol $\Pisal$ achieves any $(\tL,\tS)$ with $\tL=1, \tS < 1/2$
under \statpart.
The protocol is very similar to $\Pi_{\mathrm{live}}^q$ the liveness-favoring protocol for \sleepy \passive clients (\cref{sec:passive-ach-live-static}) but simpler, so we only describe the key difference.
Unlike \passive clients in $\Pi_{\mathrm{live}}^q$ (\cref{sec:passive-ach-live-static}), \emph{\activ} clients add transactions to their liveness queue as soon as they receive them.
This is because while \passive clients require signatures from $q > \tS n$ validators to infer that at least one honest validator received the transaction,
\activ clients can do so by gossiping the transaction themselves.
The full protocol and security proof are in \cref{sec:proofs-sync-sleepy-active-ach-live}.

\begin{theorem}
\label{thm:sleepy-active-synchrony-achievability-2}
In a synchronous network with \statpart and \sleepy \activ clients,
for all $(\tL,\tS)$ with $\tL \leq 1, \tS < 1/2$,
$\Pisal$
with Sync HotStuff as its internal protocol
achieves $(\tL,\tS)$.
\end{theorem}

%% file: algorithms/alg_freezing.tex
\begin{algorithm}[tbp]
\caption{Freezing protocol for \sleepy \activ clients}
\label{alg:freezing}
\begin{algorithmic}[1]
\scriptsize
\LineComment{Code for client}
\On{\Call{init}{$\valset,\genesis$}}
\label{loc:freezing-init}
    \State $P_{\mathrm{int}} \gets \text{new } \Piint(\valset, \genesis)$ \Comment{instantiate a new $\Piint$ client} 
    \label{loc:freezing-base-protocol}
    \State $\msgSet \gets \emptyset$ \Comment{set of valid logs seen so far}
    \State $\LOGi{}{} \gets \genesis$ \Comment{output log of the combined protocol $\PI$}
\EndOn

\On{\transcript $\transcript$ output by $P_{\mathrm{int}}.\transcribe()$ once per round or $\transcript$ received from network} \label{loc:freezing-on-witness}
    \State $\LOGint \gets \untranscribe(\transcript)$ \label{loc:freezing-verify} \Comment{extract log from \transcript}
    \State $\msgSet \gets \msgSet \cup \{\LOGint\}$ \Comment{add $\LOGint$ to set of logs seen so far}
    \label{loc:freezing-record-log}
    \State $\operatorname{gossip}(\transcript)$ \label{loc:freezing-gossip} \Comment{send the transcript to all parties} %
    \State $\operatorname{wait}(\Delta)$ \Comment{meanwhile, continue processing other events} \label{loc:freezing-wait}
    \If{$\LOGint \not\preceq \LOGi{}{}$ \textbf{and} $\forall \LOG' \in \msgSet \colon \LOGint \consistent \LOG'$} \Comment{log has grown, no conflicting logs} 
    \label{loc:freezing-check-conflict}
        \State $\LOGi{}{} \gets \LOGint$ 
        \label{loc:freezing-output}
    \EndIf
\EndOn
\end{algorithmic}
\end{algorithm}

%% file: figures/freezing_protocol.tex
\begin{figure}[tb]%
    \centering%
    \begin{tikzpicture}[
            x=1.6cm,
            y=1.6cm,
            subprotocol/.style={
                draw,
                align=center,
                inner sep=0.5em,
            },
            messages/.style={
                -latex,
            },
            party/.style={
                draw,
                dashed,
            },
        ]
        \scriptsize

        \coordinate (txsPos) at (0,0);
        \coordinate (piPos) at ($(txsPos)+(1,0)$);
        \coordinate (netPos) at ($(piPos)+(1,-1)$);
        \coordinate (waitPos) at ($(piPos)+(2,0)$);
        \coordinate (decidePos) at ($(waitPos)+(1,0)$);
        \coordinate (outputPos) at ($(decidePos)+(1.5,0)$);

        \node (txs) at (txsPos) {$\txs$};
        \node [subprotocol,minimum height=3em,minimum width=3em] (pi) at (piPos) {$\Piint$};
        \node [subprotocol,inner sep=2pt,cloud,cloud ignores aspect] (net) at (netPos) {network};
        \node [subprotocol] (wait) at (waitPos) {delay $\Delta$};
        \node [subprotocol,inner sep=3pt,regular polygon,regular polygon sides=3,shape border rotate=-90] (decide) at (decidePos) {$\consistent$};
        \node [align=left] (output) at (outputPos) {output\\log $\LOGi{}{}$};

        \draw [messages] (txs) -- (pi);
        \draw [messages] (pi) -- (wait) node [pos=0.4,above,align=center] (LOGintlabel) {internal\\log $\LOGint$};
        \draw [messages] (wait) -- (decide);
        \draw [messages] (decide) -- (output);

        \draw [messages] ([xshift=-3pt] piPos -| netPos) -- ([xshift=-3pt,yshift=-1.5pt] net.north);
        \draw [messages] ([xshift=3pt,yshift=-1.5pt] net.north) -- ([xshift=3pt] piPos -| netPos) node [right,pos=0.25] {gossip};

        \draw [messages] (txsPos -| netPos) ++ (0.35,0) |- ++(0,0.4) -| (decide.north);

        \draw [party] ([xshift=-6pt,yshift=-7pt] pi.south west) rectangle ([xshift=6pt,yshift=6pt] decide.east |- LOGintlabel.north);
        \draw [draw=none] ([xshift=-6pt,yshift=-7pt] pi.south west) -| ([xshift=6pt,yshift=6pt] decide.east |- LOGintlabel.north) node [pos=0.5,below,anchor=north east] {$\PI$};

    \end{tikzpicture}
    \caption[]{%
        \newcommand{\smallTriangle}{
            \tikz[baseline=-0.3em]{
                \node[regular polygon, regular polygon sides=3, shape border rotate=-90, draw, inner sep=0] {$\scriptscriptstyle\consistent$};
            }
        }
        The freezing protocol $\PI$ that achieves $\tL < 1/2, \tS = 1$ for \sleepy \activ clients.
        The internal protocol $\Piint$ is any certifiable \smrshort protocol that can achieve all $\tS, \tL < 1/2$.
        On seeing a log $\LOGint$ 
        from $\Piint$ or the network, the client gossips 
        $\LOGint$ (formally, the corresponding \transcript), and waits for $\Delta$ rounds.
        The \emph{conflict resolution} component \smallTriangle remembers the set $\msgSet$ of all logs it ever received at the input port on its top.
        On receiving $\LOG$ at the input port on its left, this component outputs
        $\LOG$ if there were no conflicting logs in $\msgSet$
        (see \cref{alg:freezing}~\cref{loc:freezing-check-conflict,loc:freezing-output}).
    }%
    \label{fig:freezing-protocol}%
\end{figure}%

%% file: figures/freezing_safety_proof.tex
\begin{figure}[tb]
    \centering
    \begin{tikzpicture}[y=1cm,x=1cm]
        \scriptsize

        \draw (0,0) -- (7,0) node[right] {Round}; %
        \draw[-latex] (6.5,0) -- (7,0); %

        \coordinate (r) at (4,0);
        \coordinate (rdelta) at (1,0);
        \draw (rdelta) +(0,0.1) -- +(0,-0.1) node[below, anchor=base, yshift=-1em] {$r-\Delta$};
        \draw (r) +(0,0.1) -- +(0,-0.1) node[below, anchor=base, yshift=-1em] {$r$};

        \draw [latex-,densely dotted] ([yshift=0.2cm]r.north) -- ([yshift=1cm]r.north) node [anchor=south west,align=left,xshift=-1em] {\DiagramStep{1} $p$ outputs $\LOG$};

        \draw [latex-,densely dotted] ([yshift=0.2cm]rdelta.north) -- ([yshift=0.7cm]rdelta.north) node [anchor=south west,align=left,xshift=-1em] {\DiagramStep{2} $p$ sees $\LOG$,\\gossips $\LOG$};

        \draw [latex-,densely dotted] ([yshift=0.2cm]r.north) |- ([yshift=0.6cm,xshift=0.7cm]r.north) node [anchor=west,align=left,xshift=0em] {\DiagramStep{3} All clients see $\LOG$,\\do not output $\LOG' \inconsistent \LOG$};
        
    \end{tikzpicture}
    \caption[]{Illustration for the freezing protocol's safety which is maintained during adversary majority (\cref{thm:freezing-resilience}).
    \DiagramStep{1}~Suppose that at round $r$, a client $p$ outputs a log $\LOG$.
    \DiagramStep{2}~The client must have seen $\LOG$ (and its \transcript $\transcript$) either from the internal protocol $\Pi$ or from the network
    latest by round $r - \Delta$, at which point it must have sent $\LOG$
    (and $\transcript$) to all other clients. 
    \DiagramStep{3}~Thus, by round $r$, all clients must have seen $\LOG$ 
    and thereafter will never output a log that conflicts with $\LOG$.
    }
    \label{fig:freezing-safety-proof}
\end{figure}

%% file: sections/06_alert_active.tex
\subsection{\ALert \ACtiv Clients (\cref{fig:resilience-sync-alert-active})}
\label{sec:alert-active}

\myparagraphNoDot{Achievability for \ALert \ACtiv Clients}
\label{sec:alert-active-static-ach}
Dolev and Strong~\cite{dolev-strong}, and Lamport, Shostak, and Pease~\cite{byzantine-generals}
presented protocols for the Byzantine Generals problem when all but one validator are adversary.
In their problem setting, there are no clients.
Instead, a designated leader (`general')
broadcasts a value
and all honest validators (`lieutenants') must agree on a common value, which matches the leader's value if the leader is honest.
However, the Dolev-Strong protocol can be extended to support \alert \activ clients; the client follows the same rules a validator uses to output a value.

An SMR protocol can be created
by having each validator propose a block as the leader in an instance of the Dolev-Strong protocol, allowing \alert \activ clients to agree on a unique block (possibly a default empty block) per leader.
The client's log is formed by concatenating these blocks in a set order, repeating the process to grow the log.
This approach is presented in \cite{vitalik-dolev-strong,bcube}, with optimizations in \cite{flint-adv-maj}.
We recap the protocol from \cite{vitalik-dolev-strong} in \cref{alg:smr-using-dolev-strong} and prove its security in \cref{sec:proofs-sync-alert-active-ach}.

\begin{corollary}
    \label{thm:alert-active-static-ach}
    In a synchronous network with \statpart and \alert \activ clients,
    for all $(\tL,\tS)$ with $\tL, \tS<1$,
    \cref{alg:smr-using-dolev-strong} achieves $(\tL,\tS)$.
\end{corollary}

We note that no protocol can achieve $\betaL=\betaS=1$ (\cf~\cref{sec:proofs-dynamic-alert-active-ach-impossibility}).

%% file: sections/07_sleepy_validators.tex
\section{Synchrony with \DynPart}
\label{sec:dynamic}

\subsection{\SLeepy \PAssive, \ALert \PAssive Clients (\cref{fig:resilience-dynamic-sleepy-passive,fig:resilience-dynamic-alert-passive})}
\label{sec:passive-dynamic}

\myparagraphNoDot{Impossibility for \ALert \Passive Clients}
\label{sec:passive-imp-dynamic}
This follows from~\cref{thm:alert-passive-synchrony-converse,lem:hierarchy}.

\begin{corollary}
\label{corollary:alert-passive-da-converse}
In a synchronous network with \dynpart and \alert \passive clients, no protocol can achieve resiliences $(\betaL,\betaS)$ such that $\betaL+\betaS \geq 1$.
\end{corollary}

\myparagraphNoDot{Achievability for \Sleepy \Passive Clients (Equal Resiliences)}
\label{sec:sleepy-passive-dynamic-ach}
This follows from Sleepy Consensus~\cite[Theorem 1]{sleepy} and Goldfish~\cite[Theorem 2]{goldfish}.

\begin{corollary}
\label{thm:sleeepy-passive-da-achievability-0}
In a synchronous network with \dynpart and \sleepy \passive clients,
for all $(\betaL,\betaS)$ with $\betaL < 1/2, \betaS < 1/2$,
the Sleepy Consensus protocol~\cite{sleepy} and Goldfish~\cite{goldfish} achieve $(\betaL,\betaS)$.
\end{corollary}

\myparagraphNoDot{Achievability for \Sleepy \Passive Clients (Safety-Favoring)}
\label{sec:passive-ach-dynamic}
\label{sec:passive-ach-safe-dynamic}
To achieve any resilience $(\betaL, \betaS)$ with $\betaL + \betaS < 1$ and $\betaL < \betaS$, we modify the Goldfish protocol~\cite{goldfish}.
In a nutshell, in Goldfish, voters select a block to vote for by walking down a tree of blocks, and at each fork, selecting the subtree with the largest number of votes from the previous slot.
Our key modification is to instead select the subtree with at least $\phi$ fraction of the total votes from the previous slot, where $\phi \in (\betaS, 1-\betaL]$.
The details of this protocol are in \cref{sec:proofs-dynamic-sleepy-passive-ach-safe}.

\begin{theorem}
\label{thm:sleeepy-passive-da-achievability-1}
In a synchronous network with \dynpart and \sleepy \passive clients,
for all $(\betaL,\betaS)$ with $\betaL + \betaS < 1$ and $\betaL < 1/2$,
Goldfish modified with
$\phi \in (\betaS, 1-\betaL]$
achieves $(\betaL,\betaS)$.
\end{theorem}

\myparagraphNoDot{Achievability for \Sleepy \Passive Clients (Liveness-favoring)}
\label{sec:passive-ach-live-dynamic}
For any resilience $(\betaL,\betaS)$ with $\betaL + \betaS < 1$ and $\betaL \geq \betaS$, we show a protocol $\Pi_{\mathrm{live}}^\phi$ achieving $(\betaL,\betaS)$.
The protocol is very similar to $\Pi_{\mathrm{live}}^q$, the liveness-favoring protocol for \sleepy \passive clients under \statpart (\cref{sec:passive-ach-live-static}).
The key difference is that to add a transaction to the liveness queue, clients require a fraction $\phi$ of validators to sign the transaction (unlike a fixed number $q$ in $\Pi_{\mathrm{live}}^q$).
More precisely,
at every round $r = \ell \Delta$ for some $\ell \in \mathbb{Z}$, each client calculates the number $T_\tx$ of validators that signed a transaction $\tx$.
It also calculates the number $T_{\ell-1}$ of unique validators that have either sent a `heartbeat' message for round $(\ell-1)\Delta$ or a signature on some transaction in the past.
Then, if $T_\tx / T_{\ell-1} \geq \phi$, the client adds $\tx$ to its liveness queue. The full protocol and security proof are in \cref{sec:proofs-dynamic-sleepy-passive-ach-live}.

\begin{theorem}
\label{thm:sleepy-passive-da-achievability-2}
In a synchronous network with \dynpart and \sleepy \passive clients,
for all $(\betaL,\betaS)$ with $\betaL + \betaS < 1$ and $\betaL \geq 1/2$,
the protocol $\Pi_{\mathrm{live}}^\phi$ with $\phi \in (\betaS, 1-\betaL]$
achieves $(\betaL,\betaS)$.
\end{theorem}

\subsection{\SLeepy \ACtiv Clients (\cref{fig:resilience-dynamic-sleepy-active})}
\label{sec:sleepy-active-dynamic}

\myparagraphNoDot{Impossibility for \SLeepy \ACtiv Clients}
\label{sec:sleepy-active-dynamic-imp}

\begin{theorem}
\label{thm:sleepy-active-da-converse}
In a synchronous network with \dynpart and \sleepy \activ clients, no protocol can achieve $(\betaL,\betaS)$ with $\betaL+\betaS \geq 1$ and $\betaS \geq 1/2$.
\end{theorem}

The proof is similar to \cref{thm:sleepy-active-synchrony-converse}.
Suppose a protocol can achieve $\betaL=25\%,\betaS=75\%$.
Let $P$ and $Q$ be two disjoint sets of $0.75 n$ and $0.25 n$ validators respectively, and $k_1,k_2$ be two clients awake since the start.
Consider two worlds, 1 and 2, where the $(P,k_1)$ and $(Q,k_2)$ are adversary respectively.
In both worlds, the adversary parties initially do not communicate with honest parties.
Note that liveness must hold in world 2 because only $25\%$ validators are adversary, and in world 1 because the $75\%$ adversary validators appear indistinguishable from sleepy honest validators.
Thus, in each world, client $k_i$ outputs transaction $\tx_i$ in its log.
In both worlds, a client $k_3$ awakes after round $u$ and hears from \emph{all parties} including the adversary ones.
However, the two worlds are indistinguishable for $k_3$ because $P-Q$ and $k_1-k_2$ exchange their roles, implying that its log must be the same in both worlds, and it contains $\tx_2$ as world 2 has liveness, leading to a safety violation in at least one world.
The full proof is in \cref{sec:proofs-dynamic-sleepy-active-imp}.

\myparagraphNoDot{Achievability for \Sleepy \Activ Clients (Liveness-Favoring)}
\label{sec:sleepy-active-dynamic-ach}
The protocol $\Pisal$ in \cref{sec:sleepy-active-ach-live} does not rely on the validators for its liveness, and its safety only requires safety and liveness of the internal protocol in a closed-box manner.
Therefore, the same protocol, when instantiated with an internal protocol for \dynpart (\eg, Sleepy Consensus~\cite{sleepy}), achieves the following.

\begin{theorem}
\label{thm:sleepy-active-synchrony-dynamic-achievability-2}
In a synchronous network with \dynpart and \sleepy \activ clients,
for all $(\betaL,\betaS)$ with $\betaL \leq 1, \betaS < 1/2$,
$\Pisal$ (\cref{sec:sleepy-active-ach-live})
with the Sleepy Consensus protocol~\cite{sleepy} as its internal protocol
achieves $(\tL,\tS)$.
\end{theorem}
\begin{proof}
Follows from \cite[Theorem~1]{sleepy} and \cref{lem:append-liveness,lem:append-safety}.
\end{proof}

\subsection{\Alert \Activ Clients (\cref{fig:resilience-dynamic-alert-active})}
\label{sec:alert-active-dynamic}

\myparagraphNoDot{Achievability for \Alert \Activ Clients}
\label{sec:alert-active-dynamic-ach}
We show that the SMR protocol based on Dolev-Strong (\cref{sec:alert-active-static-ach}, \cref{alg:smr-using-dolev-strong}) achieves any $\betaL<1, \betaS<1$ even under \dynpart.
Under \dynpart with \alert \activ clients, 
when a majority of the awake validators are adversary, the clients must output safe and live logs even though the validators themselves may not agree on a log (since validators are \sleepy and \activ, the impossibility in \cref{fig:resilience-dynamic-sleepy-active} applies to them).
Thus, the challenge is to design the validator's code to behave correctly even without knowing what happened while it was sleeping.

This challenge resolves itself due to the following observations.
First, while the SMR protocol (\cref{alg:smr-using-dolev-strong}) runs instances of Dolev-Strong one after the other, each instance does not depend on the previous instances.
Second, within an instance, since the Dolev-Strong protocol (\cref{alg:dolev-strong-clients}) guarantees agreement and validity
when all but one validator are adversary under \statpart(\cref{lem:dolev-strong-clients-security}),
it does so even when only one validator is honest and awake throughout the instance (all honest validators who sleep could be considered adversary).
Moreover, we don't even require the same honest validator to be awake throughout the instance but only require that for each round during the instance, \emph{some} honest validator is awake.
Finally, since each validator (including the leader) signs only one message per instance, it may sleep after it does so without affecting the protocol's remaining execution.
Thus, sleepy validators can faithfully run \cref{alg:dolev-strong-clients}.
We explain these surprising observations and prove security in \cref{sec:proofs-dynamic-alert-active-ach}.

\begin{theorem}
\label{thm:alert-active-dynamic-ach}
In a synchronous network with \dynpart and \alert \activ clients, for all $(\betaL,\betaS)$ such that $\betaL,\betaS<1$, \cref{alg:smr-using-dolev-strong} achieves $(\betaL,\betaS)$.
\end{theorem}

%% file: sections/02_related_work.tex
\section{Related Work}
\label{sec:related}

\mysubparagraph{No clients}
Much of the classic \smrshort literature~\cite{synchotstuff,abraham-sync-byz-consensus,xft,streamlet,sleepy,snowwhite,david2018ouroboros,kiayias2017ouroboros,backbone} did not explicitly consider clients.
Let's call this the `no clients' model.
In this model, validators (\aka `replicas' or `nodes') output logs,
and
in any protocol with resilience $\tL + \tS < 1$,
\sleepy \passive clients may learn the log by querying a quorum of $\tS n+1$ validators~\cite{synchotstuff,abraham-sync-byz-consensus,xft} (see \cref{fig:resilience-sync-sleepy-passive}).
However, \alert and/or \activ clients may use other means to learn the log.
For example,
\alert \activ clients can run the same `confirmation logic' that validators use.
This makes the `no clients' model equivalent to \alert \activ clients (\cref{fig:resilience-sync-alert-active}).
However, under \dynpart, the `no clients' model is equivalent to \sleepy \activ clients (\cref{fig:resilience-dynamic-sleepy-active}), possibly weaker than \alert \activ clients.
Reliable broadcast and Byzantine agreement, typically defined without clients, can also apply to different client types (\eg, \cref{def:byzantine-generals}).

\mysubparagraph{\Sleepy \Passive Clients}
\Sleepy clients have been called sleepy~\cite{sleepy}, late-spawning~\cite{shi-rethinking,bitcoin-enhanced,snowwhite}, and lazy~\cite{pye-consensus-50}.
It has been proven that no protocol with \sleepy \passive clients can achieve both $\tL \geq \frac{1}{2}$ and $\tS \geq \frac{1}{2}$~\cite{schneider-survey,sleepy,shi-rethinking}.
When safety and liveness are decoupled, \cite{aav1,mtbft} prove $\tL + \tS \geq 1$ is impossible.
But their proofs require a stronger notion of security, \emph{certifiability},
\ie, the protocol produces \emph{non-interactively} verifiable certificates (\cf \cref{def:certifiable},~\cite{roughgarden}).
Certifiable protocols are also secure for \sleepy \passive clients, but the converse is not true (\eg, longest-chain consensus~\cite{nakamoto_paper,sleepy,snowwhite,david2018ouroboros}
supports
\sleepy \passive clients, but lack certificates since clients must check for longer chains).
Thus, the impossibility results of \cite{aav1,mtbft} apply to certifiable protocols, but not necessarily to \sleepy \passive clients. We prove (in \cref{thm:alert-passive-synchrony-converse}) that $\tL + \tS \geq 1$ is impossible for both \sleepy \passive and \alert \passive clients.

\mysubparagraph{\Sleepy \activ clients}
The idea of \sleepy clients gossiping out-of-band to detect liveness or safety violations isn't new~\cite{mazieres-freezing}, although we are the first to formalize the \sleepy \activ client model and apply it to \smrshort.
Validators in some protocols~\cite{synchotstuff,streamlet} with $1/2$ resilience also wait $\Delta$ to detect conflicts, but achieving $\tS = 99\%$ requires \activ clients.
Concurrent work~\cite{leshno-economic-permissionless} proposes a variant of Bitcoin~\cite{nakamoto_paper} (which supports \dynpart) that achieves resilience $\betaL < 1/2, \betaS = 1$ in the \emph{\alert} clients model. 
However, their protocol doesn't achieve this resilience for \emph{\sleepy} clients who awaken after the adversary has corrupted a majority of validators.
In \cref{app:stubborn-nakamoto}, we show a concrete attack and prove that no protocol can achieve these resiliences under \dynpart with \emph{\sleepy} clients throughout the execution (see also \cref{fig:resilience-dynamic-sleepy-active,sec:sleepy-active-dynamic-imp}).%
\footnote{The model we use assumes a known set of validators of which $f$ are corrupted while the model in~\cite{leshno-economic-permissionless} instead uses proof-of-work and assumes limited adversary hashing power. Although these models are incomparable, our impossibility proof (\cref{thm:sleepy-active-da-converse}) applies to that model as well (see detailed discussion in \cref{app:stubborn-nakamoto}).}

\mysubparagraph{\Dynpart}
Protocols achieving $\tL<1/2,\tS<1/2$ for \dynpart appear in \cite{sleepy,mmr-da-consensus,goldfish,instant-conf-sleepy,momose-ren-sleepy,thunderella,losa-gafni-da-consensus}.
With \dynpart and `no clients', \cite{sleepy,shi-rethinking} prove that no protocol can simultaneously achieve $\betaS \geq 1/2, \betaL \geq 1/2$.

%% file: appendices.tex
\appendix
\import{./sections/}{08_discussion.tex}
\import{./sections/}{A01_model_full.tex}
\import{./sections/}{A02_proofs_sync.tex}
\import{./sections/}{A03_proofs_sleepy_validators.tex}
\import{./sections/}{09_psync.tex}
\import{./sections/}{A04_stubborn_nakamoto.tex}

%% file: sections/08_discussion.tex
\section{Discussion}
\label{sec:discussion}

\mysubparagraph{Proof-of-Stake}
\label{sec:discussion-pos}
The protocols in \cref{sec:passive-ach-live-static,sec:sleepy-active-ach-safe,sec:sleepy-active-ach-live,sec:passive-ach-live-dynamic,sec:sleepy-active-dynamic-ach} 
are independent of the internal protocol's validator selection mechanism, and can be applied to proof-of-stake.
The protocol in \cref{sec:passive-ach-safe-static} is based on Sync HotStuff and \cref{sec:passive-ach-safe-dynamic} on Goldfish~\cite{goldfish}, and similar protocols are deployed on proof-of-stake Ethereum. The protocol in \cref{sec:alert-active-static-ach} supports proof-of-stake~\cite{vitalik-dolev-strong}, but the one in \cref{sec:alert-active-dynamic-ach} 
doesn't as validators lack knowledge of the log and stake distribution.

\mysubparagraph{Preserving Security under Partial Synchrony}
\label{sec:discussion-partial-synchrony}
Partially-synchronous PBFT-style protocols (\eg, Casper~\cite{casperffg}, Tendermint~\cite{tendermint}, HotStuff~\cite{hotstuff} or HotStuff2~\cite{hotstuff2})
maintain safety under asynchrony with resilience $\tL<1/3, \tS<1/3$.
Applying our freezing gadget (\cref{sec:sleepy-active-ach-safe}) to such a protocol preserves this safety under asynchrony, while making it always safe under synchrony.
Safety is only compromised if \emph{both} the network is asynchronous, and the adversary exceeds $1/3$.

\mysubparagraph{Adaptive Corruption}
\label{sec:discussion-adaptive}
Our impossibility results only use static corruption, but all protocols inherit security under adaptive corruption from
Sync HotStuff variants~\cite{synchotstuff,mtbft}, Dolev-Strong~\cite{dolev-strong}, and Goldfish~\cite{goldfish}.

\mysubparagraph{Heterogeneous Clients}
\label{sec:discussion-flexible}
Heterogeneous clients 
are described
in~\cite{fbft,fbft2,oflex}.
In the Dolev-Strong-based SMR protocol (\cref{sec:alert-active-static-ach}), 
\alert \activ clients may output the resulting log achieving any $\tL,\tS<1$, 
\passive clients may query $\frac{n}{2} + 1$ validators achieving $\tL,\tS<1/2$,
and \sleepy \activ clients may further use the freezing gadget (\cref{sec:sleepy-active-ach-safe}) achieving $\tL<1/2, \tS=1$ or the liveness gadget (\cref{sec:sleepy-active-ach-live}) achieving $\tL=1, \tS<1/2$ simultaneously.

\mysubparagraph{Liveness-favoring Protocols}
Optimistic rollups can benefit from layer 1 blockchains that prioritize liveness since the liveness of the layer 1 is necessary for the timely inclusion of fraud proofs.
Although layer 1's safety is also often needed for the rollup's safety, rollups that settle on a different chain than the one they use for ordering the blocks~\cite{celestiums_blog} would need only liveness from the settlement chain.

%% file: sections/A01_model_full.tex
\section{Model}
\label{sec:app-model-full}

In this section, we describe the model and notation used in the rest of the paper.
A \smr (\smrshort) consensus protocol is a distributed protocol run among two types of parties: \emph{validators} and \emph{clients}.
Validators take inputs called \emph{transactions} and enable clients to agree on
a sequence of confirmed transactions called the \emph{log}.

\myparagraph{Time}
Time proceeds in logical units called \emph{rounds} indexed by $r = 0, 1, 2, \ldots$.
We assume the maximum clock offset between any two parties is bounded (any bounded clock offset can be absorbed into the network delay bound~\cite{sleepy}).

\myparagraph{Cryptography}
We assume an ideal random oracle as a common source of randomness for the protocols.
We assume probabilistic polynomial-time (PPT) adversaries
and the existence of collision-resistant hash functions and unforgeable digital signatures.

\myparagraph{Validators}
There is a fixed set $\valset$ of parties called \emph{validators} known to all parties.
We denote the number of validators by $n = |\valset|$.
We assume a public key infrastructure (PKI): each validator has a public key and a secret key with which they may sign their messages, and all parties know the public keys of all validators.
In this work, we adopt the permissioned setting~\cite{pye-roughgarden-permissionless,budish-pye-roughgarden-economic,sleepy} where the set of validators is fixed and known to all parties.
Our impossibility results also apply to the weaker proof-of-stake (in the literature also called ``quasi-permissionless''~\cite{pye-roughgarden-permissionless,budish-pye-roughgarden-economic}) setting where the set of validators is known but may change over time.
We discuss in \cref{sec:discussion} how our protocols can be adapted to the proof-of-stake setting.

\myparagraph{Clients}
Unlike validators, the number and identities of the clients is not known to all parties, and clients do not have public keys.
Thus, messages sent by clients cannot be authenticated by other parties and the adversary can impersonate an honest client (\cf `identity theft'~\cite{okun09}).

\myparagraph{Adversary}
For simplicity, we assume a \emph{static} adversary,
so that
at the beginning of the execution,
before any randomness is drawn,
the PPT adversary $\Adv$ corrupts
a set of validators $\corruptset{} \subseteq \valset$.
These validators are called \emph{adversary}
and are controlled by $\Adv$ (also known as Byzantine faults).
We also use \emph{adversary validator} to denote a corrupted validator.
Parties that are not adversary are called \emph{honest}.
The number of adversary validators is denoted by $f = |\corruptset{}|$.
Since our model allows clients to influence the execution by sending messages, we also allow $\Adv$ to corrupt any number of clients.
While a protocol's security may be conditioned on the number of adversary validators $f$, it must be independent of the number of adversary clients because the adversary may impersonate even honest clients due to the lack of authentication.
The adversary has access to the internal state of all adversary parties, including private keys.
Modeling static corruption makes our impossibility results stronger.
Protocols that we build as closed-box transformations of an existing protocol inherit the latter's security under adaptive corruption (details in \cref{sec:discussion}).

\myparagraph{Validator models}
We use the sleepy model of consensus~\cite{sleepy} to model intermittently crashed honest validators.
At every round $r$, a subset $\awakeset{r} \subseteq \valset$ of validators are awake while the rest are \emph{asleep}.
When asleep, validators behave like temporarily crashed nodes: they do not run computation or send messages.
Whenever a validator is awake, it knows the current round (\ie, it wakes up with a synchronized clock).
Based on the \wokeness of validators, we define two validator models:
\begin{itemize}
    \item \textbf{\Statpart}: All validators are always awake, ($\forall r \colon \awakeset{r} = \valset$).
    \item \textbf{\Dynpart}: At the start of each round $r$, $\Adv$ selects $\awakeset{r}$. Adversary validators are always awake. Other parties do not know whether a validator is awake or asleep.
\end{itemize}

\myparagraph{Client Models}
We classify clients along two orthogonal criteria:
\begin{itemize}
    \item \textbf{\Activ} clients may send messages to other parties, \textbf{\passive} clients do not.
    \item \textbf{\Alert} clients are always awake while \textbf{\sleepy} clients may be put to sleep by $\Adv$.
    When asleep, clients do not perform any computation, send messages, or output new logs.
\end{itemize}

This results in four client models shown in \cref{fig:models-hierarchy}.
For example, the \emph{\sleepy \activ} model means that all clients are \sleepy and \activ.
\Activ clients are newly introduced in this work. In deviation from previous literature, we allow clients to send messages and other parties to act on such messages.
However, due to the lack of a PKI for clients, their messages cannot be authenticated, so really the best they can do is relay messages received from validators to other clients.
On the one hand, this is a more accurate depiction of blockchain implementations in which communication is facilitated by a non-eclipsed gossip network comprised of both clients and validators.
On the other hand, this seemingly insignificant change in the network model enables us to circumvent impossibility results~\cite{aav1,mtbft}, thus solving \smrshort under higher adversary resilience than with \passive clients (\cref{fig:resilience-plots}).

\myparagraph{Network delay models}
Parties can send messages to each other and call a functionality $\operatorname{gossip}(.)$ to send a message to all other parties.
We consider two standard models of network communication: the synchronous model and the partially (eventually) synchronous model.
\begin{itemize}
    \item \textbf{Synchronous model:}
    There is a known constant $\Delta$ such that if an honest party sends a message at round $r$, then every honest party receives the message by round $r + \Delta$.
    Within this bound, the adversary chooses when each honest party receives each message.%
    \footnote{Gossip networks have been shown to maintain connectivity, and thus synchrony, even under adversary majority~\cite{scuttlebutt,bftflooding,bftflooding2}.}
    \item \textbf{Partially synchronous model:} there exists a round $\GST$ (unknown to honest parties) and a known constant $\Delta$ such that if an honest party sends a message at round $r$, then every honest party receives the message by round $\max\{r,\GST\} + \Delta$.
    The adversary chooses $\GST$ adaptively and decides when each honest party receives each message within these bounds.
\end{itemize}

The adversary may also inject some of its own messages, and a party does not know the time at which other parties received the messages.
Note that in both models, messages are delivered to the inbox of asleep parties but they can process the messages only after awakening
(like the model in \cite{sleepy}).
In practice, equivalent behavior can be achieved by having the awakening party query online parties who reply with the `important' past messages the awakening party then processes (\eg, `initial block download'~\cite{btcdevp2pnetworkheadersfirst}), with the weak extra assumption that at least one (potentially different each round) honest party stays awake between consecutive rounds to relay the messages.
Thus, while a \sleepy party still receives all messages an \alert party receives, \sleepy parties are strictly less powerful because they cannot
record when they received a message.

\myparagraph{Notation} We use $A \preceq B$ to denote that sequence $A$ is a (not necessarily strict) prefix of the sequence $B$. We use $A \consistent B$ (`$A$ is consistent with $B$') as a shorthand for $A \preceq B \,\lor B\, \preceq A$.

\myparagraph{Definition of \smrshort}
At the start of each round, each awake party may receive some transactions as input.
At the end of every round $r$, each awake honest client $k$ outputs a log (sequence of transactions) $\LOGi{k}{r}$.
For a client $k$ asleep at round $r$, let $\LOGi{k}{r} = \LOGi{k}{r-1}$.
For all clients $k$, $\LOGi{k}{0} = \genesis$.
We define the following properties for an \smrshort protocol:

\begin{definition}[Safety]
    \label{def:safety-full}
    An \smrshort protocol $\Pi$ is \emph{safe}
    iff
    for all rounds $r,s$
    and all honest clients $k,k'$,
    $\LOGi{k}{r} \consistent \LOGi{k'}{s}$.
\end{definition}

\begin{definition}[Liveness]
    \label{def:liveness-full}
    An \smrshort protocol $\Pi$ is \emph{live} with latency $u$
    iff
    for all rounds $r$
    ($r \geq \GST$ for partial synchrony),
    if a transaction $\tx$ was received by an awake honest validator or \activ client before round $r - u$,
    then for all honest clients $p$
    awake during rounds $[r-u,r]$,
    $\tx \in \LOGi{p}{r}$.%
    \footnote{Clients may not output new logs for a few rounds after awakening. We use a single parameter $u$ for the maximum of such delay and the protocol's latency.}
\end{definition}

\begin{definition}[Resilience]
    \label{def:resilience-full}
    For \statpart, a family of \smrshort protocols $\Pi(n)$ achieves \emph{safety resilience} $\tS \in [0,1]$ and \emph{liveness resilience} $\tL \in [0,1]$ if
    for all $n$,%
    \footnote{The number of parties is constrained to be polynomial in the security parameter.}
    $\Pi(n)$ is safe with overwhelming probability over executions with $f \leq \tS n$ and live with overwhelming probability over executions with $f \leq \tL n$.
    For \dynpart, denote the adversary fraction $\frac{f}{\min_{r} \awakeset{r}}$ by $\beta$.
    Then, a protocol $\Pi$ achieves safety resilience $\betaS \in [0,1]$ and liveness resilience $\betaL \in [0,1]$ if
    $\Pi$ is safe with overwhelming probability over executions with $\beta \leq \betaS$
    and live with overwhelming probability over executions with $\beta \leq \betaL$.
\end{definition}

Note that it is trivial to get protocols with $\betaS = 1, \betaL = 0$ (never output) and $\betaL = 1, \betaS = 0$ (output everything in any order), so we don't consider these edge cases going forward.

\myparagraph{Hierarchy of Models}
We group the models defined above into four categories as shown in \cref{fig:models-hierarchy}.
Within each category, one model is more powerful than the other.
For example, \activ clients are more powerful than \passive clients because \activ clients can simulate \passive clients by staying silent.
Thus, any protocol designed for \passive clients will work equally well for \activ clients.
Therefore, solving \smrshort for \activ clients is at least as easy as solving \smrshort for \passive clients.
This is formalized in the lemma below.

\begin{lemma}[(\cf \cref{fig:models-hierarchy})]
\label{lem:hierarchy}
    Define the four pairs of models:
    validator activity models $(A_0,A_1) = (\iconvalidatoralert{},\iconvalidatorsleepy{})$,
    network delay models $(B_0,B_1) = (\iconvalidatorsync{},\iconvalidatorpsync{})$,
    client communication models $(C_0,C_1) = (\iconclientactiv{},\iconclientpassive{})$,
    client activity models $(D_0,D_1) = (\iconclientalert{},\iconclientsleepy{})$.
    For all $i_1,j_1,k_1,l_1 \in \{0,1\}$
    and $i_2 \leq i_1, j_2 \leq j_1, k_2 \leq k_1, l_2 \leq l_1$,
    if some \smrshort protocol $\Pi$ achieves resilience $(\tL_1,\tS_1)$ in the model ($A_{i_1}$,$B_{j_1}$,$C_{k_1}$,$D_{l_1}$)
    and $(\tL_2,\tS_2)$ in the model ($A_{i_2}$,$B_{j_2}$,$C_{k_2}$,$D_{l_2}$),
    then $\tL_2 \geq \tL_1$ and $\tS_2 \geq \tS_1$.
\end{lemma}
\begin{proof}
    It is sufficient to prove the statement in the four cases when from $(i_1,j_1,k_1,l_1)$ to $(i_2,j_2,k_2,l_2)$, exactly one index is decreased and the other three remain the same. We now prove these four cases.

    If $\Pi$ achieves $(\betaL,\betaS)$ under \dynpart, in particular, $\Pi$ is safe (live) when $n$ validators are all awake and up to $\betaS n$ ($\betaS n$) of them are adversary.
    If $\Pi$ is safe (live) under partial synchrony, in particular, it is safe (live) with the same number of adversary validators when the adversary sets $\GST = 0$ which is equivalent to synchrony.
    If $\Pi$ is safe (live) under \passive clients, it is also safe (live) when the \activ clients do not send any messages.
    If $\Pi$ is safe (live) under \sleepy clients, in particular, it is safe (live) when no client sleeps.
\end{proof}

The above lemma implies that any resilience pair achievable in a `harder' model in \cref{fig:models-hierarchy} is also achievable in an `easier' model.
Conversely, any impossibility result proven for an `easier' model also holds for a `harder' model.

%% file: sections/A02_proofs_sync.tex
\section{Proofs for Synchrony with \StatPart}
\label{sec:proofs-sync-static}

\subsection{\SLeepy \PAssive Clients, \ALert \PAssive Clients}
\label{sec:proofs-sync-passive}

\subsubsection{Impossibility for \ALert \PAssive Clients}
\label{sec:proofs-sync-alert-passive-imp}

\begin{proof}[Proof of \cref{thm:alert-passive-synchrony-converse}]
    Proof is by contradiction.
    Suppose there exists a protocol with safety resilience $\tS$ and liveness resilience $\tL$ such that $\tL+\tS \geq 1$.
    Then, there are numbers $f^S$ and $f^L$ such that $f^S+f^L = n$, and the protocol is safe and live in the presence of $f^S$ and $f^L$ adversary validators respectively.
    Let $P$ and $Q$ be disjoint sets of $f^S = n-f^L$ and $f^L$ validators respectively.
    Consider the following three worlds:
    
    \smallskip
    \noindent
    \textbf{World 1:}
    Validators in $P$ are honest, and those in $Q$ are adversary.
    There is a single client $k_1$.
    The environment inputs a single transaction $\tx_1$ to the validators in $P$ at time $0$.
    The adversary validators do not communicate with those in $P$ and ignore their messages.
    Towards $k_1$, they simulate the behavior of honest validators that have not received any transaction from the environment and that cannot communicate with those in $P$.
    Due to liveness under $f^L$ adversary validators, $k_1$ outputs $\tx_1$ (and no other transaction from the environment) as part of its log by time $u$.
    
    \smallskip
    \noindent
    \textbf{World 2:}
    Validators in $P$ are honest, and those in $Q$ are adversary.
    There is a single client $k_2$.
    The environment inputs a single transaction $\tx_2$ to the validators in $P$ at time $0$.
    The adversary validators do not communicate with those in $P$ and ignore their messages.
    Towards $k_2$, they simulate the behavior of honest validators that have not received any transaction from the environment and that cannot communicate with those in $P$.
    By liveness, $k_2$ outputs $\tx_2$ (and no other transaction from the environment) as part of its log by time $u$.
    
    \smallskip
    \noindent
    \textbf{World 3:}
    World 3 is a hybrid world.
    Validators in $Q$ are honest, and those in $P$ are adversary.
    The environment inputs the transactions $\tx_1$ and $\tx_2$ to the validators in $P$ at time $0$.
    Validators in $P$ simulate the execution in world 1 towards $k_1$ and the execution in world 2 towards $k_2$ via a \emph{split-brain attack}.
    They do not communicate with the validators in $Q$ and ignore their messages.
    Since the worlds 1 and 3 are indistinguishable in $k_1$'s
    view, it outputs $\tx_1$ (but not $\tx_2$) as part of its log by time $u$.
    Since the worlds 2 and 3 are indistinguishable in $k_2$'s
    view, it outputs $\tx_2$ (but not $\tx_1$) as part of its log by time $u$.
    However, this implies a safety violation in the presence of $f^S$ adversary validators, which is a contradiction.
\end{proof}

\subsubsection{Achievability for \SLeepy \PAssive Clients (Liveness-Favoring)}
\label{sec:proofs-sync-sleepy-passive-ach-live}

\begin{proof}[Proof of \cref{thm:sleepy-passive-synchrony-achievability-2}]
    Let $f$ be the number of adversary validators.
    
    \textbf{Liveness:} Suppose $f / n \leq \tL$, \ie, $f \leq n-q$, and consider a transaction $\tx$ input to an honest validator at some round $r$.
    Now, $\tx$ gathers signatures from all honest validators by round $r+\Delta$, and all clients observe these signatures by round $r+2\Delta$.
    Then, since there are $q$ or more honest validators, all honest clients add $\tx$ to their liveness queues by round $r+2\Delta$.
    Every transaction added to the liveness queue of a client at some round $r'$ is output as part of its log by round $r'+u_\mathrm{int}$.
    Therefore, $\tx$ is output as part of all honest clients' logs by round $r+u_\mathrm{int} + 2\Delta$, implying that $\Pi_{\mathrm{live}}^q$ satisfies liveness with latency $u_\mathrm{int} + 2\Delta$ and resilience $n-q$.
    
    \textbf{Safety:} Suppose $f / n \leq \tS$, \ie, $f < q$.
    Then, the internal protocol is safe and live with latency $u_\mathrm{int}$ as $q \leq n/2$.
    Any transaction $\tx$ added to the liveness queue of an honest client $k$ at some round $r$ must have been signed by $q$ validators before round $r$, one of which is honest.
    Thus, $\tx$ would be input to the internal protocol $\Piint$ by round $r$, and by liveness, output as part of $k$'s internal log $\LOGint$ by round $r+u_\mathrm{int}$.
    Since $k$ attempts to append $\tx$ to $\LOGint$ for the first time at round $r+u_\mathrm{int}$, and $\tx$ appears as part of $\LOGint$ by round $r+u_\mathrm{int}$, $\tx$ is not added to the tip of $\LOGint$.
    By the same logic, if $f < q$, no transaction added to the liveness queue of an honest client is appended to the tip of its internal log, implying that each honest client outputs its internal log as it is.
    Finally, safety follows from the safety of the internal protocol.
    \end{proof}

\subsection{\SLeepy \ACtiv Clients}
\label{sec:proofs-sync-sleepy-active}

\subsubsection{Impossibility for \SLeepy \ACtiv Clients}
\label{sec:proofs-sync-sleepy-active-imp}

\begin{proof}[Proof of \cref{thm:sleepy-active-synchrony-converse}]
    Proof is by contradiction.
    Suppose there exists a protocol with resiliences $\tS$ and $\tL$ such that $\tS,\tL \geq n/2$. Then, the protocol is safe and live in the presence of $f = \lceil n/2 \rceil$ adversary validators.
    Let $P$, $Q$ and $R$ denote disjoint sets of $n-f$, $n-f$ and $2f-n$ validators respectively.
    Consider the following four worlds:
    
    \textbf{World 1:} Validators in $P$ are honest, and those in $Q$ and $R$ have crashed.
    There is a single client $k_1$.
    The environment inputs a single transaction $\tx_1$ to the validators in $P$ at time $0$.
    Since $|Q \cup R| \leq f$, by liveness, $k_1$ outputs $\tx_1$ (and no other transaction from the environment) as part of its log by time $u$.
    
    \textbf{World 2:} Validators in $Q$ are honest, and those in $P$ and $R$ have crashed.
    There is a single client $k_2$.
    The environment inputs a single transaction $\tx_2$ to the validators in $Q$ at time $0$.
    Since $|P \cup R| \leq f$, by liveness, $k_2$ outputs $\tx_2$ (and no other transaction from the environment) as part of its log by time $u$.
    
    \textbf{World 3:}
    Validators in $P$ are honest, and those in $Q \cup R$ are adversary.
    Validators in $R$ have crashed.
    There are two honest clients, $k_1$ and $k_3$, and the adversary simulates a client $k_2$.
    Client $k_3$ joins the protocol at round $u$.
    The environment inputs a single transaction $\tx_1$ to the validators in $P$ at time $0$.
    
    Client $k_2$ and the validators in $Q$ do not communicate with the client $k_1$ and the validators in $P$.
    Thus, for $k_1$, world 3 is indistinguishable from world 1, and it outputs $\tx_1$ (and no other transaction from the environment) as part of its log by time $u$.
    In the meanwhile, $k_2$ and the validators in $Q$ start with transaction $\tx_2$, and emulate the execution in world 2 until round $u$.
    
    Once $k_3$ joins the protocol at round $u$, $k_2$ and the validators in $Q$ emulate towards $k_3$ the behavior of the honest validators (in $Q$) and the client $k_2$ in world 4.
    In other words, they pretend like honest validators and an honest client who have been shunned by the validators in $P \cup R$ and client $k_1$.
    Since $|Q \cup R| \leq f$, by liveness, $k_3$ outputs $\tx_1$ as part of its log by time $2u$.
    
    \textbf{World 4:}
    Validators in $Q$ are honest, and those in $P \cup R$ are adversary.
    Validators in $R$ have crashed.
    There are two honest clients, $k_2$ and $k_3$, and the adversary simulates a client $k_1$.
    Client $k_3$ joins the protocol at round $u$.
    The environment inputs a single transaction $\tx_2$ to the validators in $Q$ at time $0$.
    
    Client $k_1$ and the validators in $P$ do not communicate with the client $k_2$ and the validators in $Q$.
    Thus, for $k_2$, world 4 is indistinguishable from world 2, and it outputs $\tx_2$ (and no other transaction from the environment) as part of its log by time $u$.
    In the meanwhile, $k_1$ and the validators in $P$ start with transaction $\tx_1$, and emulate the execution in world 1 until round $u$.
    
    Once $k_3$ joins the protocol at round $u$, $k_1$ and the validators in $P$ emulate towards $k_3$ the behavior of the honest validators (in $P$) and the client $k_1$ in world 3.
    In other words, they pretend like honest validators and an honest client who have been shunned by the validators in $Q \cup R$ and client $k_2$.
    Since $|P \cup R| \leq f$, by liveness, $k_3$ outputs $\tx_2$ as part of its log by time $2u$.
    
    Finally, note that worlds 3 and 4 are indistinguishable by $k_3$ with overwhelming probability, since the validators and clients send the same messages in both worlds.
    Therefore, $k_3$ outputs the same log, containing $\tx_1$ and $\tx_2$, in both worlds by time $2u$.
    However, this implies a safety violation either in world 3, where $k_1$ outputs the log $[\tx_1]$ by round $u$, or in world 4, where $k_2$ outputs the log $[\tx_2]$ by round $u$.
    This is a contradiction as the protocol must have been safe in the presence of $f$ adversary validators.
\end{proof}

\subsubsection{Achievability for \SLeepy \ACtiv Clients (Safety-Favoring)}
\label{sec:proofs-sync-sleepy-active-ach-safe}

\begin{definition}[{Certifiable protocol}]
    An \smrshort protocol $\Pi$
    is certifiable if there exists
    a computable functionality $\transcribe$ (the \emph{\transcript producer})
    and
    a computable deterministic non-interactive function
    $\untranscribe$ (the \emph{\transcript consumer})
    such that
    when a client $p$ invokes $\transcribe()$ at round $r$, it produces a \transcript $\transcript$ such that $\untranscribe(\transcript) = \LOGi{p}{r}$.
\end{definition}

\begin{definition}[Certifiable safety]
    A certifiable protocol $\Pi$ 
    is certifiably safe if
    $\Pi$ is safe,
    and moreover,
    if at any round $r$,
    the adversary outputs a \transcript $\transcript$ such that $\untranscribe(\transcript) = \LOG$,
    then for all clients $q$, for all rounds $s$,
    $\LOG \consistent \LOGi{q}{s}$.
    A certifiable protocol $\Pi$ achieves certifiable safety resilience $\tC$ if $\Pi$ is certifiably safe with overwhelming probability over executions with $f \leq \tC n$.
\end{definition}

A more formal pseudocode of the protocol based on \cref{def:certifiable,def:certifiable-safety} is in \cref{alg:freezing}.

\begin{lemma}[Safety]
\label{lem:freezing-safety}
Suppose the network is synchronous and the clients are \sleepy and \activ.
Then, $\PI$ has safety resilience $\tS = 1$.
\end{lemma}

\begin{proof}
See \cref{fig:freezing-safety-proof} for reference.
    Towards contradiction, let $r$ be the smallest round such that for some $s \geq r$, and some honest clients $p,q$, $\LOGi{p}{r} \inconsistent \LOGi{q}{s}$.
    For shorthand, let $\LOG = \LOGi{p}{r}$.
    Then, at round $r-\Delta$, client $p$ must have 
    seen a \transcript $\transcript$ such that $\untranscribe(\transcript) = \LOG$.
    Client $p$ also gossiped 
    $\transcript$
    at round $r-\Delta$, which means that before the end of round $r$, client $q$ must have 
    seen $\transcript$.
    Thus, client $q$ added $\LOG$
    to its set $\msgSet$ before the end of round $r$.
    However, since client $q$ output $\LOGi{q}{s} \inconsistent \LOG$ at round $s \geq r$, this is a contradiction to the freezing (\cref{alg:freezing}~\cref{loc:freezing-check-conflict}).
\end{proof}

\begin{lemma}[Liveness]
\label{lem:freezing-liveness}
If $\Piint$ has certifiable safety resilience $\tC_{\mathrm{int}}$ and liveness resilience $\tL_{\mathrm{int}}$,
then $\PI$ has liveness resilience $\min\{\tL_{\mathrm{int}}, \tC_{\mathrm{int}}\}$.
\end{lemma}

\begin{proof}
    Let $u = u_{\mathrm{int}} + \Delta$ where $u_\mathrm{int}$ is the latency of $\Piint$.
    Let $r < \rmaj$ be any arbitrary round.
    Suppose that a transaction $\tx$ is received by all honest validators before round $r - u$.
    Consider an honest client $p$ that wakes up before $r - u$.
    Due to liveness of $\Piint$, at round $s = r - u + u_{\mathrm{int}}$, 
    $\tx \in {\LOGint}_{p}^{s}$ (the log output by the internal protocol $\Piint$).
    At round $s$, client $p$ runs $\transcript \gets \transcribe()$ and adds $L = \untranscribe(\transcript)$ to its set $\msgSet$(\cref{alg:freezing}~\cref{loc:freezing-on-witness,loc:freezing-verify}).
    Recall from \cref{def:certifiable} that $\LOG = {\LOGint}_{p}^{s}$.
    Due to certifiable safety, 
    the set $\msgSet$
    of client $p$, 
    contains only logs that are consistent with $\LOG$.
    Therefore, at round $s + \Delta = r$, $\LOGi{p}{r} \succeq \LOG \ni \tx$ (due to \cref{alg:freezing}~\cref{loc:freezing-check-conflict,loc:freezing-output}).
\end{proof}

\begin{proof}[Proof of \cref{thm:freezing-resilience}]
    From \cref{lem:freezing-safety,lem:freezing-liveness}.
\end{proof}

\subsubsection{Achievability for \SLeepy \ACtiv Clients (Liveness-Favoring)}
\label{sec:proofs-sync-sleepy-active-ach-live}

Protocol pseudocode: \cref{alg:sleepy-active-static-live}, block diagram: \cref{fig:always-live-protocol}.

\import{./algorithms/}{alg_liveness_sleepy_active.tex}

\import{./figures/}{always_live_protocol.tex}

\begin{lemma}
\label{lem:append-liveness}
    Suppose the network is synchronous, and the clients are \sleepy and \activ.
    Then, the protocol $\Pisal$ has liveness resilience $\tL = 1$.
\end{lemma}
\begin{proof}
    Consider a transaction $\tx$ received by an honest validator at round $r$. The validator gossips $\tx$ to all clients. 
    All clients receive $\tx$ by round $r+\Delta$ and add it to their liveness queues.
    For any client $k$, by round $r' = r + \Delta + u_{\mathrm{int}} + \Delta$, either $\tx$ is in the internal log of $k$ or $k$ appends $\tx$ in its output log, therefore, $\tx \in \LOGi{k}{r'}$.
    The protocol $\Pisal$ is thus live with resilience $\tL = 1$ and latency $u_{\mathrm{int}} + 2\Delta$.
\end{proof}

\begin{lemma}
\label{lem:append-safety}
    Suppose the network is synchronous, and the clients are \sleepy and \activ.
    If the internal protocol $\Piint$ has resilience $(\tL, \tS)$, then the protocol $\Pisal$ has safety resilience $\min\{\tL,\tS\}$.
\end{lemma}
\begin{proof}
    Suppose the number of adversary validators is $f \leq n \min\{\tL, \tS\}$.
    Thus, $\Piint$ is safe and live.
    We will show that for every client $k$ and round $r$, $\LOGi{k}{r} = {\LOGint}_k^r$, \ie, the output log is identical to the internal log.
    Then, $\Pisal$ is safe due to the safety of $\Piint$.

    To show that $\LOGi{k}{r} = {\LOGint}_k^r$, consider any transaction $\tx$ that client $k$ adds to its liveness queue at some round $r'$.
    Since client $k$ gossips $\tx$, all honest validators receive $\tx$ by round $r' + \Delta$.
    Due to liveness of $\Piint$, $\tx \in {\LOGint}_{k}^{r'+\Delta+u_{\mathrm{int}}}$.
    Therefore, client $k$ does not append $\tx$ to $\LOGi{}{}$ and simply drops it from the liveness queue.
    Since this holds for all transactions, $\LOGi{k}{r} = {\LOGint}_k^r$ for all $r$.
\end{proof}

\begin{proof}[Proof of \cref{thm:sleepy-active-synchrony-achievability-2}]
From \cref{lem:append-liveness,lem:append-safety}.
\end{proof}

\subsection{\ALert \ACtiv Clients}
\label{sec:proofs-sync-alert-active}

\subsubsection{Achievability for \ALert \ACtiv Clients}
\label{sec:proofs-sync-alert-active-ach}

We first define a variant of the Byzantine Generals problem in which clients (not validators) output values and then recap the Dolev-Strong protocol with \alert \activ clients.

\begin{definition}
\label{def:byzantine-generals}
Let $\CV$ be a predefined set of values and let $\bot \notin \CV$ be a predefined default value.
In the Byzantine Generals problem, a leader $\ell$ \emph{\BGbroadcasts} a value $v_{\ell} \in \CV$ at a known start round $R$ and each client $k$ \emph{\BGoutputs} a value $v_k \in \CV\cup \{\bot\}$ with the following properties:
\begin{itemize}
    \item Termination: For some $u$, all clients output a value by round $R_{\ell} + u$.
    \item Agreement: For all honest clients $k, k'$, $v_{k} = v_{k'}$.
    \item Validity: If $\ell$ is honest, then for all clients $k$, $v_{k} = v_{\ell}$.
\end{itemize}
\end{definition}

The Dolev-Strong protocol (with clients) achieving termination, agreement and validity when up to $n-1$ validators are adversary is shown in \cref{alg:dolev-strong-clients} (\cf~\cite{vitalik-dolev-strong,dec-thoughts-dolev-strong}).
Since we will use our Byzantine Generals protocol to build an SMR protocol over $n$ validators, we consider the leader to be one of the validators, although, in general, the leader could be any party with a public key.
In \cref{alg:dolev-strong-clients}, $\signedmsg{p}{m}$ denotes messages $m$ signed by party $p$ and the protocol uses an instance identifier $\id$ to enable running multiple instances in the SMR protocol.
As in the classic Dolev-Strong protocol~\cite{dolev-strong,dec-thoughts-dolev-strong}, validators build a signature chain in which the leader signs its value, the second validator signs the leader's signed message, and so on (\cref{line:dolev-strong-val-sign}), and a signature chain is considered valid if it arrives within a timeout (\cref{line:dolev-strong-val-round-check,line:dolev-strong-clients-round-check}).
The key differences from classic Dolev-Strong are that clients broadcast messages they receive as-is (without signing) to all parties and the timeouts are twice as long to accommodate for the round-trip delay between clients and validators ($2k\Delta$ in \cref{alg:dolev-strong-clients} v.s. $k\Delta$ in \cite{dolev-strong}).

\import{./algorithms}{alg_dolev_strong.tex}

\begin{lemma}
\label{lem:dolev-strong-clients-security}
    For any $f < n$, \Cref{alg:dolev-strong-clients} satisfies agreement and validity when $f$ validators are adversary.
\end{lemma}
\begin{proof}
    Termination: All clients output a value after $(2n-1)\Delta$ rounds.

    Validity:
    Suppose the leader $\ell$ is honest and \BGbroadcasts value $v_{\ell}$.
    By round $\Delta$, all clients receive $\signedmsg{\ell}{v}$.
    Thus, the condition in \cref{line:dolev-strong-clients-round-check} is true, all clients add $v$ to their set of candidate output values $\Vout$.
    Moreover, since the leader is honest and signatures are unforgeable, no party ever receives $\signedmsg{\ell}{v'}$ for $v' \neq v$. Therefore, $\Vout = \{v\}$ and the client \BGoutputs $v$ at the end of round $(2n-1)\Delta$.

    Agreement:
    Let's refer to a message of the form $\signedmsg{j_k}{\signedmsg{j_{2}}{\signedmsg{j_1}{\id,\ell,v}}...}$ where $j_1,...,j_k$ are distinct validators and $j_1=\ell$ as a $k$-signature chain on $v$.
    First, we show that if a client $c$ adds a value $v$ to $\Vout$, then all other clients do so too.
    Since client $c$ added $v$ to $\Vout$, for some $k \leq n$, it received a $k$-signature chain $m$ on $v$ by round $(2k-1)\Delta$.
    If at least one of the validators $j_1,...,j_k$ who signed this message is honest,
    then due to \cref{line:dolev-strong-val-round-check}, for some $k' \leq n$, this validator signed the $k'$-th signature in the chain by round $2(k'-1)\Delta$, so client $c'$ receives a $k'$-signature chain by round $(2k'-1)\Delta$ and thus also adds $v$ to $\Vout$.
    If no signatory of $m$ is honest, then $k \leq n-1$.
    In this case, client $c$ sends $m$ to all parties, so all validators receive $m$ by round $2k\Delta$.
    Since the condition in \cref{line:dolev-strong-val-round-check} is satisfied, at least one honest validator (who has not yet signed by assumption) signs $m$, and so client $c'$ receives a $(k+1)$-signature chain by round $(2k+1)\Delta \leq (2n-1)\Delta$. So, client $c'$ also adds $v$ to $\Vout$.
    Agreement follows since we have established that all clients have the same set $\Vout$ at the end of round $(2n+1)\Delta$.
\end{proof}

Next, we build an SMR protocol using the Dolev-Strong protocol with clients (\cref{alg:dolev-strong-clients}).
A similar construction is already found in \cite{vitalik-dolev-strong,bcube} but we recap it in \cref{alg:smr-using-dolev-strong} for completeness.
Every $2n\Delta$ rounds (which \cref{alg:dolev-strong-clients} takes to terminate), each validator starts a new instance of the Dolev-Strong protocol (\cref{alg:dolev-strong-clients}) as the leader.
At the end of $2n\Delta$ rounds, clients agree on one block (possibly empty) from each validator and add them to their logs in a predetermined ordering over the validators.

\import{./algorithms}{alg_smr_using_dolev_strong.tex}

\begin{proof}[Proof of \cref{thm:alert-active-static-ach}]
    Safety follows from the agreement property of the Dolev-Strong protocol with clients (\cref{lem:dolev-strong-clients-security}).
    Assuming honest parties broadcast all transactions they receive, an honest validator will \BGbroadcast a block containing all valid transactions sent to any honest party.
    By validity of the Dolev-Strong protocol with clients (\cref{lem:dolev-strong-clients-security}), all clients will append this block to their logs.
\end{proof}

%% file: algorithms/alg_liveness_sleepy_active.tex
\begin{algorithm}[btp]
    \caption{\smrshort protocol $\Pisal$ achieving $\tL = 1, \tS < 1/2$}
    \label{alg:sleepy-active-static-live}
    \begin{algorithmic}[1]
    \scriptsize
    \On{\Call{init}{$\valset,\genesis$}}
    \label{loc:append-init}
        \State $P \gets \text{new } \Piint(\valset, \genesis)$ \Comment{instantiate a new $\Piint$ client} \label{loc:append-base-protocol}
        \State $Q \gets \emptyset$ \Comment{liveness queue: txs seen so far}
        \State $\LOGi{}{} \gets \genesis$ \Comment{output log of the combined protocol $\Pisal$}
    \EndOn
    \On{transaction $\tx$ from the network at round $r$}
        \State $Q.\operatorname{enqueue}((\tx,r))$ \Comment{add $\tx$ to the liveness queue}
        \State $\operatorname{gossip}(\tx)$
    \EndOn
    \On{$\LOG$ output by $P$ at round $r$} \label{loc:append-on-witness}
        \For{$(\tx,r') \in Q$ such that $r' \leq r - u_{\mathrm{int}} - \Delta$ and $\tx \notin \LOG$}
            \State $\LOG \gets \LOG || \tx$ 
        \EndFor
        \State $\LOGi{}{} \gets L$ \Comment{output log}
    \EndOn
    \end{algorithmic}
\end{algorithm}

%% file: figures/always_live_protocol.tex
\begin{figure}[tb]%
    \centering%
    \begin{tikzpicture}[
            x=2cm,
            y=1cm,
            subprotocol/.style={
                draw,
                align=center,
                inner sep=0.5em,
            },
            messages/.style={
                -latex,
            },
            party/.style={
                draw,
                dashed,
            },
        ]
        \scriptsize

        \coordinate (txsPos) at (0,0);
        \coordinate (netPos) at ($(txsPos)+(0.75,-1)$);
        \coordinate (piPos) at ($(txsPos)+(1.75,0)$);
        \coordinate (waitPos) at ($(txsPos)+(1.75,1)$);
        \coordinate (decidePos) at ($(waitPos)+(1.625,0)$);
        \coordinate (concatPos) at ($(piPos)+(2.5,0)$);
        \coordinate (outputPos) at ($(concatPos)+(1,0)$);

        \node (txs) at (txsPos) {$\txs$};
        \node [subprotocol,inner sep=2pt,cloud,cloud ignores aspect] (net) at (netPos) {network};
        \node [subprotocol] (pi) at (piPos) {$\Piint$\\\tiny$\tS = \tL < 1/2$};
        \node [subprotocol] (wait) at (waitPos) {delay $u_{\mathrm{int}}+\Delta$};
        \node [subprotocol,inner sep=2pt,diamond,aspect=2] (decide) at (decidePos) {$\txs \subseteq \LOGint$?};
        \node [subprotocol] (concat) at (concatPos) {$\LOGint||\txs$};
        \node (output) at (outputPos) {$\LOGi{}{}$};

        \draw [messages] (txs) -- (pi);
        \draw [messages] (pi) -- (concat) node [pos=0,below,anchor=north west] {$\LOGint$};
        \draw [messages] (concat) -- (output);
        \draw [messages] (decide) -| (concat) node [pos=0,above,anchor=south west] {no} node [pos=0.75,right] {$\txs$};
        \draw [messages] ([yshift=3pt] wait.east) -- ([yshift=3pt,xshift=4.5pt] decide.west);
        \draw [messages] (pi.east) -- ++(0.3,0) |- ([yshift=-3pt,xshift=4.5pt] decide.west);
        \draw [messages] ([xshift=-3pt] txsPos -| netPos) -- ([xshift=-3pt] net.north);
        \draw [messages] ([xshift=3pt] net.north) -- ([xshift=3pt] txsPos -| netPos);
        \draw [messages] ([xshift=1em] txsPos -| netPos) |- (wait);
        
    \end{tikzpicture}
    \caption[]{%
        A protocol that achieves $\tL = 1, \tS < 1/2$ for \sleepy \activ clients.
        The internal protocol $\Piint$ is any SMR protocol safe and live under honest majority, achieving any $\tS = \tL < 1/2$.
        On receiving any transaction $\tx$, parties gossip the transaction to the network (all parties).
        Clients add $\tx$ to a local liveness queue.
        After $u_{\mathrm{int}} + \Delta$ rounds (where $u_{\mathrm{int}}$ is the maximum latency of $\Pi$), if $\tx$ is not included in the log $\LOGint$ output by the client from $\Piint$, the client appends $\tx$ to $\LOGint$ to output the final confirmed log $\LOGi{}{}$.
    }%
    \label{fig:always-live-protocol}%
\end{figure}%

%% file: algorithms/alg_dolev_strong.tex
\begin{algorithm}[tbp]
\caption{Dolev--Strong protocol with clients}
\label{alg:dolev-strong-clients}
\begin{algorithmic}[1]
\scriptsize
\LineComment{Each instance of the protocol is identified by $(\id, \ell)$ where $\ell$ is the leader.}
\LineComment{All parties know the starting round $R_{\id}$ for each $\id$}
\LineComment{Code for leader $\ell$ (who is also a validator)}
\On{\Call{\BGbroadcast}{$\id,\ell,v$} at round $R_{\id}$}
    \State Send $\signedmsg{\ell}{\id,\ell,v}$ to all parties
\EndOn
\LineComment{Code for validator $i$}
\On{receiving $m = \signedmsg{j_k}{\signedmsg{j_{2}}{\signedmsg{j_1}{(\id,\ell,v}}...}$ where $j_1,...,j_k \neq i$ are distinct validators and $j_1 = \ell$}
    \If{current round $\leq R_{\id} + 2k\Delta$} \label{line:dolev-strong-val-round-check}
        \State Send $\signedmsg{i}{m}$ to all parties \label{line:dolev-strong-val-sign}
    \EndIf
\EndOn
\LineComment{Code for client}
\State $\Vout \gets \emptyset$ \Comment{Set of candidate output values}
\On{receiving $m = \signedmsg{j_k}{\signedmsg{j_{2}}{\signedmsg{j_1}{\id,\ell,v}}...}$ where $j_1,...,j_k$ are distinct validators and $j_1 = \ell$}
    \If{current round $\leq R_{\id} + (2k-1)\Delta$} \label{line:dolev-strong-clients-round-check}
        \State $\Vout \gets \Vout \cup \{v\}$
        \State Send $m$ to all parties
    \EndIf
\EndOn
\At{the end of round $R_{\id} + (2n-1)\Delta$}
    \If{$|\Vout| = 1$}
        \State $\Call{\BGoutput}{\id,\ell,v}$ where $\Vout = \{v\}$
    \Else
        \State $\Call{\BGoutput}{\id,\ell,\bot}$
    \EndIf
\EndAt
\end{algorithmic}
\end{algorithm}

%% file: algorithms/alg_smr_using_dolev_strong.tex
\begin{algorithm}[tbp]
\caption{SMR protocol achieving any $\tL,\tS < 1$}
\label{alg:smr-using-dolev-strong}
\begin{algorithmic}[1]
\scriptsize
\State $R_{\id} \gets \id \cdot 2n\Delta$ for all $\id = 0,1,2,...$ \Comment{known to all parties}
\LineComment{Code for validator $i$}
\At{round $R_{\id}$ for $\id=0,1,2,...$}
    \State $\Call{\BGbroadcast}{\id,\ell,B}$ where $B$ is a block of transactions received so far
\EndAt
\LineComment{Code for client}
\State $\LOGi{}{} \gets [\;]$, $\id \gets 1$
\On{$\Call{\BGoutput}{\id,\ell,B_{\ell}}$ for all $\ell \in \valset$}
    \For{$\ell$ \textbf{in} $\operatorname{enumerate}(\valset)$} \Comment{all validators in a predetermined order}
        \State $\operatorname{append}(\LOGi{}{}, B_{\ell})$ \Comment{append $B_{\ell}$ to log; treat invalid $B_{\ell}$ or $\bot$ as empty block}
    \EndFor
    \State $\id \gets \id+1$
    \State Output $\LOGi{}{}$
\EndOn
\end{algorithmic}
\end{algorithm}

%% file: sections/A03_proofs_sleepy_validators.tex
\section{Proofs for Synchrony with \DynPart}
\label{sec:proofs-dynamic}

\subsection{\SLeepy \PAssive Clients, \ALert \PAssive Clients}
\label{sec:proofs-dynamic-sleepy-passive}

\subsubsection{Achievability for \SLeepy \PAssive Clients (Safety-Favoring)}
\label{sec:proofs-dynamic-sleepy-passive-ach-safe}

\myparagraph{Recap of Goldfish (\cf \cite{goldfish})}
Goldfish is an SMR protocol secure under synchrony and \dynpart for any resilience $\betaS = \betaL <1/2$.
Goldfish divides time into slots of length $3\Delta$.
Each slot $t$ has a leader, which proposes a block, and a set of validators called voters that cast slot $t$ votes,
all selected via a verifiable random function (VRF). 
Each message contains a slot number and because of the VRF, all parties ignore messages from validators who were not a leader or voter for the claimed slot.
Each validator and client maintains a \emph{buffer} and a tree of blocks and votes called the \emph{bvtree}.
Upon receiving a valid message (block or vote), each validator echoes the message and adds it to its buffer, but does not add it to the bvtree immediately (\emph{message buffering}).
The crux of Goldfish is the mechanism through which messages are added to and removed from the bvtree as guided by two principles: \emph{message buffering} and \emph{vote expiry}.

At the beginning of each slot $t$, \ie, time $3\Delta t$,
the leader momentarily combines its buffer and bvtree without merging them permanently and runs the \emph{GHOST-Eph} fork-choice rule on the combined bvtree using the slot $t-1$ votes.
The rule outputs a canonical chain within the combined bvtree.
It iteratively moves down the tree, starting at the genesis block, and, at each block $B$, observing the subtrees rooted at $B$'s children.
It then selects the child block with the largest number of slot $t-1$ votes by unique validators for the blocks in that tree.
This process is repeated until reaching a leaf, which identifies the canonical chain.%
\footnote{Note that sleepy parties can run the \emph{GHOST-Eph} fork-choice rule because each message specifies the slot it belongs to.}
Finally, the leader extends the tip of the identified canonical chain with a block, and proposes the new block along with the combined bvtree. 
Note that \emph{vote expiry} is in action here, since the GHOST-Eph rule considers only the votes from the previous slot $t-1$, but not the earlier slots.

At time $3\Delta t + \Delta$, each slot $t$ voter merges the bvtree proposed by the leader with its local bvtree.
Subsequently, it votes for the tip of the canonical GHOST-Eph chain, again identified using only the slot $t-1$ votes on the merged bvtree.
Note that the voter does not use its buffer at this stage, instead adding the leader's bvtree to its own before voting (vote buffering).

At time $3\Delta t + 2\Delta$, each validator and client permanently adds the messages in its buffer into its bvtree.
Each client finds the canonical GHOST-Eph chain, this time by running the \emph{GHOST-Eph} fork-choice rule on its bvtree (after buffer is added) using the \emph{slot $t$ votes}. 
It outputs the sequence of blocks on this canonical chain from slots $t-\kappa$ or older as the log, where $\kappa$ is a security parameter.

\import{./algorithms/}{alg_modified_goldfish_forkchoice.tex}

\textbf{Our modification of Goldfish.}
To achieve any resilience $\betaS \leq \phi$ and $\betaL < 1-\phi$, we modify the GHOST-Eph fork-choice rule used by Goldfish as follows (\Cref{alg:goldfish-fcr}).
Within any iteration of the GHOST-Eph fork-choice, validators do not simply select a child block with the largest number of slot $t$ votes (or slot $t-1$ votes, depending on which slot's votes are being considered).
They instead inspect the number of slot $t$ votes by unique validators on each tree rooted at the children of a block $B$ (\ie, the weight of the sub-trees) as well as the total number of slot $t$ votes by unique validators observed so far (\ie, the total weight).
Then, if the fraction of the weight of one of the subtrees (rooted at a child of $B$) over the total weight is at least $\phi$, the validator moves to that child and repeats this process.
If none of the subtrees have sufficient weight or $B$ does not have any children (\ie, a leaf block), then the validator terminates the fork-choice rule at $B$ and returns $B$ along with its prefix.
We denote the Goldfish protocol using the modified GHOST-Eph rule by $\Pi_{\mathrm{live}}^\phi$.
Finally, to support clients, we stipulate that upon becoming awake\footnote{Through Goldfish's joining procedure~\cite{goldfish}}, each client maintains a buffer and bvtree, and outputs a log at rounds $3\Delta t + 2\Delta$ of every slot $t$ via the same rule as validators, \ie, after merging its buffer and bvtree.

\begin{proof}[Proof of~\Cref{thm:sleeepy-passive-da-achievability-1}]
Proof is a minor modification of the proof of the Goldfish protocol~\cite[Appendix B]{goldfish}.
We consider an execution of our modified Goldfish in a synchronous network with \dynpart and \sleepy, \passive clients.
Recall the definition of adversary fraction $\beta$ from~\Cref{def:resilience}\footnote{An adversary fraction of $\beta$ as defined in~\Cref{def:resilience} implies a $(\beta, 3\Delta)$-compliant execution of Goldfish (\cf~\cite[Definition 2]{goldfish}), enabling us to replace the notion of compliant executions with $\beta$ fraction in the proof of the modified protocol.}.

We first note that a generalized form of~\cite[Lemma 1]{goldfish} is still true for our modified Goldfish execution as our modifications do not affect the voter selection.
Therefore, 
w.o.p., for every slot $t$, the number of adversary slot $t$ voters at round $3\Delta(t + 1) + \Delta$ is less than a $\beta$ fraction of the slot $t$ voters awake at round $3\Delta t + \Delta$. 
Also w.o.p., all slot intervals of length $\kappa$ have at least one slot $t$, where an honest validator is recognized as the slot $t$ leader by all awake honest validators at round $3\Delta t$.

Furthermore, \cite[Lemma 2]{goldfish} is also true as its proof is also not affected by our modification:
If a validator $v$ is recognized as the leader of a slot $t$ by all awake honest validators at some round $3\Delta t + \Delta$, then, all honest slot $t$ voters awake at round $3 \Delta t + \Delta$ vote for $v$'s proposal.

We next split~\cite[Lemma 3]{goldfish} into two parts to help with the liveness and safety proofs respectively.
Proof of the lemma closely resembles that of~\cite[Lemma 3]{goldfish}:
\begin{lemma}
\label{lemma:lemma-3-remastered}
We consider two cases for a slot $t$:
\begin{enumerate}
    \item Suppose \emph{all} honest slot $t$ voters awake at round $3\Delta t + \Delta$ vote for a descendant of some block $B$.
    Then, given any $\beta \leq 1-\phi$, 
    w.o.p., \emph{all} honest slot $t+1$ voters awake at round $3\Delta (t+1)+\Delta$ vote for a descendant of $B$.
    \item Suppose \emph{no} honest slot $t$ voter awake at round $3\Delta t + \Delta$ votes for any descendant of some block $B$ (including $B$ itself).
    Then, given any $\beta < \phi$,
    w.o.p., \emph{no} honest slot $t+1$ voter awake at round $3\Delta (t+1)+\Delta$ vote for a descendant of $B$.
\end{enumerate}
\end{lemma}
\begin{proof}[Proof of~\Cref{lemma:lemma-3-remastered}]
Consider an honest slot $t+1$ voter $v$ awake at round $3\Delta (t+1)+\Delta$. 
Since $v$ must have been awake at least since round $3\Delta t + 2\Delta$ due to the joining procedure of Goldfish~\cite[Section 3.1]{goldfish}, its bvtree at round $3\Delta t + 2\Delta$ contains all votes broadcast by honest slot $t$ voters awake at round $3\Delta t + \Delta$.
The same is true for its bvtree at round $3\Delta(t+1)+\Delta$ after merging it with the bvtree in any proposal.
Moreover, when $\beta \leq 1-\phi$
the number of adversary slot $t$ voters at round $3\Delta(t + 1) + \Delta$ is at most a $\beta$ fraction of the slot $t$ voters awake at round $3\Delta t + \Delta$ (\cite[Lemma 1]{goldfish}).
Hence, in the first case, the number of slot $t$ votes for $B$'s descendant in $v$'s bvtree is larger than a $\phi$ fraction of the total number of slot $t$ votes by unique validators in $v$'s bvtree at round $3\Delta(t+1)+\Delta$ (\cite[Lemma 1]{goldfish}).
Consequently, upon invoking the GHOST-Eph fork-choice rule at round $3\Delta(t+1)+\Delta$, $v$ selects $B$'s descendant over all blocks conflicting with $B$ and moves down the tree until at least reaching a child of $B$. 
Thus, at round $3\Delta(t + 1) + \Delta$, the fork choice rule returns a descendant of $B$, and $v$ votes for it.

Now, in the second case, the number of slot $t$ votes for $B$'s descendant in $v$'s bvtree is smaller than a $1-\phi$ fraction of the total number of slot $t$ votes by unique validators in $v$'s bvtree at round $3\Delta(t+1)+\Delta$.
Hence, upon invoking the GHOST-Eph fork-choice rule at round $3\Delta(t+1)+\Delta$, $v$ does not select $B$ over blocks conflicting with $B$.
Thus, at round $3\Delta(t + 1) + \Delta$, fork choice does not return $B$ or any of its descendants, and $v$ does not vote for $B$ or its descendants.
\end{proof}

\textbf{Safety:}
To prove safety, we modify the proof of~\cite[Theorem 1]{goldfish}.
Suppose $\beta < \phi$, and an honest validator $v$ with proposed block $B$ is accepted as the leader of some slot $t$ by all awake honest validators at round $3\Delta t + \Delta$.
From~\cite[Lemmas 1 and 2]{goldfish} and~\Cref{lemma:lemma-3-remastered} part (ii), it follows by induction that w.o.p., for any $t' \geq t$, no honest slot $t'$ voter awake at round $3\Delta t' + \Delta$ votes for a block that is \emph{not} consistent with $B$.

By synchrony, the honest votes of slot $t'$ reach all honest validators (and clients) awake at round $3\Delta t' + 2\Delta$ by then, when they also merge the votes into their bvtrees. 
The number of honest slot $t'$ voters awake at round $3\Delta t' + 2\Delta$ is greater than a $1-\phi$ fraction of the total number of slot $t'$ voters at round $3\Delta (t'+1) + 2\Delta$ (by Lem. 1). 
Upon invoking the GHOST-Eph rule at rounds $3\Delta t' + 2\Delta$, $3\Delta (t'+1)$ and $3\Delta (t'+1) + \Delta$, respectively, an awake honest validator, or client (who must have been awake since at least $3\Delta t' + 2\Delta$) observes that at every iteration of the fork choice, every block that conflicts with $B$ has less slot $t'$ votes in its subtree (and on itself) than a $\phi$ fraction of the total number of slot $t'$ votes in the bvtree. 
Thus, the fork choice rule returns a block that is consistent with $B$.

Now, let $\ch_1$ and $\ch_2$ denote the two chains confirmed by some clients $k_1$ and $k_2$ at slots $t_1$ and $t_2 \geq t_1$ respectively.
Note that the slot interval $[t_1-\kappa,t_1]$ has at least one slot $t$, where an honest validator with proposed block $B$ is recognized as the slot leader
by all awake honest validators at round $3\Delta t + \Delta$, and, by the arguments above, no block that is not consistent with $B$ is ever identified by any awake honest validator's or client's fork choice rule in rounds $r \geq 3\Delta t + 2\Delta$.
Now, as $t \geq t_1 - \kappa$, but by Goldfish’s confirmation rule, blocks in $\ch_1$ are from no
later than $t_1 - \kappa$, $\ch_1$ is in the prefix of $B$.
Moreover, by the earlier argument, $\ch_2$ is consistent with $B$.
Therefore, $\ch_1$ and $\ch_2$ are consistent.

\textbf{Liveness:}
Suppose $\beta < 1-\phi$.
Then, liveness follows from~\cite[Theorems 1, 2 and 3]{goldfish}, which hold given~\cite[Lemmas 1 and 2]{goldfish} and~\Cref{lemma:lemma-3-remastered} part (i), the latter implying the same result as~\cite[Lemmas 3]{goldfish}.
\end{proof}

\subsubsection{Achievability for \SLeepy \PAssive Clients (Liveness-Favoring)}
\label{sec:proofs-dynamic-sleepy-passive-ach-live}

Similar to~\Cref{sec:passive-ach-live-static}, we describe a family $\Pi_{\mathrm{live}}^\phi$ of protocols,
$\phi \in (0,1/2]$, such that $\Pi_{\mathrm{live}}^\phi$ is live with resilience $\betaL \leq 1-\phi$ and safe with resilience $\betaS < \phi$.
The protocol $\Pi_{\mathrm{live}}^\phi$ is very similar to its counterpart in~\Cref{sec:passive-ach-live-static}, consisting of an \emph{internal protocol} $\Piint$ and a \emph{liveness queue}.
The internal protocol can be any SMR protocol that provides all resiliences $\betaS < 1/2,\ \betaL < 1/2$ under synchrony and \dynpart (\eg,~\cite[Section 4]{sleepy}).
To determine whether a transactions $\tx$ should be added to the liveness queue, validators 
observe the number of signatures on $\tx$ and of validators that are believed to be awake; either because they signed $\tx$, or recently announced they are awake.
Validators add $\tx$ to be liveness queue if their fraction is $\phi$ or more.

Let $u_\mathrm{int}$ be the liveness parameter of $\Piint$.
Each honest awake validator $v$ participates in the internal protocol.
At every round $r$ that is a multiple of $\Delta$, \ie, $r = \ell \Delta$ for some $\ell \in \mathbb{Z}$, $v$ also does the following:
If it has received a transaction $\tx$ from the environment (or the other validators) for the first time within the rounds $((\ell-1)\Delta, \ell \Delta]$, it sends $\tx$ and a signature on it to all parties (including clients).
It also signs and sends the number `$\ell$' as a heartbeat message.

Each client locally maintains a liveness queue and an internal log $\LOGint$.
At every round $r = \ell \Delta$ for some $\ell \in \mathbb{Z}$, each client $k$ calculates the tally $T_\tx$ of signatures observed for each transaction $\tx$.
It also calculates the number $T_{\ell-1}$ of unique validators that have either sent a signature on the number $\ell-1$ or a signature on some transaction in the past.
Then, if $T_\tx / T_{\ell-1} \geq \phi$, $k$ adds $\tx$ to its liveness queue.
To output its log at a round $r$, the client $k$ appends its liveness queue to its internal log with the delay $u_\mathrm{int}$ as in~\Cref{sec:passive-ach-live-static}.

\begin{proof}[Proof of~\Cref{thm:sleepy-passive-da-achievability-2}]
Recall the definition of $\beta$ from~\Cref{def:resilience}.

\textbf{Liveness:} Suppose $\beta \leq 1-\phi$, and consider a transaction $\tx$ input to an honest validator for the first time at some round $r \in ((\ell-1)\Delta, \ell\Delta]$.
At round $(\ell+1)\Delta$, $\tx$ gathers signatures from all honest validators awake at round $(\ell+1)\Delta$, and all clients observe these signatures by round $(\ell+2)\Delta$.
Then, for any $T_\tx$ and $T_{(\ell+1)\Delta}$ in a client's view at round $(\ell+2)\Delta$, it holds that $T_\tx / T_{(\ell+1)\Delta} \geq 1-\beta \geq \phi$.
Therefore, all clients awake at round $(\ell+2)\Delta$ add $\tx$ to their liveness queues.
Every transaction added to the liveness queue of a client at some round $r'$ is output as part of its log by round $r'+u_\mathrm{int}$.
Hence, $\tx$ is output as part of all clients' logs by round $u_\mathrm{int} + (\ell+2)\Delta$, implying that $\Pi_{\mathrm{live}}^\phi$ satisfies liveness with parameter $u_\mathrm{int} + 3\Delta$ and resilience $1-\phi$.

\textbf{Safety:} Suppose $\beta < \phi$.
Then, the internal protocol is safe and live with parameter $u_\mathrm{int}$ as $\phi \leq 1/2$.
Any transaction $\tx$ added to the liveness queue of a client $k$ at some round $(\ell+1)\Delta$ must have been signed by $\phi T_\ell$ or more validators for the value of $T_\ell$ in $k$'s view.
Now, $T_\ell$ is the same or larger than the size of the set that contains all honest validators awake at round $\ell \Delta$ and all adversary validators whose signatures on $\tx$ were received by $k$.
Let $H$, $A$, $\tilde{A}$ respectively denote the numbers of (i) the honest validators awake at round $\ell \Delta$, (ii) the adversary validators whose signatures on $\tx$ were received by $k$, and (iii) the remaining adversary validators.
Then, $\phi T_\ell > \beta T_\ell \geq \beta (H+A) = \beta (H+A+\tilde{A}) - \beta \tilde{A} \geq A + (1-\beta) \tilde{A}$, since $\beta (H+A+\tilde{A}) \geq A + \tilde{A}$ by the definition of $\beta$.
This implies $\phi T_\ell - A > 0$, \ie, one of the signatures on $\tx$ received by $k$ is by an honest validator.
Thus, $\tx$ would be input to the internal protocol $\Piint$ by round $(\ell+1)\Delta$, and by liveness, output as part of the internal log $\LOGint$ by round $(\ell+1)\Delta+u_\mathrm{int}$.
As $k$ attempts to append $\tx$ to $\LOGint$ for the first time at round $(\ell+1)\Delta+u_\mathrm{int}$, and $\tx$ appears as part of $\LOGint$ by round $(\ell+1)\Delta+u_\mathrm{int}$, $\tx$ is not added to the tip of $\LOGint$.
Therefore, if $\beta < \phi$, no transaction added to the liveness queue of an honest client is appended to the tip of its internal log, implying that each honest client outputs its internal log as it is.
Finally, safety follows from the safety of the internal protocol.
\end{proof}

\subsection{\SLeepy \ACtiv Clients}
\label{sec:proofs-dynamic-sleepy-active}

\subsubsection{Impossibility for \SLeepy \ACtiv Clients}
\label{sec:proofs-dynamic-sleepy-active-imp}

\begin{proof}[Proof of~\Cref{thm:sleepy-active-da-converse}]
    Proof is by contradiction.
    Suppose there exists a protocol with resiliences $\betaS = \beta$ and $\betaL = 1 - \beta \leq \betaS$ for some $\beta \in [1/2,1]$.
    Let $P$ and $Q$ denote disjoint sets of $\beta n$ and $(1-\beta) n$ validators.
    Consider the following four worlds:
    
    \textbf{World 1:} Validators in $P$ are honest and awake, and those in $Q$ are asleep.
    There is a single client $k_1$.
    The environment inputs a single transaction $\tx_1$ to the validators in $P$ at time $0$.
    By liveness, $k_1$ outputs $\tx_1$ (and no other transaction from the environment) as part of its log by time $u$.
    
    \textbf{World 2:} Validators in $Q$ are honest and awake, and those in $P$ are asleep.
    There is a single client $k_2$.
    The environment inputs a single transaction $\tx_2$ to the validators in $Q$ at time $0$.
    By liveness, $k_2$ outputs $\tx_2$ (and no other transaction from the environment) as part of its log by time $u$.
    
    \textbf{World 3:}
    Validators in $P$ are honest and awake, and those in $Q$ are adversary.
    There are two honest clients, $k_1$ and $k_3$, and the adversary simulates a client $k_2$.
    Client $k_3$ joins the protocol at round $u$.
    The environment inputs a single transaction $\tx_1$ to the validators in $P$ at time $0$.
    
    Client $k_2$ and the validators in $Q$ do not communicate with the client $k_1$ and the validators in $P$.
    Thus, for $k_1$, world 3 is indistinguishable from world 1, and it outputs $\tx_1$ (and no other transaction from the environment) as part of its log by time $u$.
    In the meanwhile, $k_2$ and the validators in $Q$ start with transaction $\tx_2$, and emulate the execution in world 2 until round $u$.
    
    Once $k_3$ joins the protocol at round $u$, $k_2$ and the validators in $Q$ emulate towards $k_3$ the behavior of the honest validators (in $Q$) and the client $k_2$ in world 4.
    In other words, they pretend like honest validators and an honest client who have been shunned by the validators in $P$ and client $k_1$.
    Since $|Q| / n \leq \betaL$, by liveness, $k_3$ outputs $\tx_1$ as part of its log by time $2u$.
    
    \textbf{World 4:}
    Validators in $Q$ are honest and awake, and those in $P$ are adversary.
    There are two honest clients, $k_2$ and $k_3$, and the adversary simulates a client $k_1$.
    Client $k_3$ joins the protocol at round $u$.
    The environment inputs a single transaction $\tx_2$ to the validators in $Q$ at time $0$.
    
    Client $k_1$ and the validators in $P$ do not communicate with the client $k_2$ and the validators in $Q$.
    Thus, for $k_2$, world 4 is indistinguishable from world 2, and it outputs $\tx_2$ (and no other transaction from the environment) as part of its log by time $u$.
    In the meanwhile, $k_1$ and the validators in $P$ start with transaction $\tx_1$, and emulate the execution in world 1 until round $u$.
    
    Once $k_3$ joins the protocol at round $u$, $k_1$ and the validators in $P$ emulate towards $k_3$ the behavior of the honest validators (in $P$) and the client $k_1$ in world 3.
    In other words, they pretend like honest validators and an honest client who have been shunned by the validators in $Q$ and client $k_2$.
    
    Finally, note that worlds 3 and 4 are indistinguishable by $k_3$ with overwhelming probability, since the validators and clients send the same messages in both worlds.
    Therefore, $k_3$ outputs the same log as in world 3, which contains $\tx_1$.
    Now, if the first transaction in $k_3$'s log is $\tx_1$, this implies a safety violation in world 4, since $k_2$ outputs the log $[\tx_2]$ by round $u$ in world 4.
    This is a contradiction since the protocol must have been safe, as $|P| / n = \beta = \betaS$.
    On the other hand, if the first transaction in $k_3$'s log is not $\tx_1$, this implies a safety violation in world 3, since $k_1$ outputs the log $[\tx_1]$ by round $u$ in world 3.
    This is a contradiction again, since the protocol must have been safe, as $|Q| / n = 1-\beta \leq \betaS$.
\end{proof}

\subsection{\ALert \ACtiv Clients}
\label{sec:proofs-dynamic-alert-active}

\subsubsection{Achievability for \ALert \ACtiv Clients}
\label{sec:proofs-dynamic-alert-active-ach}

We show that the SMR protocol based on Dolev-Strong (\cref{sec:alert-active-static-ach}, \cref{alg:smr-using-dolev-strong}) achieves any $\betaL<1, \betaS<1$ even under \dynpart.
Under \dynpart with \alert \activ clients, 
when a majority of the awake validators are adversary, the clients must output safe and live logs even though the validators themselves may not agree on a log (since validators are \sleepy and \activ, the impossibility in \cref{fig:resilience-dynamic-sleepy-active} applies to them).
Thus, the challenge is to design the validator's code to behave correctly even without knowing what happened while it was sleeping.

This challenge resolves itself due to the following observations.
First, while the SMR protocol (\cref{alg:smr-using-dolev-strong}) runs instances of Dolev-Strong one after the other, each instance does not depend on the previous instances.
Second, within an instance, since the Dolev-Strong protocol (\cref{alg:dolev-strong-clients}) guarantees agreement and validity
when all but one validator are adversary under \statpart(\cref{lem:dolev-strong-clients-security}),
it does so even when only one validator is honest and awake throughout the instance (all honest validators who sleep could be considered adversary).
Moreover, we don't even require the same honest validator to be awake throughout the instance but only require that for each round during the instance, \emph{some} honest validator is awake.
Finally, since each validator (including the leader) signs only one message per instance, it may sleep after it does so without affecting the protocol's remaining execution.
Thus, sleepy validators can faithfully run \cref{alg:dolev-strong-clients}.

\begin{proof}[Proof of~\Cref{thm:alert-active-dynamic-ach}]
    For any $\betaL, \betaS< 1$, we know that at any given round $r$, there is at least one honest node awake.
    First, we will prove that \cref{alg:dolev-strong-clients} satisfies agreement and validity (\cref{def:byzantine-generals}), where the definition of validity is modified to require an honest \emph{and awake} leader.
    Then, using that, we will show that \cref{alg:smr-using-dolev-strong} satisfies safety and liveness.
    
    Validity:
    Suppose the leader $\ell$ is awake and honest and \BGbroadcasts value $v_{\ell}$.
    By round $\Delta$, all clients receive $\signedmsg{\ell}{v}$.
    Thus, the condition in \cref{line:dolev-strong-clients-round-check} is true, all clients add $v$ to their set of candidate output values $\Vout$.
    Moreover, since the leader is honest and signatures are unforgeable, no party ever receives $\signedmsg{\ell}{v'}$ for $v' \neq v$. Therefore, $\Vout = \{v\}$ and the client \BGoutputs $v$ at the end of round $(2n+1)\Delta$.
    
    Agreement:
    Let's refer to a message of the form $\signedmsg{j_k}{\signedmsg{j_{2}}{\signedmsg{j_1}{v}}...}$ as a $k$-signature chain on $v$.
    First, we show that if a client $c$ adds a value $v$ to $\Vout$, then all other clients do so too.
    Since client $c$ added $v$ to $\Vout$, for some $k \leq n$, it received a $k$-signature chain $m$ on $v$ by round $(2k-1)\Delta$.
    If at least one of the validators $j_1,...,j_k$ who signed this message is honest,
    then due to \cref{line:dolev-strong-val-round-check}, for some $k' \leq n$, this validator signed the $k'$-th signature in the chain by round $2(k'-1)\Delta$ (when it was awake), so client $c'$ receives a $k'$-signature chain by round $(2k'-1)\Delta$ and thus also adds $v$ to $\Vout$.
    If no signatory of $m$ is honest, then $k \leq n-1$.
    In this case, client $c$ sends $m$ to all parties, so all validators receive $m$ by round $2k\Delta$.
    Note that a validator can check the condition in \cref{line:dolev-strong-val-round-check} using $m$ and knowledge of the leader, validator set, and current round, even if it has been sleeping earlier.
    Since the condition is satisfied, an honest awake validator (who exists and has not yet signed by assumption) signs $m$, and so client $c'$ receives a $(k+1)$-signature chain by round $(2k+1)\Delta \leq (2n-1)\Delta$. So, client $c'$ also adds $v$ to $\Vout$.
    Agreement follows since we have established that all clients have the same set $\Vout$ at the end of round $(2n+1)\Delta$.
    
    For the \smrshort protocol (\cref{alg:smr-using-dolev-strong}), safety again follows immediately from agreement.
    Liveness follows from validity since an honest awake validator will \BGbroadcast a block containing all valid transactions it has received.
\end{proof}

\subsubsection{Impossibility for \ALert \ACtiv Clients}
\label{sec:proofs-dynamic-alert-active-ach-impossibility}

We conclude by showing that it is impossible to achieve safety and liveness resiliences of exactly $1$ simultaneously.

\begin{theorem}
In a synchronous network with \alert validators and \alert \activ clients, no protocol can achieve resiliences $(\tL, \tS)$ such that $\tL = \tS = 1$.
\end{theorem}

\begin{proof}
Proof is by contradiction. 
Suppose there exists a protocol $\Pi$ with resiliences $\tS = \tL = 1$.
Consider the world (called world 1), where all validators are adversary and crashed, and there are $n'$ clients (at least two of which are honest).
By assumption, the protocol satisfies safety and liveness, even though the adversary can simulate any (polynomial) number of clients.

Next, consider a world (called world 2) with a synchronous network and $n'$ \alert validators that are connected by authenticated channels.
These validators simulate $n$ other crashed validators in their head, and run the protocol $\Pi$ above, assuming the simulated validators are crashed.
Even though they are connected via authenticated channels, they can run the protocol $\Pi$; since the communication among the $n'$ clients in world 1 can be emulated by the $n'$ validators in world 2.
By the assumption above, safety and liveness are satisfied for these validators in world 2, even though the adversary can simulate $n+f$ validators for any constant $f \geq n/3$.
However, this contradicts with the well-known FLM'85 impossibility result~\cite{FLM85}, implying that the protocol cannot be safe and live in world 1.
\end{proof}

%% file: algorithms/alg_modified_goldfish_forkchoice.tex
\begin{algorithm}[tb]
    \caption{Modified GHOST-Eph fork-choice rule.}
    \label{alg:goldfish-fcr}
    \begin{algorithmic}[1]
        \scriptsize
        \CommentLine{The blocktree is denoted by $\mathcal{T}$.}
        \CommentLine{$\textsc{Children}(\mathcal{T},B)$ returns the set of $B$'s children within $\mathcal{T}$.}
        \CommentLine{$\textsc{Votes}(\mathcal{T}, B,t)$ returns the number of slot $t$ votes by unique validators in the subtree defined by a block $B$ within $\mathcal{T}$.}
        \Function{\sc GHOST-Eph}{$\mathcal{T}, t$}
            \Let{B}{B_0}
            \Comment{Start fork-choice at genesis block}
            \While{$\TRUE$}
                \CommentLine{Choose the subtree that has $\phi$ fraction of the slot $t$ votes from unique validators.}
                \Let{\mathsf{Finish}}{\TRUE}
                \For{$B'\in\operatorname{\textsc{Children}}(\mathcal{T},B)$}
                    \CommentLine{The original Goldfish rule~\cite{goldfish} would have selected the child with the heaviest subtree.}
                    \If{$\textsc{Votes}(\mathcal{T}, B',t) \geq \phi \cdot \textsc{Votes}(\mathcal{T}, B_0,t)$}
                        \Let{B}{B'}  \label{line:vote-count}
                        \Let{\mathsf{Finish}}{\FALSE}
                    \EndIf
                \EndFor
                \If{$\mathsf{Finish}$}
                    \Return $B$
                \EndIf
            \EndWhile
        \EndFunction
    \end{algorithmic}
\end{algorithm}

%% file: sections/09_psync.tex
\section{Partial Synchrony}%
\label{sec:partial-synchrony}

\begin{corollary}
\label{thm:psync-converse}
    Suppose the network is partially synchronous with \statpart. Then for any type of clients, no protocol achieves $(\tL,\tS)$ such that $2\tL+\tS \geq 1$.
\end{corollary}

\begin{corollary}
\label{thm:psync-achievability}
    Suppose the network is partially synchronous with \statpart. Then for any type of clients, for all $q \in (n/2,n]$, HotStuff~\cite{hotstuff} with a quorum size $q$ achieves all $(\tL,\tS)$ with $\tL \leq \frac{n-q}{n}$ and $\tS < \frac{2q-n}{n}$.
\end{corollary}

\begin{corollary}
\label{thm:psync-dynamic-converse}
    Suppose the network is partially synchronous with \dynpart. Then for any $\tS > 0, \tL > 0$, no protocol can achieve $(\tL,\tS)$.
\end{corollary}

\Cref{thm:psync-converse} follows from the `split brain proof'~\cite[Theorem~3.1]{mtbft}, in turn inspired by~\cite{dls88,bkl19}. 
The impossibility is proven for the `no client' model, which as discussed in \cref{sec:related}, is equivalent to \alert \passive clients. Due to \cref{lem:hierarchy}, the impossibility result holds for all client types.
\Cref{thm:psync-achievability} follows from \cite[Theorems~2 and~4]{hotstuff} by replacing the quorum sizes with $q \in (n/2,n]$. 
The protocol 
is proven
secure for \sleepy \passive clients, thus for all other clients too.
Other protocols Streamlet~\cite{streamlet}, Casper FFG~\cite{casperffg}, and Tendermint~\cite{tendermint} can also be used to achieve the same result.
Finally, \cref{thm:psync-dynamic-converse} follows from the `blockchain CAP theorem'~\cite{pye-roughgarden-cap-theorem,aa}.

%% file: sections/A04_stubborn_nakamoto.tex
\section{\emph{Stubborn Nakamoto}}
\label{app:stubborn-nakamoto}

\subsection{A Concrete Attack on the Protocol}

There have been attempts at creating protocols that maintain safety against \emph{all}
adversaries ($\betaS = 1$) and liveness against $1/2$ adversaries ($\betaL = 1/2$) in the setting of Bitcoin.
It has been posited that \activ clients can achieve safety resilience $\betaS=1$. The \emph{Stubborn Nakamoto} protocol, put forth 
in a recent preprint~\cite{leshno-economic-permissionless},
is akin to 
\cref{alg:freezing}.
The model is synchronous, with \dynpart and \sleepy \activ clients.
The protocol~\cite[Def.~3]{leshno-economic-permissionless} is largely identical to \cref{alg:freezing}, but uses Bitcoin as its internal protocol $\Piint$.
The \activ clients, upon receiving a new candidate ledger to be confirmed, in the form of a $k$-deep block in a longest chain,
gossip it and wait $2\Delta$ rounds
before they output it.
However, because the protocol aims to work in the \sleepy validator setting,
the \emph{internal} protocol is not\footnote{Certifiable safety is proven impossible in the `unsized' sleepy validator setting such as proof-of-work~\cite{roughgarden}.} \emph{certifiable}. Concretely, 
the paper's
``certificates''
are proof-of-work blockchains starting at the genesis block and attesting to the confirmation of transactions that have been buried under $k$ blocks. But such ``certificates'' do not satisfy \emph{certifiable safety} in
\cref{def:certifiable,def:certifiable-safety}. The reason 
is that a block's transactions should be output in the log only if the block is $k$-deep \emph{in the longest chain},
but the ``certificate'' only guarantees that the block is $k$-deep \emph{in some chain} and
does not rule out the existence of longer chains.
A first draft of the \emph{Stubborn Nakamoto} paper made claims of security that were not exactly correct, and this is due to the uncertifiability of
$\Piint$.

For a concrete attack, consider a majority mining adversary, and \sleepy \activ clients\footnote{This attack works even on \emph{\alert} \activ clients.}.
Suppose that the adversary mining rate is very high, and consider two honest clients $P_1$ and $P_2$,
initially agreeing on the genesis block. The adversary performs a \emph{balancing attack}. She keeps mining two independent and eternally-growing
chains $C_1$ and $C_2$, aiming for $P_1$ to output $C_1$ and $P_2$ to output $C_2$. Initially, the adversary mines
$k$ blocks on $C_1$ and another $k$ blocks on $C_2$ (before the honest miners manage to mine any $k$-long chain).
At time $t_0$, the adversary simultaneously sends the first $k$ blocks of $C_1$ to $P_1$ and the first $k$ blocks of $C_2$ to $P_2$.
When $P_1$ receives the $k$ blocks of $C_1$, he will gossip them and wait $2\Delta$ rounds before confirming them.
In the meantime, $P_2$ also receives the $k$ blocks of $C_2$ and does the same.
From that point on, the adversary will mine blocks on top of both chains and disseminate one new block of $C_1$ to $P_1$
and one new block of $C_2$ to $P_2$ every $\Delta / 2$ rounds (assume she either has sufficient mining power to keep mining,
or she has premined them in advance).
Before $P_1$ has received $P_2$'s gossiped blocks, the adversary has mined another block on top of $C_1$ and
made it known to $P_1$ at time $t_0 + \Delta / 2$, thereby causing $P_1$ to grow the longest chain in its view.
When $P_1$ receives $P_2$'s gossiped blocks at time $t_0 + \Delta$, the chain received from $P_2$ is no longer candidate
for confirmation, as it is not a longest chain. Therefore, the message from $P_2$ to $P_1$
does not stop $P_1$ from outputting $C_1$. The process continues with both clients having different ever-growing longest chains,
without ever halting. As a result, no matter what confirmation depth $k$ is used, the protocol is unsafe.

Trying to 
patch the protocol to achieve the desired resiliences $\tL<1/2, \tS=1$ for \sleepy \activ clients cannot work due to the impossibility
shown in \cref{thm:sleepy-active-da-converse}.
We have communicated this attack on the previous version of their paper to the authors, and they have acknowledged it. In the latest version of the paper, the security theorem has been corrected in response to our remark to state that security is provided to online clients with a safety resilience 100\% only as long as no majority adversary has made attempts to break safety prior to the sleepy client waking up.

\subsection{Impossibility for Proof-of-Work}

\Cref{thm:sleepy-active-da-converse} proves that with \dynpart and \sleepy \activ clients, for any $\epsilon \in (0,1/2)$, resilience $\tL = \epsilon, \tS \geq 1 - \epsilon$ are impossible.
In particular, $\tL > 0, \tS = 1$ is impossible.
Our model assumed a fixed known set of validators, a number $f$ of which are corrupted.
However, Bitcoin's validator model has two key differences.
First, Bitcoin uses proof-of-work and assumes that each validator has a limited hashing power~\cite{backbone}.
Second, the number of validators, in this case, the total hash rate, is not known (to start with, we may consider it fixed as in the static difficulty Bitcoin model~\cite{backbone}).

On one hand, having unknown number of validators makes Bitcoin's model harder to solve \smrshort than in the \dynpart model.
On the other hand, the adversary's power being determined by its hashing power makes
Bitcoin's model incomparable to our model.
In particular, some impossibility proofs (\eg, \cref{thm:sleepy-active-synchrony-converse}) that use a split-brain attack would not hold in Bitcoin's model because the adversary cannot simultaneously use its hashing power to simulate two different executions (``mine two chains'').

However, the impossibility result in \cref{thm:sleepy-active-da-converse} holds even in Bitcoin's model because it does not use a split-brain attack.
Rather, while honest miners run one execution, the adversary simulates an alternate execution without communicating with honest parties, and 
an adversary with $1-\epsilon$ fraction of the hashing power can simulate an alternate live execution that appears as if it was run by honest miners.
The crux of the proof is that 
a \sleepy client who awakens later in the execution cannot distinguish whether $1-\epsilon$ fraction of hashing power is adversary or $\epsilon$ fraction is adversary.

A model for Bitcoin's setting is described below.
The proof of \cref{thm:sleepy-active-da-converse} follows in exactly the same way under this model too.
The model is the same as that in \cref{sec:app-model-full}, except for the following modifications. 
The proof-of-work is modeled using a permitter oracle~\cite{backbone,pye-roughgarden-permissionless,pye-roughgarden-cap-theorem}.
At every round, each validator can call the permitter oracle at most once\footnote{Recall that a round is an arbitrarily small unit of time. Moreover, without loss of generality, we may assume that each validator has the same hashing power since a validator with higher hashing power may be considered as multiple validators.} with a message $m$ and the oracle responds with $\mathcal{R}(m)$ where $\mathcal{R}$ is a random oracle~\cite{pye-roughgarden-permissionless}.
Following \cite{backbone}, each honest party is allowed unlimited ``verification'' queries to $\mathcal{R}$.